\documentclass[twocolumn]{aastex631}

\begin{document}

\title{2.5-D MHD Simulation of the Formation and Evolution of Plasmoids in Coronal Current Sheets}

\author{Sripan Mondal}

\author{Abhishek K Srivastava}
\affiliation{Department of Physics, Indian Institute of Technology (BHU), Varanasi-221005, India. Email:- asrivastava.app@iitbhu.ac.in}

\author{David I. Pontin}
\affiliation{School of Information and Physical Sciences, University of Newcastle, Australia.}

\author{Ding Yuan}
\affiliation{ Shenzhen Key Laboratory of Numerical Prediction for Space Storm, Institute of Space Science and Applied Technology, Harbin Institute of Technology, Shenzhen, Guangdong, China. Email:- yuanding@hit.edu.cn}

\affiliation{Key Laboratory of Solar Activity and Space Weather, National Space Science Center, Chinese Academy of Sciences, Beijing, China.}

\author{Eric R.~Priest}
\affiliation{Mathematics Institute, St Andrews University, KY16 8QR, St Andrews, UK.}




\begin{abstract}
In the present paper, using \texttt{MPI-AMRVAC}, we perform a 2.5-D numerical MHD simulation of the dynamics and associated thermodynamical evolution of an initially force-free Harris current sheet subjected to an external velocity perturbation under the condition of uniform resistivity.  The amplitude of the magnetic field is taken to be 10 Gauss, typical of the solar corona. We impose a Gaussian velocity pulse across this current sheet mimicking the interaction of fast magnetoacoustic waves with a current sheet in corona. This leads to a variety of dynamics and plasma processes in the current sheet, which is initially quasi-static. The initial pulse interacts with the current sheet and splits into a pair of counter-propagating wavefronts, which forms a rarefied region and leads to inflow and a thinning of the current sheet. The thinning results in Petschek-type magnetic reconnection followed by tearing instability and plasmoid formation. The reconnection outflows containing outward-moving plasmoids have accelerated motions with velocities ranging from 105-303 km \(\mathrm{s^{-1}}\). The average temperature and density of the plasmoids are found to be 8 MK and twice the background density of the solar corona, respectively. These estimates of velocity, temperature and density of plasmoids are similar to values reported from various solar coronal observations. Therefore, we infer that the external triggering of a quasi-static current sheet by a single velocity pulse is capable of initiating magnetic reconnection and plasmoid formation in the absence of a localized enhancement of resistivity in the solar corona.
\end{abstract}



\section{Introduction} \label{sec:1}
Reconnection is a fundamental physical process of a magnetized plasma. It can convert stored magnetic energy into thermal energy,  bulk fluid kinetic energy and charged particle energy at astronomical, space and laboratory scales. (See review articles by \citet{2000mrp..book.....B,2007rmfm.book.....B,2009ARA&A..47..291Z,2011SSRv..160...45U,2015A&ARv..23....4T,2016PPCF...58a4021L,2016PhPl...23e5402Y,2022LRSP...19....1P} and references cited therein for details). Magnetic reconnection takes place when magnetic fields of differing orientations are brought together in close proximity, with a rapid rotation of the magnetic vector across a thin layer of high electric current density  termed as a current sheet (CS) \citep{1953sun..book..532C,1986JGR....91.5579P,2014masu.book.....P,2022LRSP...19....1P}. Due to the presence of high resolution observatories, the large cohort of observational signatures of magnetic reconnection such as bi-directional flows, heating and associated plasma dynamics are ubiquitously seen in the solar corona. However, the detailed nature of CS formation and its evolution resulting in magnetic reconnection and solar eruptions are yet to be fully understood. Therefore, an important way to extract the detailed physics behind magnetic reconnection under various plasma conditions, such as those in the solar corona, is to numerically simulate or model the formation and evolution of a CS. To model the evolution of a CS and the resulting magnetic reconnection, either internal perturbations such as local resistivity enhancement or external perturbations in the form of MHD waves are needed. Use of MHD waves to perturb the CS gives rise to externally driven or forced reconnection. The onset and features of different types of magnetic reconnection and the advances in  modeling  the reconnection scenario are discussed in various review articles cited above.

Consider a thin, elongated CS in the solar corona. When its aspect ratio (i.e., the ratio of length to width of the CS) is sufficiently large ($\gtrsim 100$), then beyond a critical value of Lundquist number \((S_{L}= LV_{A}/\eta > 3 \times 10^{4}\))  three distinct phases of nonlinear evolution of reconnection can be characterized \citep{2009PhPl...16k2102B,2012PhPl...19i2110B}. In the initial stage, the CS system evolves to a quasi-steady but transient state having Sweet-Parker like characteristics with an extended, thin CS. After this stage, a rapid secondary instability sets in to produce plasmoids or magnetic islands along the fragmented CS. Lastly, the system evolves to a nonlinear phase in which it exhibits rapid and impulsive reconnection. In this third and final stage, some of the smaller islands coalesce to form larger islands that are eventually convected outwards through the CS. The formation of plasmoids results in a higher reconnection rate under coronal conditions than the one predicted by the Sweet-Parker model. It is similar to the normalized reconnection rate obtained in the Petschek model, i.e., \(\pi/[8\ln S]\). Since the solar corona has a high value of \(S_{L}\) (typically $10^{12}-10^{14}$), the growth of the plasmoid instability in an elongated CS is expected during the onset of solar flares. Therefore, the plasmoid instability is considered as a viable mechanism to explain the high reconnection rate reported from the observations of flares. As a result, the plasmoid instability has been studied extensively in the literature via numerical simulations \citep[e.g.,][]{2009PhPl...16k2102B,2010PhPl...17f2104H,2012PhPl...19i2110B,2012PhRvL.109z5002H,2013PhPl...20e5702H,2015shin.confE..26H,2016ApJ...818...20H}. At the same time,  the advent of high-resolution observational facilities has given rise to several observations of moving plasma blobs along an elongated current sheet. Also, jets and flows, which are the dynamical response of the magnetic reconnection, have been observed. They have, therefore, confirmed observationally the formation and motion of the plasmoids found in the simulations \citep[e.g.,][]{2001ApJ...555L..65A,2009ApJ...698L..51L,2012ApJ...745L...6T,2014A&A...561A.134Z,2016SoPh..291..859Z,2018ApJ...866...64C,2022FrASS...8..238C,2022NatCo..13..640Y}. \citet{2018ApJ...868..148L} inferred that superarcade downflows in elongated current sheet may well be evidence of plasmoids. But later on, \citet{2022NatAs...6..317S} suggested that these dark downflows are basically self-organized structures generated during the interaction of reconnection outflows with the flare arcade.

In the dynamic solar corona, a CS can be easily perturbed by photospheric footpoint motions or MHD waves, potentially leading to magnetic reconnection. For example, a flare can be initiated by another distant flare via superthermal particle beams and shock waves propagating from the earlier flare site \citep{1978A&A....68..145N,1981sfmh.book...47S,1983SoPh...89..355F}. Observations of this kind of sympathetic flare initiated by MHD waves or shock waves have also been reported in the literature \citep{2001ApJ...559.1171W,2018ApJ...860...54O,2020ApJ...905..150Z}. A few numerical simulations have studied the external forcing of reconnection and tearing mode instability by fast-mode MHD waves or shocks \citep{1982ApJ...258..823S,1983JPlPh..30..109S,1997AdSpR..19.1895O,2019A&A...623A..15P}. However, all of these studies have implemented an anomalous resistivity to facilitate fast reconnection. Therefore the role of external forcing on the different stages of reconnection in the absence of a localized  resistivity enhancement has not yet been studied in a comprehensive manner, which is what we are attempting here. Now both current sheets and EUV waves are likely to be commonly present in the solar corona \citep[e.g.,][]{2014MNRAS.444.1119Z,2017SCPMA..60b9631C,2018ApJ...866...64C,2020ApJ...900..192F}. Therefore it is important to study this aspect of external forcing more carefully with specific initial physical conditions such as typical solar coronal current sheets and fast external velocity disturbances.

In our present paper, we investigate  the physical role of transient external velocity perturbations on the dynamics of an initially quasi-stable CS. We do not incorporate a localized enhancement of resistivity  \citep[as in][]{1997A&A...326.1252O, 2019A&A...623A..15P}. Rather, we adopt a uniform resistivity throughout the simulation domain. Since the generation of transient velocity disturbances is likely to be impulsive rather than periodic in practice, we prescribe an initial, uni-directional velocity pulse rather than imposing periodic perturbations. The detailed physical scenario for the different phases of the dynamics of a coronal CS is investigated. The formation, evolution and physical characteristics of the plasmoids are studied. We present the numerical setup of the model in section 2 followed by results in section 3. In section 4, we summarize the results and discuss a few key points related to physical feasibility of the dynamics. We compare with  observations and consider the effect of the velocity amplitude of the disturbances on CS dynamics. In section 5, we conclude our work, emphasizing  its importance as a comprehensive study of externally driven reconnection under solar coronal conditions.

\begin{figure*}

\mbox{
\hspace{-1.5 cm}
\includegraphics[height=5.5 cm,trim={11cm 0 11cm 0},clip]{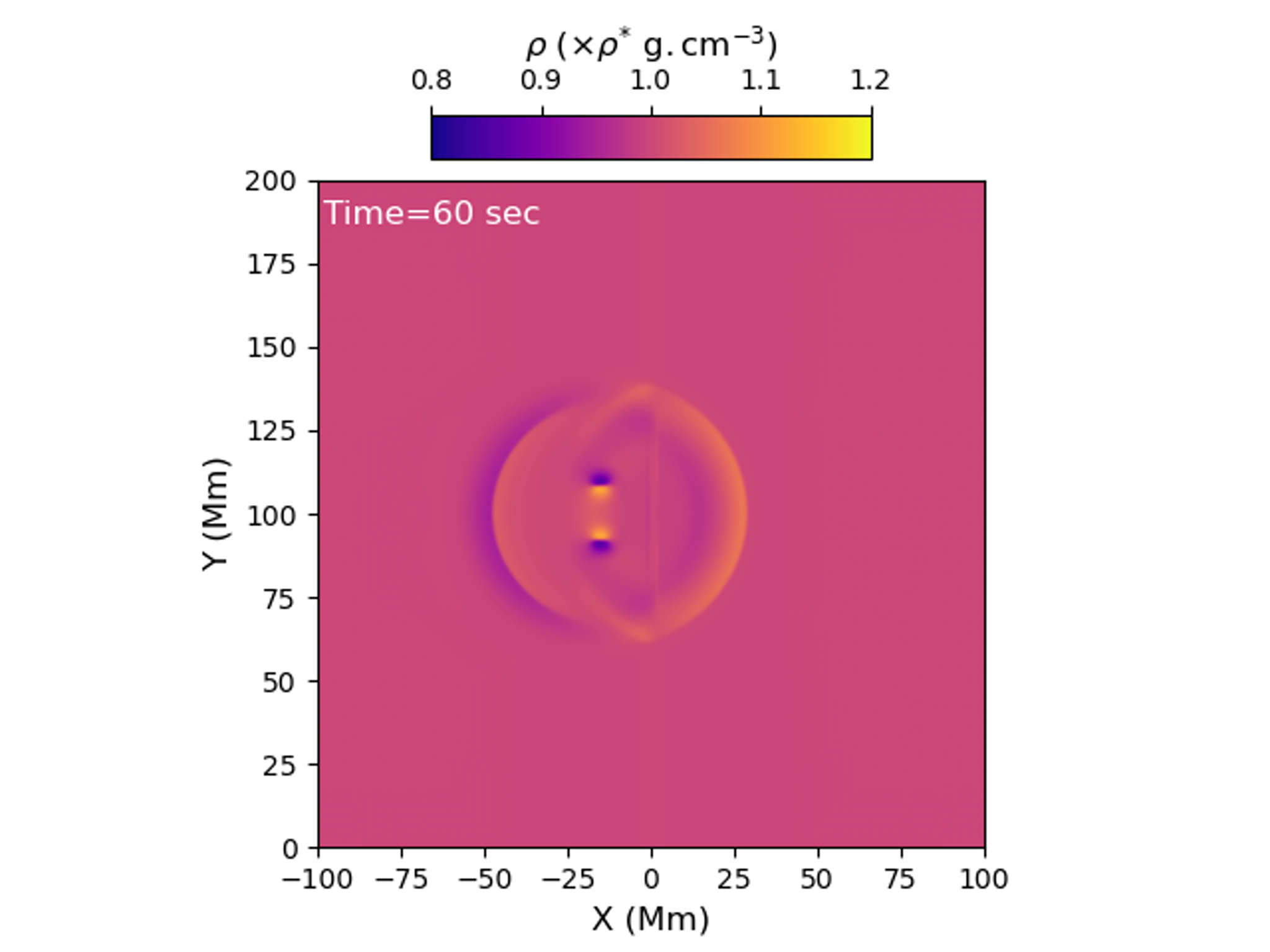}
\includegraphics[height=5.5 cm,trim={11cm 0 11cm 0},clip]{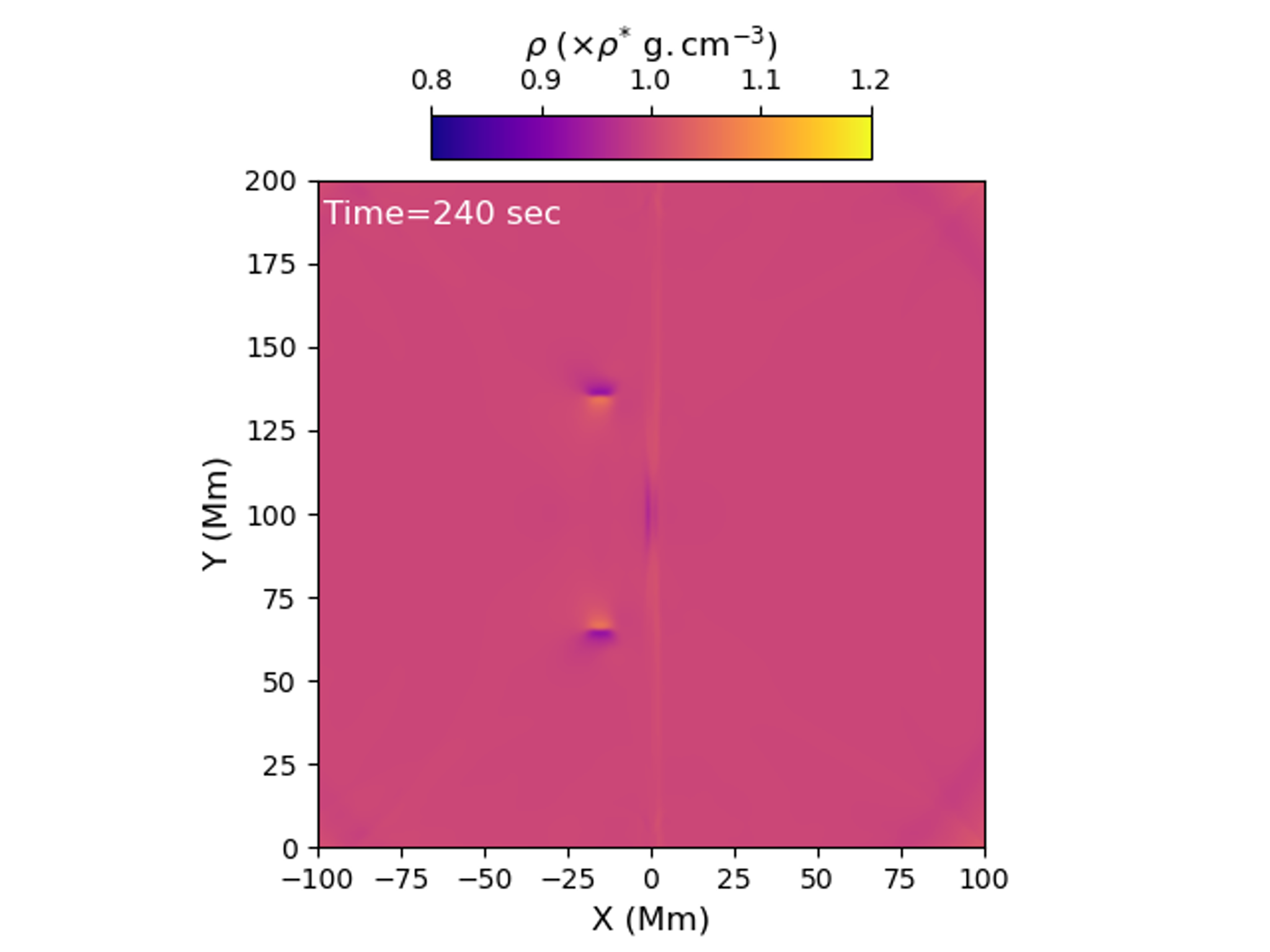}
\includegraphics[height=5.5 cm,trim={11cm 0 11cm 0},clip ]{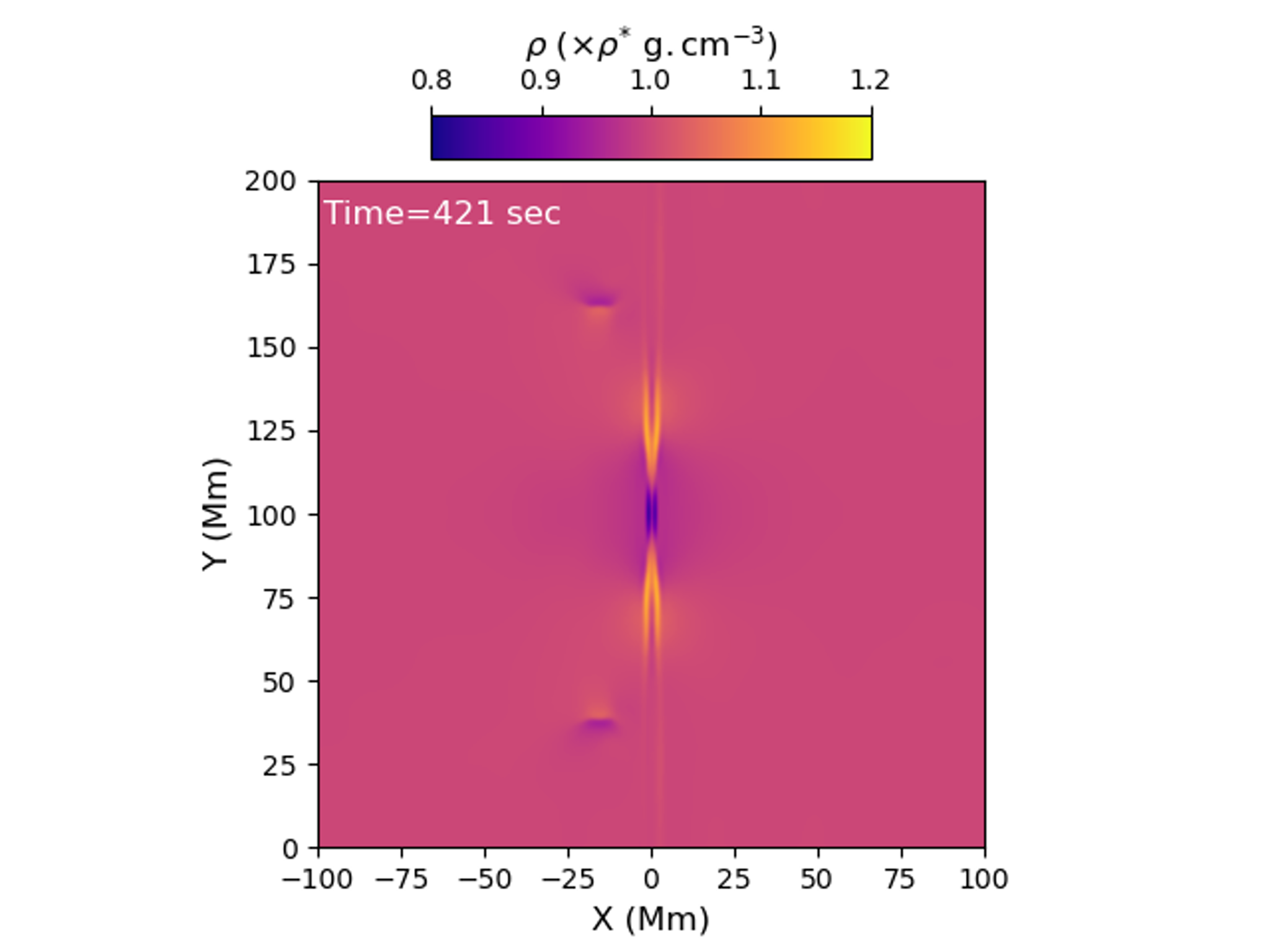}
\includegraphics[height=5.5 cm,trim={11cm 0 11cm 0},clip ]{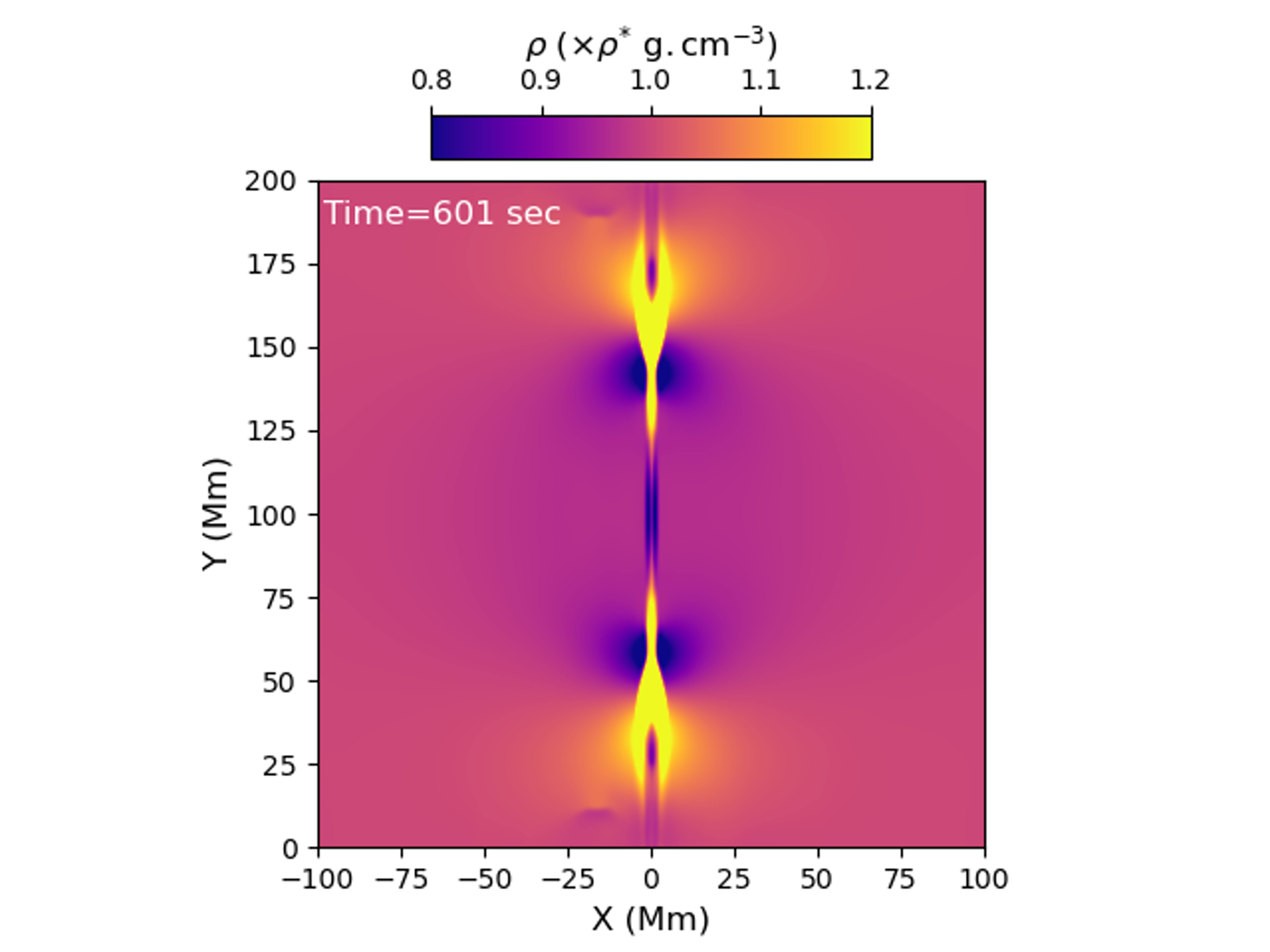}
}

\mbox{
\hspace{-1.5 cm}
\includegraphics[height=5.5 cm,trim={11cm 0 11cm 0},clip]{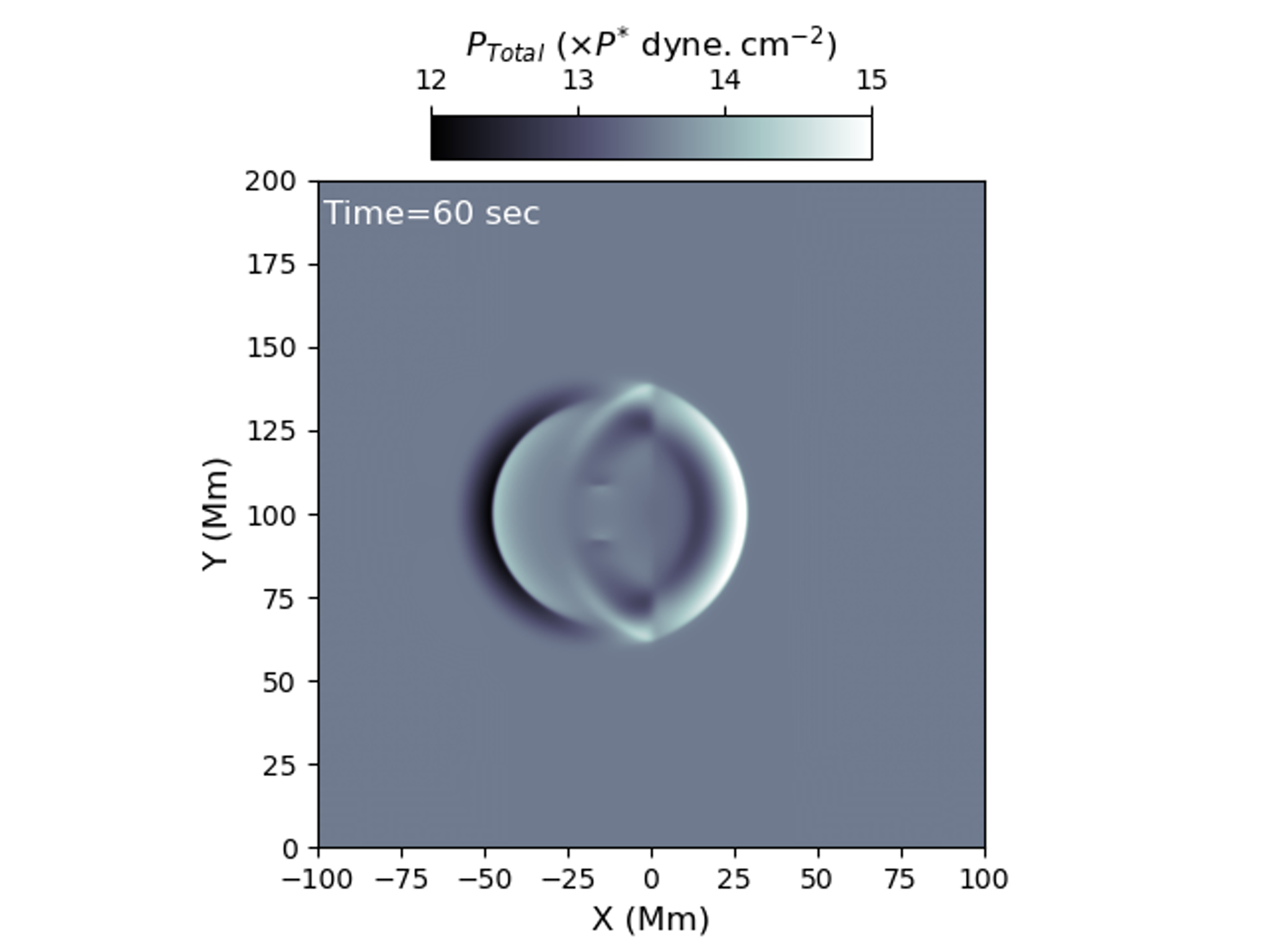}
\includegraphics[height=5.5 cm,trim={11cm 0 11cm 0},clip]{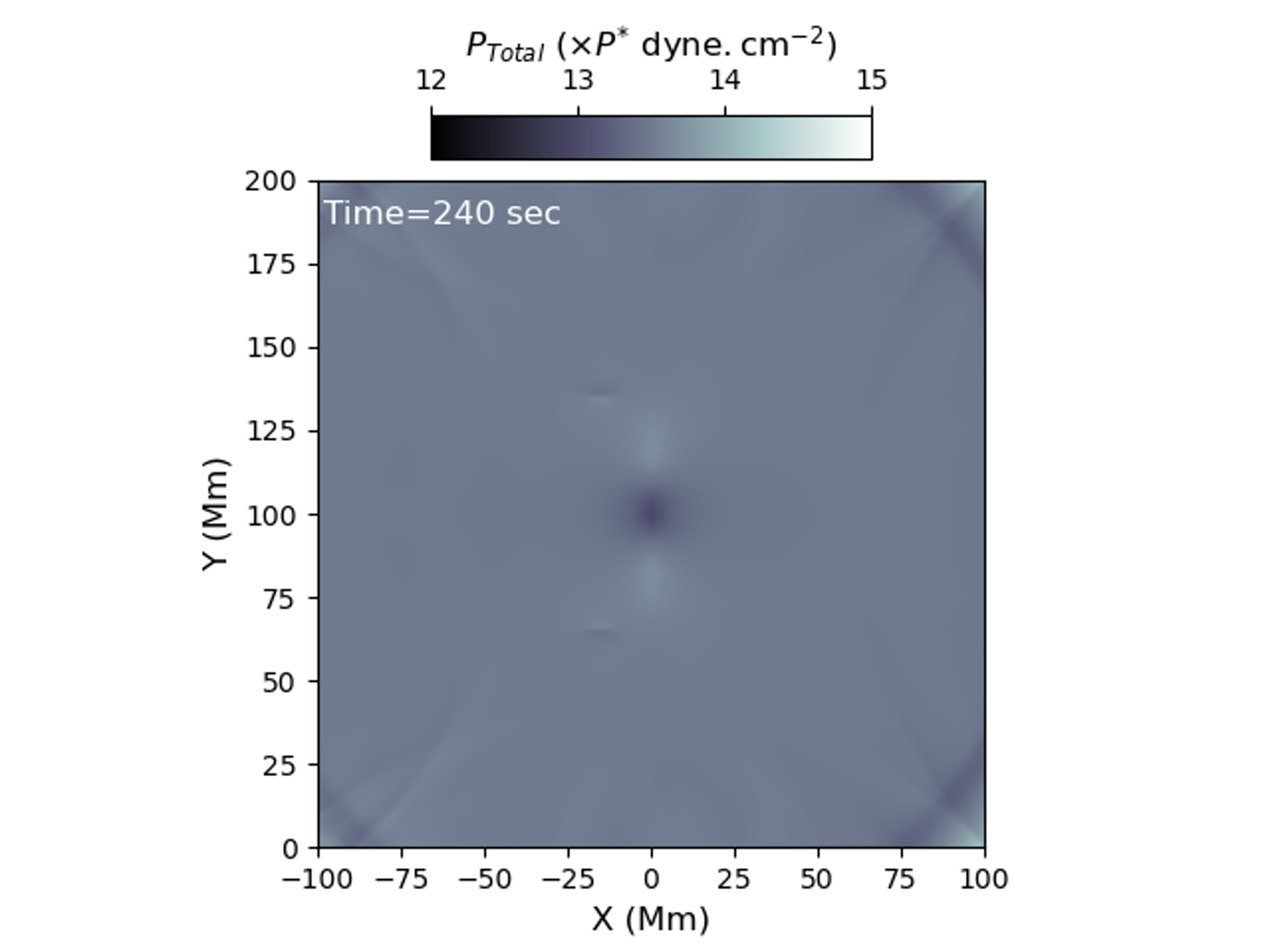}
\includegraphics[height=5.5 cm,trim={11cm 0 11cm 0},clip]{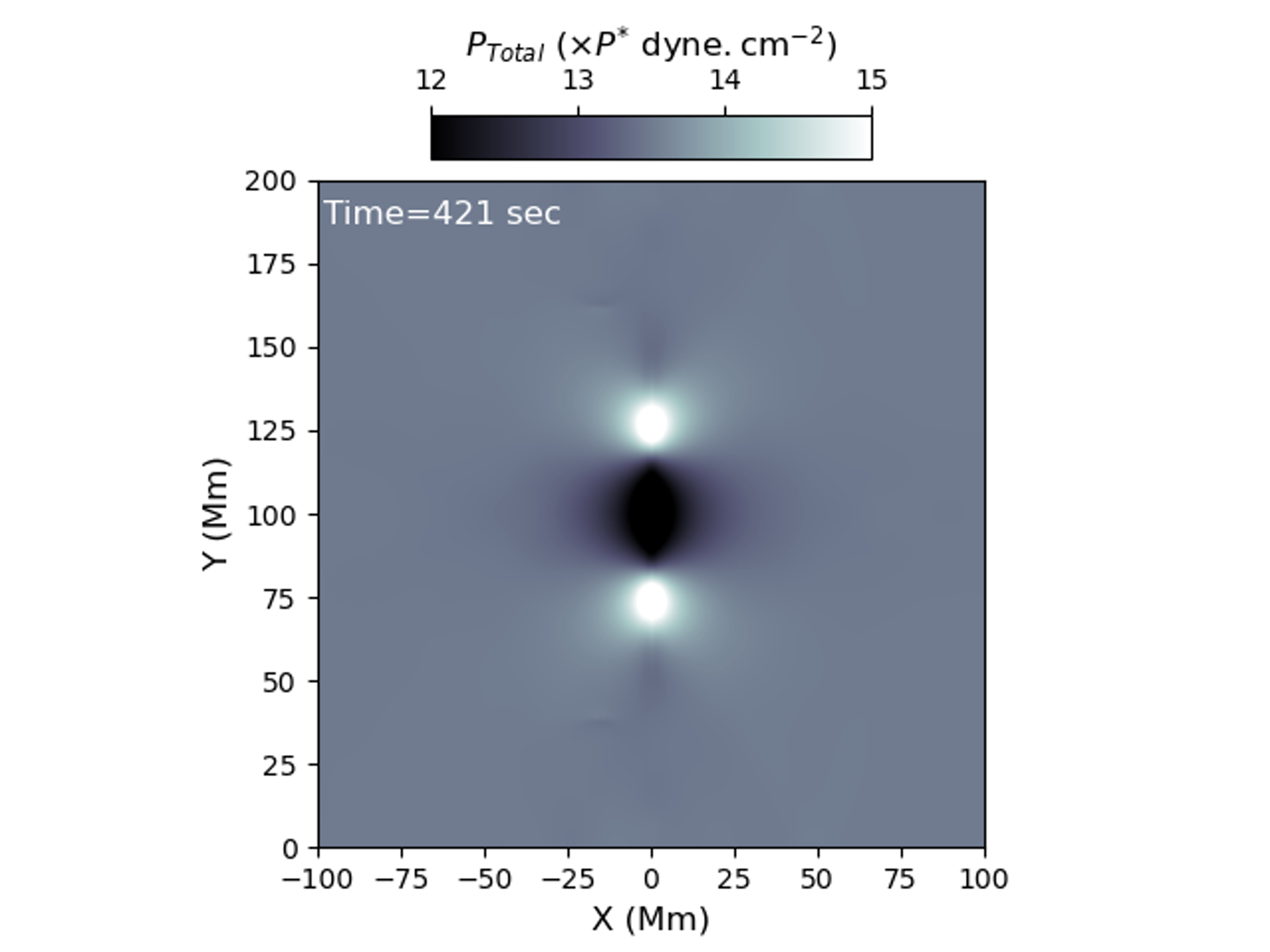}
\includegraphics[height=5.5 cm,trim={11cm 0 11cm 0},clip]{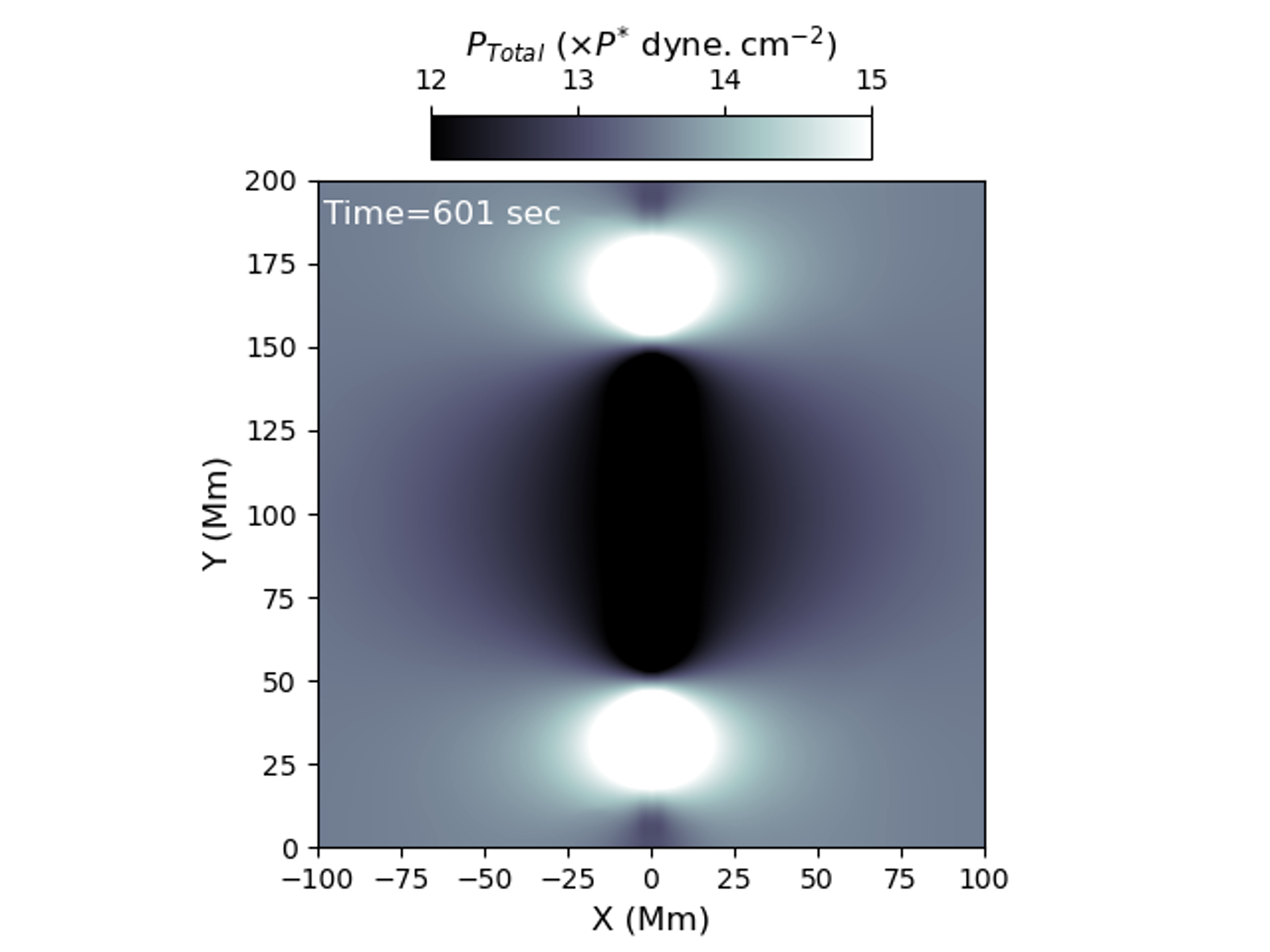}
}

\mbox{
\hspace{-1.5 cm}
\includegraphics[height=5.5 cm,trim={11cm 0 11cm 0},clip]{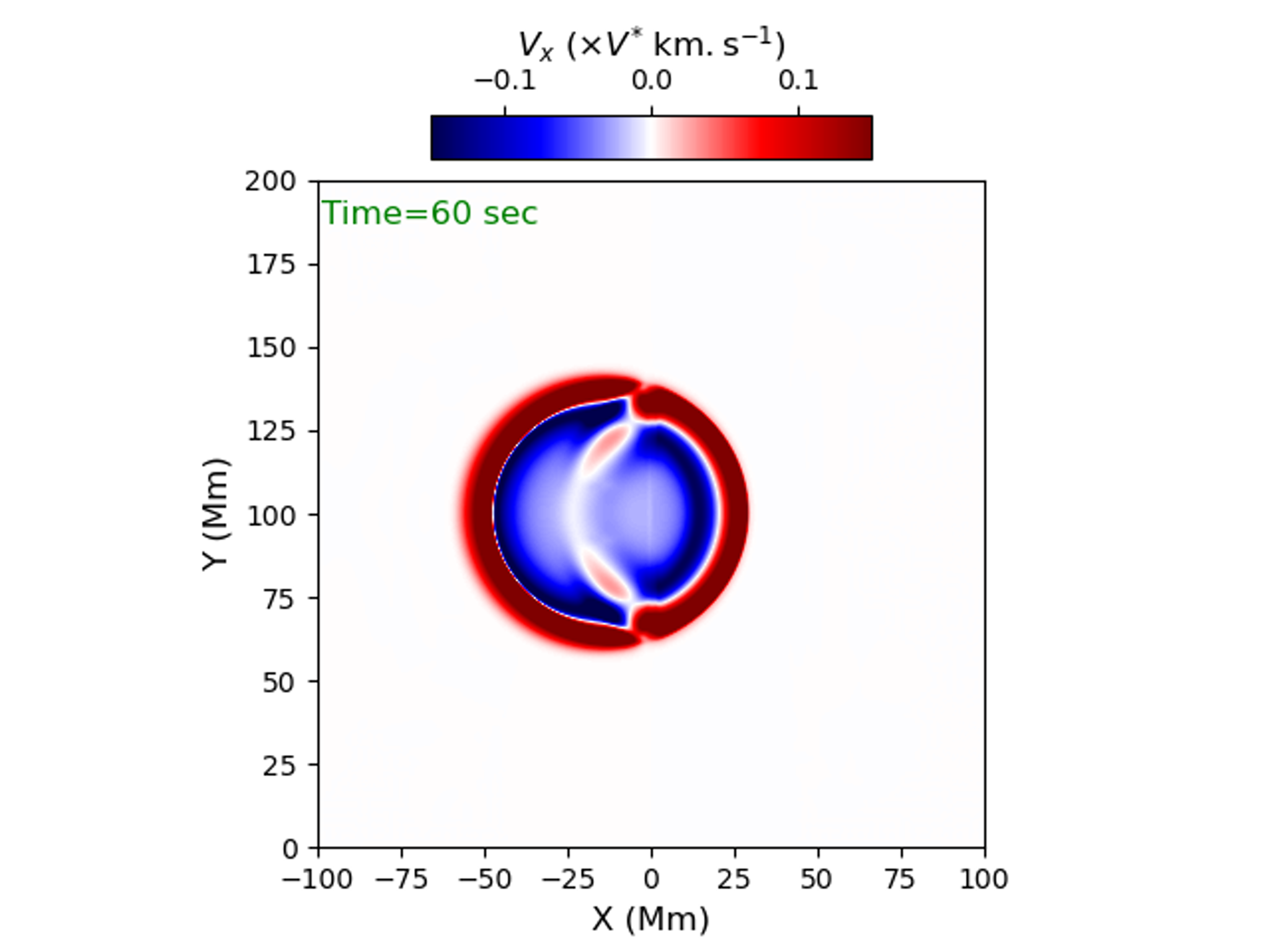}
\includegraphics[height=5.5 cm,trim={11cm 0 11cm 0},clip]{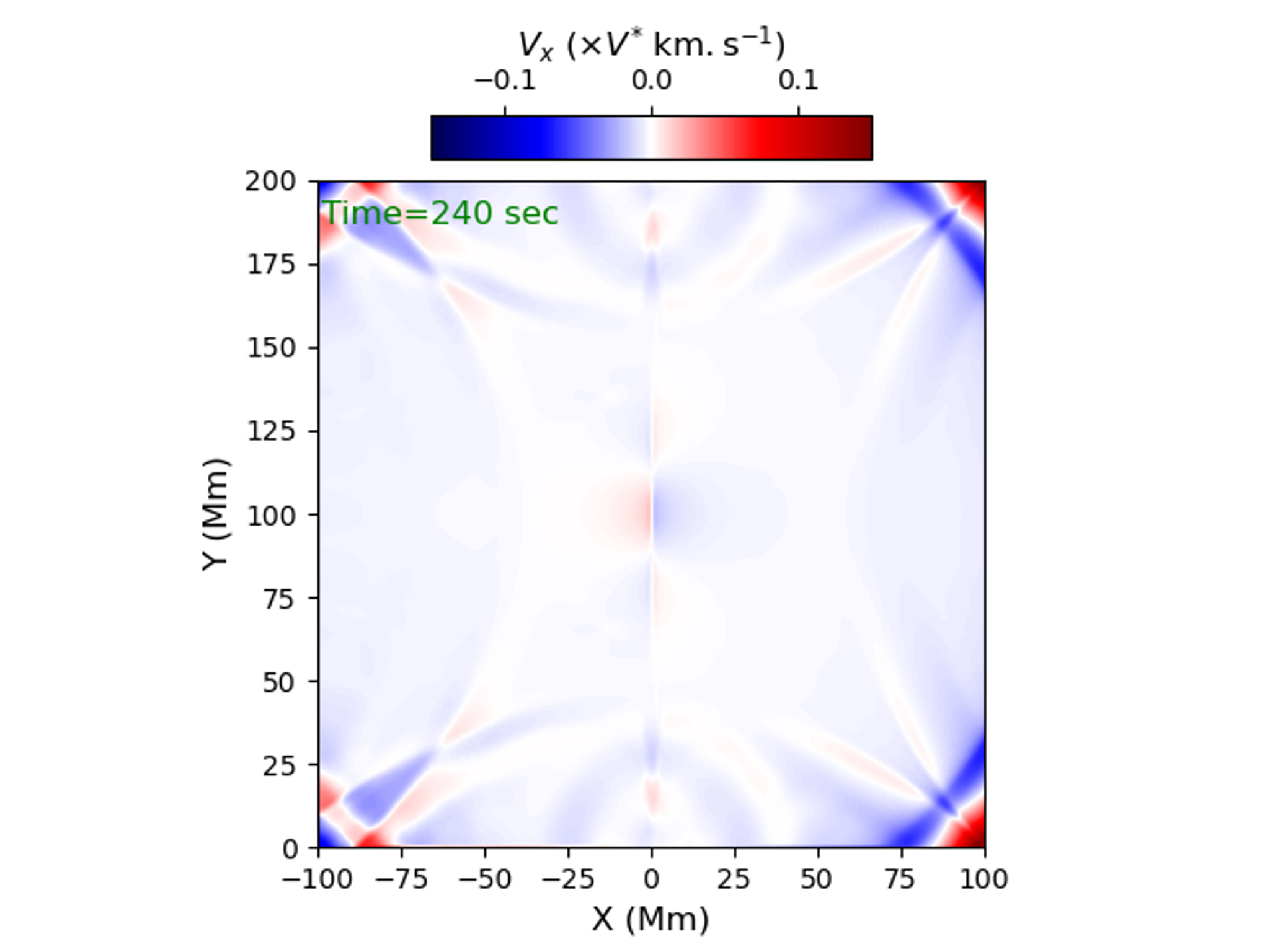}
\includegraphics[height=5.5 cm,trim={11cm 0 11cm 0},clip ]{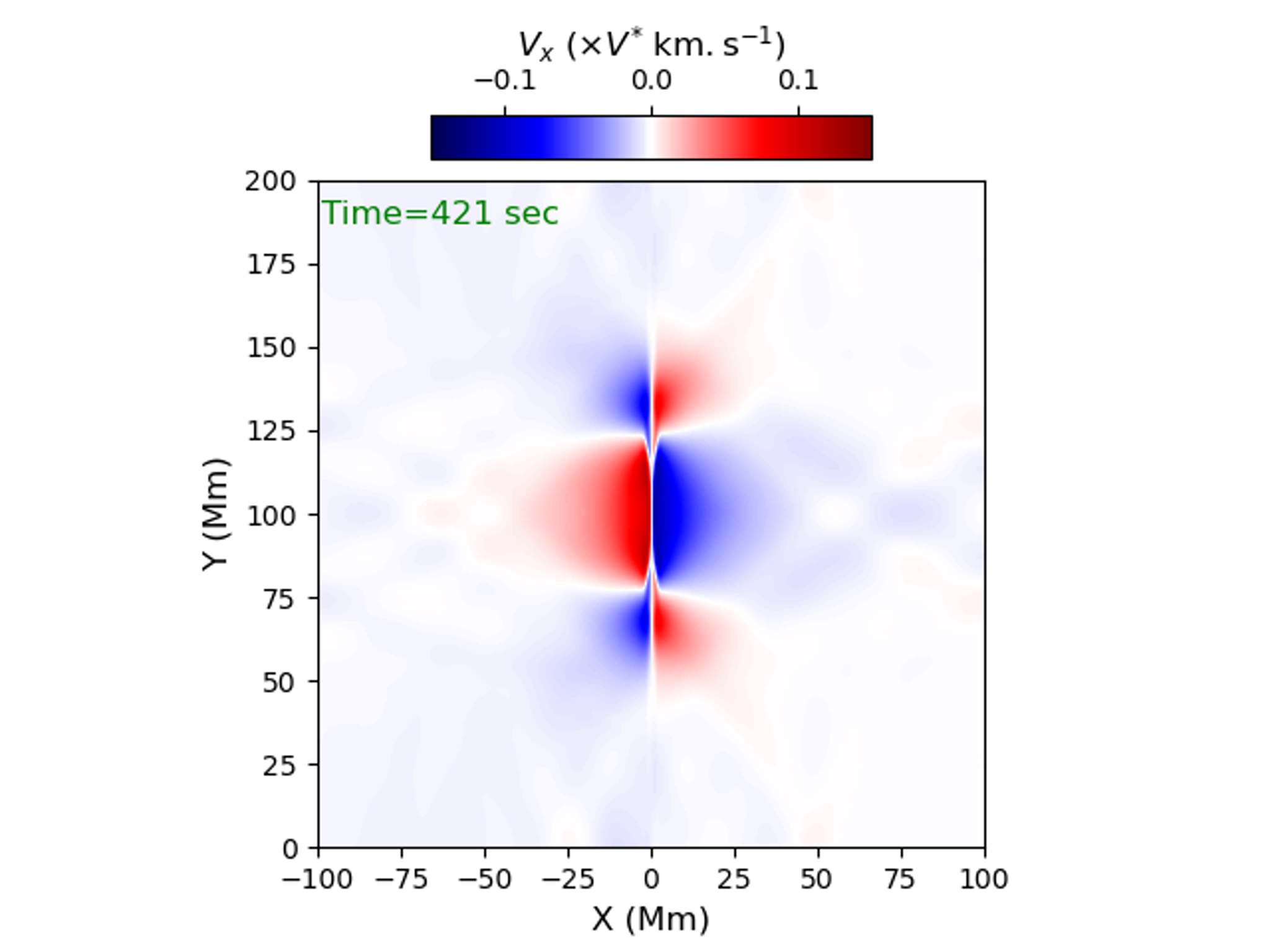}
\includegraphics[height=5.5 cm,trim={11cm 0 11cm 0},clip ]{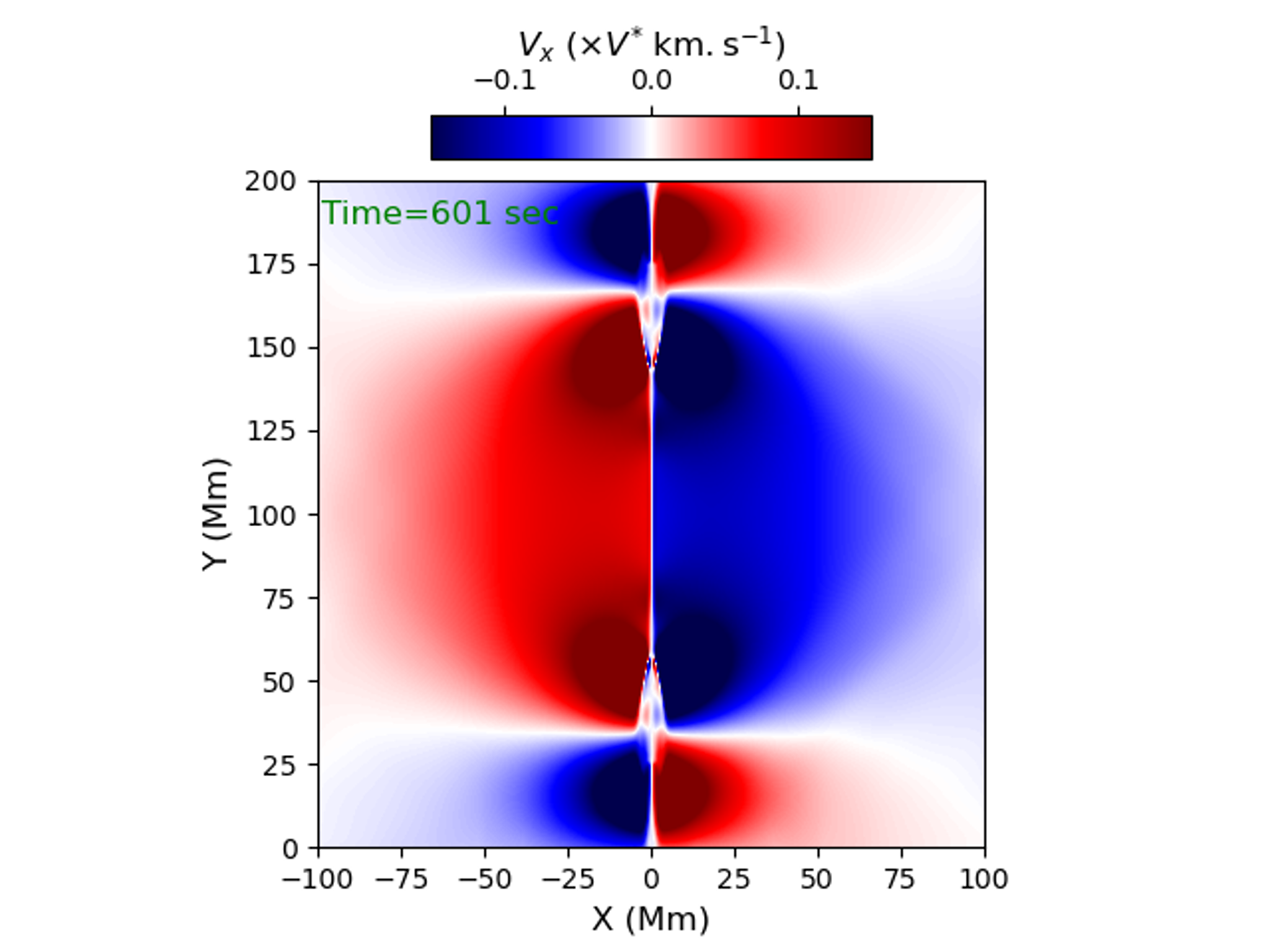}
}

\caption{Snapshots of various quantities at times 60, 240, 421 and 601 seconds are shown from left to right. Top: The density distribution maps. A slow mode shock propagates along the magnetic field lines parallel to the CS and causes the density in the middle of the CS to decrease.  Middle: The total pressure distribution (kinetic+magnetic) maps. At 240 seconds, the total pressure in the CS region has become smaller than in its surroundings. As time progresses, this total pressure gradient increases. Bottom: The $x$-component of velocity maps. The inward total pressure gradient and density gradient towards CS causes an inflow towards the CS which drives magnetic reconnection. The  evolution of these quantities from the start of the simulation to 601 seconds is available as animations in the online HTML version. The real-time animation duration is 5 seconds. The start of the thinning phase, bi-directional outflows and fragmentation of CS are annotated in the animation of the density map.}
\label{label 1}
\end{figure*}

\section{Numerical setup of the MHD model} \label{sec:2}
Using the open source Message Passing Interface-Adaptive Mesh Refinement-Versatile Advection Code (MPI-AMRVAC) \footnote{\url{http://amrvac.org}} \citep{2012JCoPh.231..718K,2012ascl.soft08014V,2014ApJS..214....4P,2018ApJS..234...30X}, we construct a numerical model of the dynamics and reconnection process in a coronal current sheet. The spatial dimensions are -100 Mm to 100 Mm in the $x$-direction and 0 to 200 Mm in the $y$-direction. We run the simulation for 1178 seconds (around 20 minutes). The spatial resolution before any refinement is 1.56 Mm in both the directions which is then subjected to a maximum of four levels of adaptive mesh refinement (AMR). This results in a smallest cell of size 97.5 km. To examine the effect of an external velocity perturbation on the dynamics of a force-free Harris current sheet (CS), we solve the following set of well known magnetohydrodynamics (MHD) equations numerically \citep[e.g.,][]{2018ApJS..234...30X}. We use a `twostep' method for temporal integration and `HLL' scheme to calculate the flux at cell interfaces. 

\begin{equation} 
\frac{\partial \rho}{\partial t} + \nabla \cdot ( \rho \vec{v} ) = 0,
\end{equation}

\begin{equation}
  \frac{\partial}{\partial t}(\rho \vec{v}) + \nabla \cdot \left [ \rho \vec{v}\vec{v}  + \left ( p + \vec{B}^2/2 \right ) \vec{I} - \vec{B}\vec{B} \right ] = 0 ,
\end{equation}

\begin{equation}
  \frac{\partial e}{\partial t} +  \nabla \cdot ( e\vec{v} + \left ( p + \vec{B}^2/2 \right ) \vec{v} -\vec{B}\vec{B} \cdot \vec{v}) = \eta \vec{J^{2}}-
   \vec{B} \cdot \nabla \times (\eta \vec{J}) 
\end{equation}

\begin{equation}
  \frac{\partial \vec{B}}{\partial t} + \nabla \cdot (\vec{v}\vec{B} - \vec{B}\vec{v})+ \nabla \times (\eta \vec{J}) = 0,
  \label{equation 4}
\end{equation}

\begin{equation}
 \nabla \cdot \vec{B} =0,   
\end{equation}

\begin{equation}
 \vec{J} = \nabla \times \vec{B}.
\end{equation}
Here \(\rho\), \(\vec{B}\), \(\vec{v}\), \(\vec{J}\) and \(\eta\) are mass density, magnetic field vector, velocity vector, current density vector and resistivity, respectively. Initially, \(\rho\) and $T$ are set to  uniform values of \(2.34 \times 10^{-15}\) g \(\mathrm{cm^{-3}}\) (\(\rho^{*})\) and 1 MK ($T^{*}$) respectively, which are appropriate for the solar corona. Initially, the thermal pressure $p$ is calculated to be 0.32 dyne \(\mathrm{cm^{-2}}\) ($P^{*}$) using the ideal gas law. The factors used to normalize length, magnetic field, velocity and current density are \(L^{*}= 10^{9}\) cm, \(B^{*}\)= 2 Gauss, $V^{*}= 116.45~$km s$^{-1}$ and \(J^{*}\)= 4.77 statA~cm$^{-2}$, respectively. $e$ is the total energy density defined as
\begin{equation}
 e = \frac{p}{\gamma -1} + \frac{\rho v^{2}}{2} + \frac{B^{2}}{2}
\end{equation}
with \(\gamma= \frac{5}{3}\) being the ratio of specific heats for a monatomic gas. Gravity is not included in our simulations as we are modelling a localized segment of a coronal current sheet and we ignore stratification. Note that we use continuous boundary specifications for all the boundaries in which the variables are extrapolated using the closest inner mesh cell value to all of the ghost cells, i.e., the gradient is kept zero for all the variables. We choose the background parameters density, temperature, and magnetic field in such a way that our simulation and its corresponding outputs mimic the observations of reconnection in the Sun's corona with a plasma \(\beta\) of 0.079. Moreover, we discuss the detailed physics of reconnection that may trigger solar eruptions at coronal heights, which is important for interpreting the observations. In the energy equation, we have not included energy losses via thermal conduction or radiative cooling in our primary model. In addition, no background heating term has been added, but an Ohmic heating term has been included for consistency with the resistive term in the magnetic induction equation (Equation \ref{equation 4}). The initial magnetic field is given by \citep[e.g.,][]{2018ApJS..234...30X}
\begin{equation}
   B_{x} = 0
\end{equation}
\begin{equation}
   B_{y} = - B_{0}~\tanh \left(\frac{x}{l}\right)
\end{equation}
\begin{equation}
   B_{z} = B_{0}~{\rm sech} \left(\frac{x}{l}\right), 
\end{equation}

where \(B_{x}\) and \(B_{y}\) are the magnetic field components in the plane of a force-free CS. \(B_{z}\) is  the guide field essential for providing force balance to maintain the system in initial equilibrium in the absence of a plasma pressure gradient. The relevance of using a guide field is discussed in Appendix B. Basically, we use a similar initial magnetic field configuration to that described in section 3.2.4 of \citet{2018ApJS..234...30X} but with different values of the physical parameters. We set the magnetic field amplitude ($B_{0}$) and CS half-width ($l$) to be 10 G and 1.5 Mm, respectively. We adopt a uniform resistivity with dimensionless value  \(2 \times 10^{-4}\) (corresponding to magnetic diffusivity of \(2.4 \times 10^{8}~\mathrm{m^{2}s^{-1}}\) in physical units) throughout the simulation domain, which gives a Lundquist number of \(4.8 \times 10^{5}\). This is reasonable for coronal applications since the limitations of current numerical simulations preclude much higher values. \citet{2022A&A...666A..28S} used a similar dimensionless resistivity in their numerical simulation of coronal reconnection. It is also similar to those used in the study of plasmoid instability by various authors \citep[e.g.,][]{2009PhPl...16k2102B,2010PhPl...17f2104H,2012PhRvL.109z5002H,2013PhPl...20e5702H,2015shin.confE..26H,2016ApJ...818...20H,2022A&A...666A..28S}.

To disturb the CS which is initially in quasi-static equilibrium, we impose an external velocity perturbation in the $x$-direction in the form of a Gaussian pulse given by
\begin{equation}
    v_{x} = v_{0}~\mathrm{exp} \left(- \frac{(x-x_{0})^{2}+(y-y_{0})^{2}}{w^{2}}\right).
\end{equation}
Here $x_{0}$ is set at 15 Mm away from the centre of the CS at $x=0$ and $y_{0}$ is set at a height of 100 Mm in the middle of the domain in order to minimize the  finite boundary effects. The width of this Gaussian pulse ($w$) is taken to be 4 Mm. The amplitude ($v_{0}$) of velocity pulse is 350 km \(\mathrm{s^{-1}}\) which is $0.6\mathrm{v_{A}}$, where $v_{A}=580\ \mathrm{km\ s^{-1}}$ is the Alfv\'en speed for our choice of parameters. This transient velocity pulse differs from the periodic disturbance used in \citet{2019A&A...623A..15P}. Solar eruptions are accompanied by the generation and propagation of EUV waves away from the eruption site in a transient manner \citep[e.g.,][]{2014MNRAS.444.1119Z,2018ApJ...868L..33L,2018ApJ...866...64C,2018ApJ...858L...1Z,2022ApJ...929L...4Z}. They may sometimes in turn trigger reconnection, oscillations, eruptions and resonant energy transfer in other regions, especially where prominences, coronal loops, or ambient localized coronal magnetic structures are located. In our model we mimic such a process by a Gaussian velocity pulse. Observational studies  report the velocity of fast MHD waves or EUV waves to be of  order \(350~\mathrm{km\ s^{-1}}\) or even higher. So for solar coronal conditions where the fast MHD wave speed is around \(1000~\mathrm{km\ s^{-1}}\), the amplitude of the velocity pulse taken in this work is appropriate. To explore the detailed physical behaviour of reconnection driven by such fast-mode perturbations, we choose to impose: (i) a straight local segment of the coronal current sheet without curvature; (ii) uniform coronal plasma conditions without structuring; (iii) no energy loss or heating terms in the energy equation. These complexities will be included in future works. 

\begin{figure*}
\mbox{

\includegraphics[height=6 cm, width=6 cm]{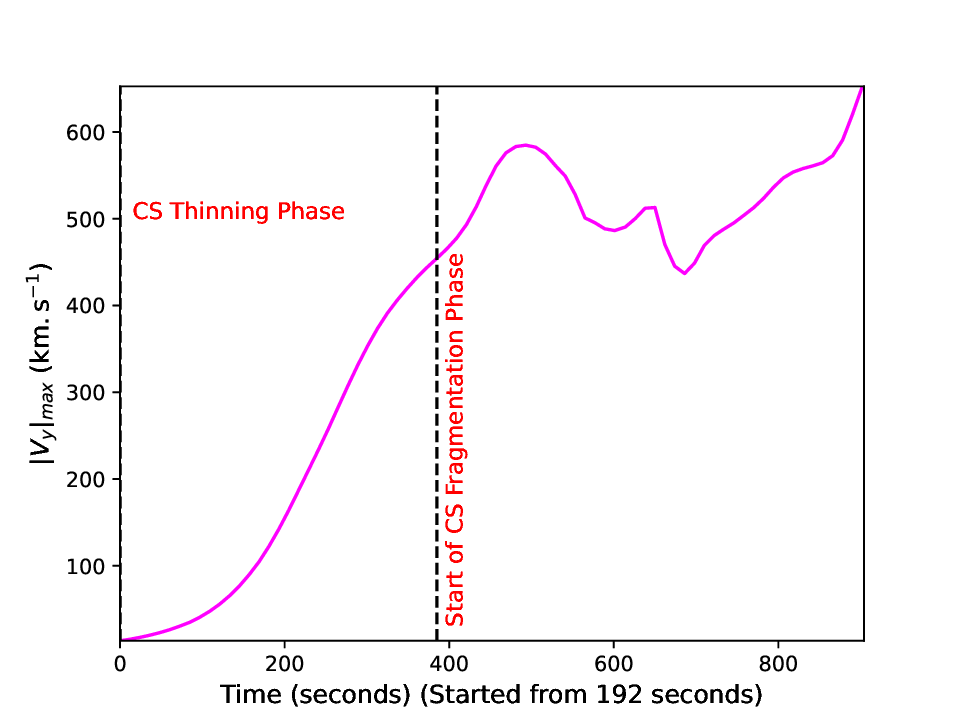}
\includegraphics[height=6 cm, width=6 cm]{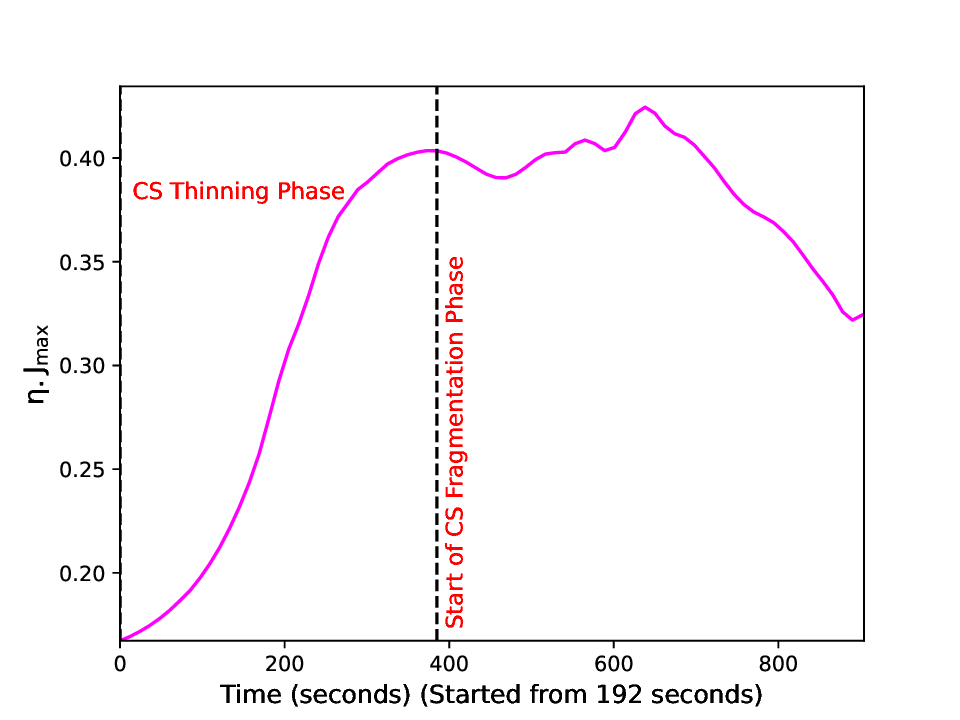}
\includegraphics[height=6 cm, width=6 cm]{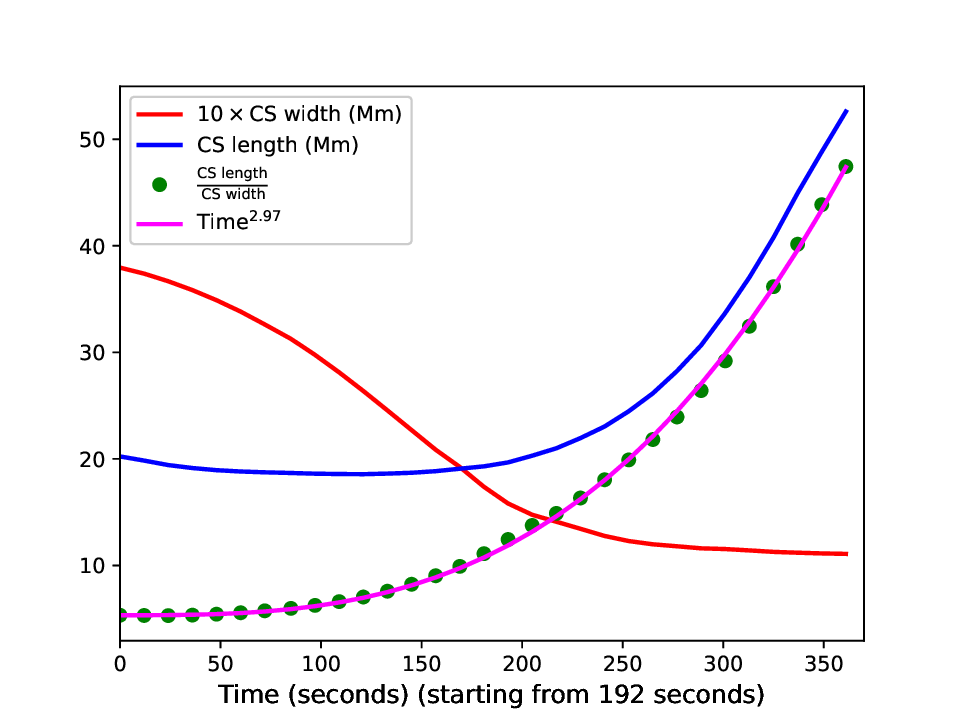}}
\caption{Left: The temporal evolution of the maximum of the outflow velocity \(v_{y}\) along the CS, which shows a gradual increase followed by an impulsive bursty part due to the plasmoid instability. Middle: The temporal evolution of \(\eta J_{Max}\) along the CS, which is an another proxy of the reconnection rate. Right: Temporal evolution of the CS width at a height of 100 Mm, CS length, and the aspect ratio of the CS, i.e., the ratio of its length to width. The temporal evolution of the aspect ratio of the CS follows a power law  with exponent 2.97.}
\label{label 2}
\end{figure*}
\begin{figure*}
\mbox{
\includegraphics[height=6 cm, width=6 cm]{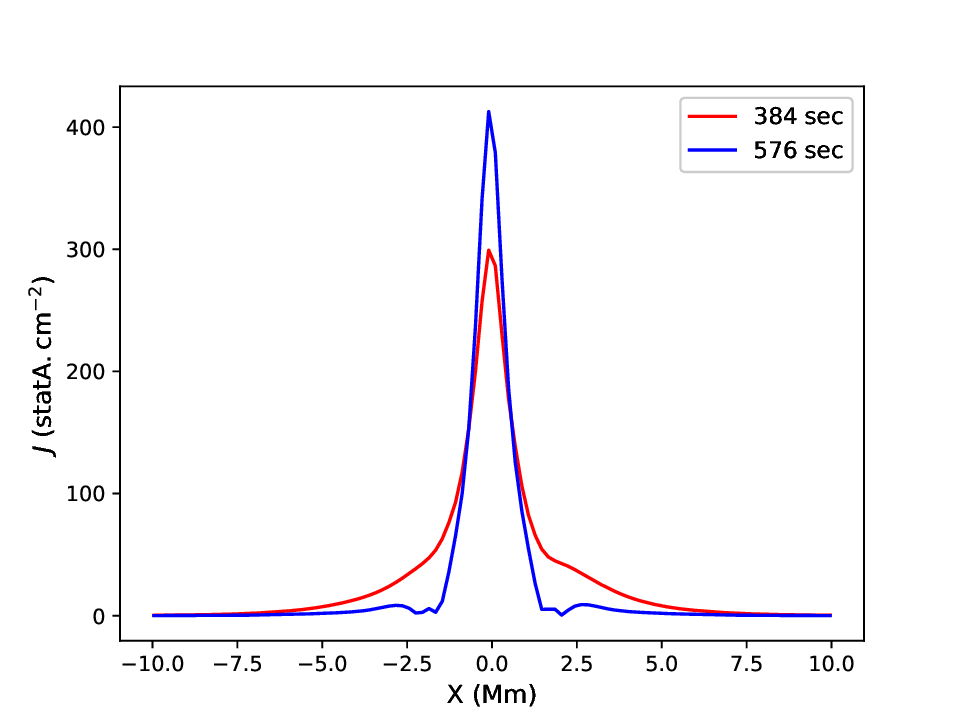}
\includegraphics[height=6 cm, width=6 cm]{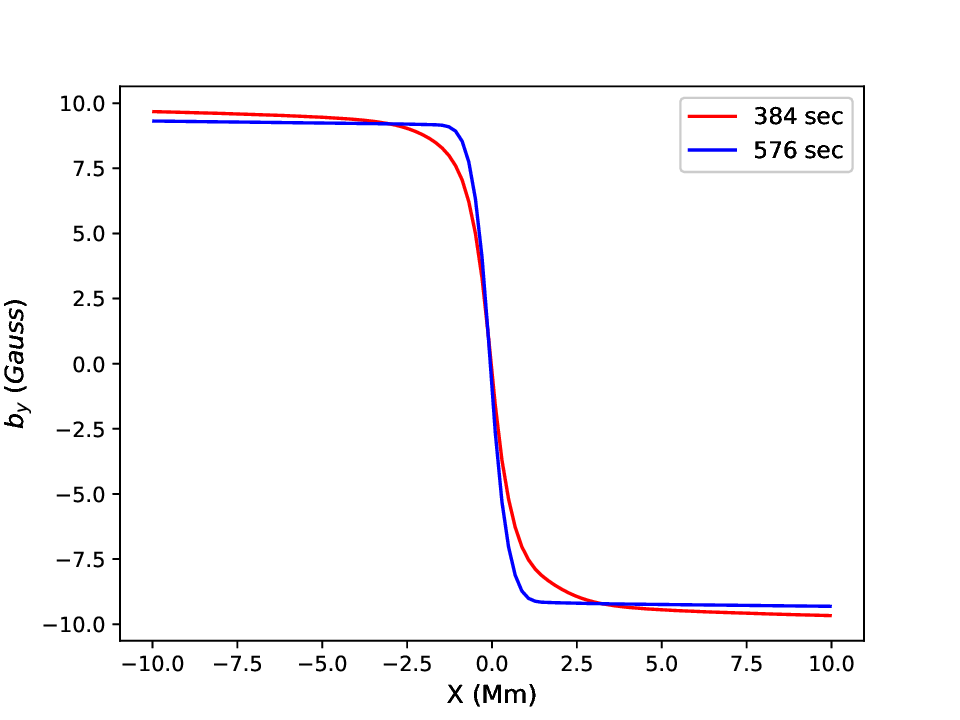}
\includegraphics[height=6 cm, width=6 cm]{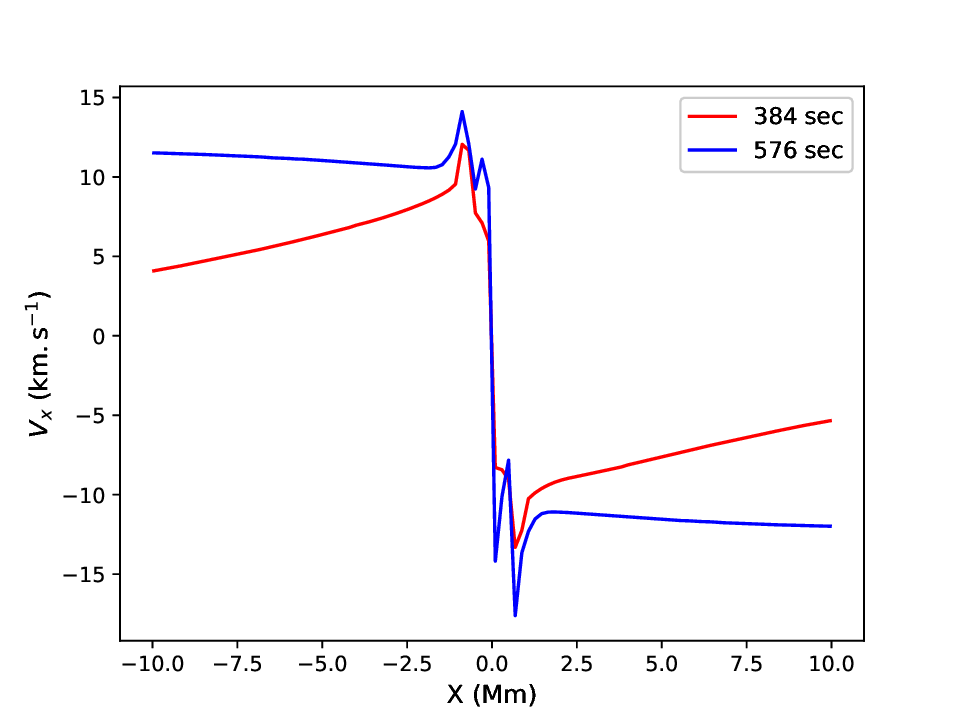}
}
\caption{The distribution of total current density ($J$), magnetic field along the CS (\(b_{y}\)) and inflow velocity (\(v_{x}\)) (from left to right) across the CS at $y$=100 Mm. The gradual decrease of \(b_{y}\), increase of $J$ and \(v_{x}\) towards the CS from the surroundings at 384 seconds, just at the beginning of bi-directional outflows, are signatures of Petschek-type reconnection. Also, the steepness of the gradients of these quantities at 576 seconds just before the fragmentation starts, are a signature of Sweet-Parker reconnection.}
\label{label 3}
\end{figure*}
\begin{figure*}

\mbox{
\hspace{1.75 cm}
\includegraphics[height=5 cm, width=7 cm]{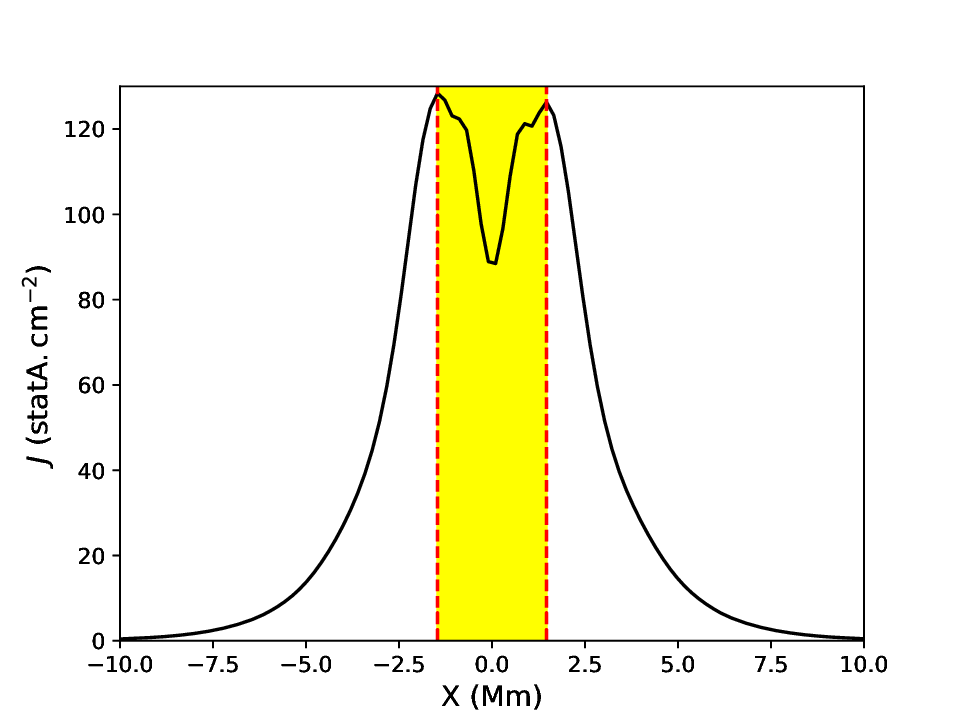}
\includegraphics[height=5 cm, width=7 cm]{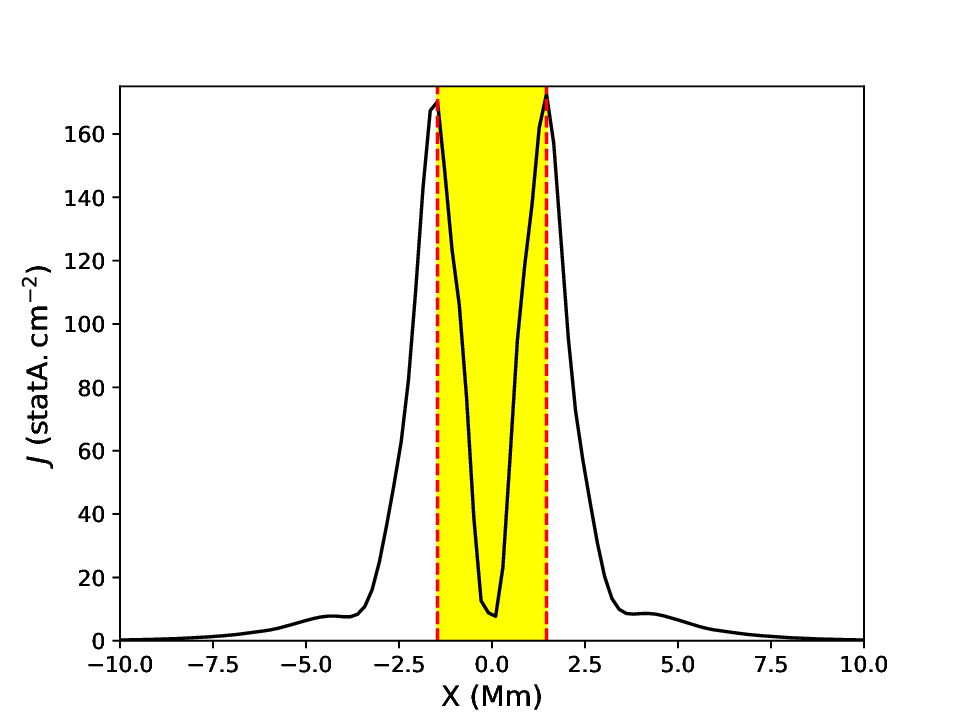}
}
\mbox{
\hspace{1.75 cm}
\includegraphics[height=5 cm, width=7 cm]{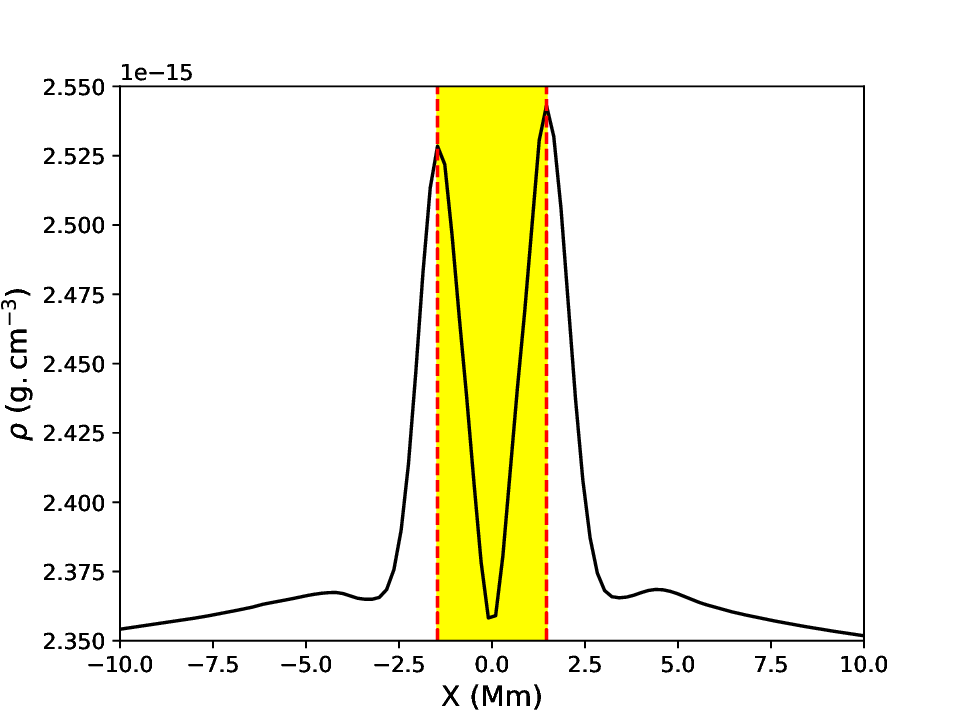}
\includegraphics[height=5 cm, width=7 cm]{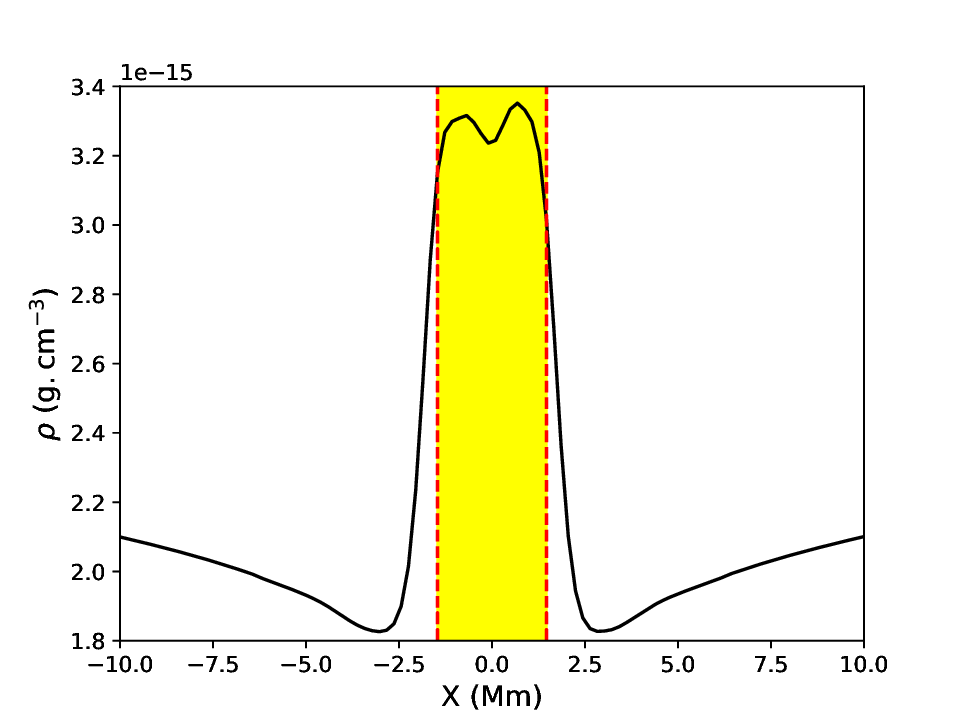}
}
\mbox{
\hspace{1.75 cm}
\includegraphics[height=5 cm, width=7 cm]{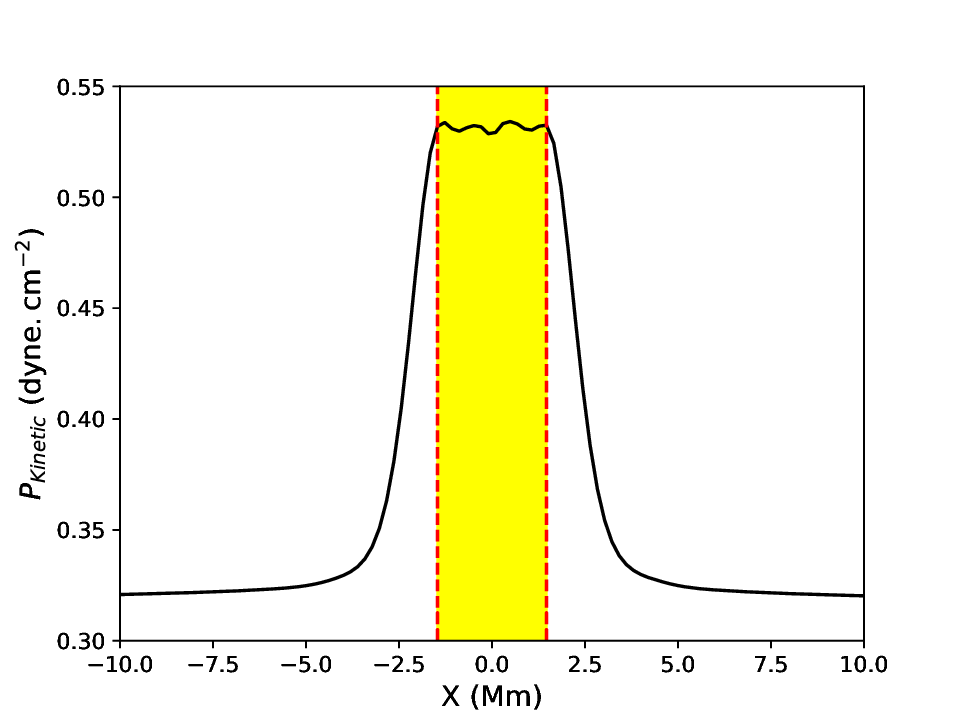}
\includegraphics[height=5 cm, width=7 cm]{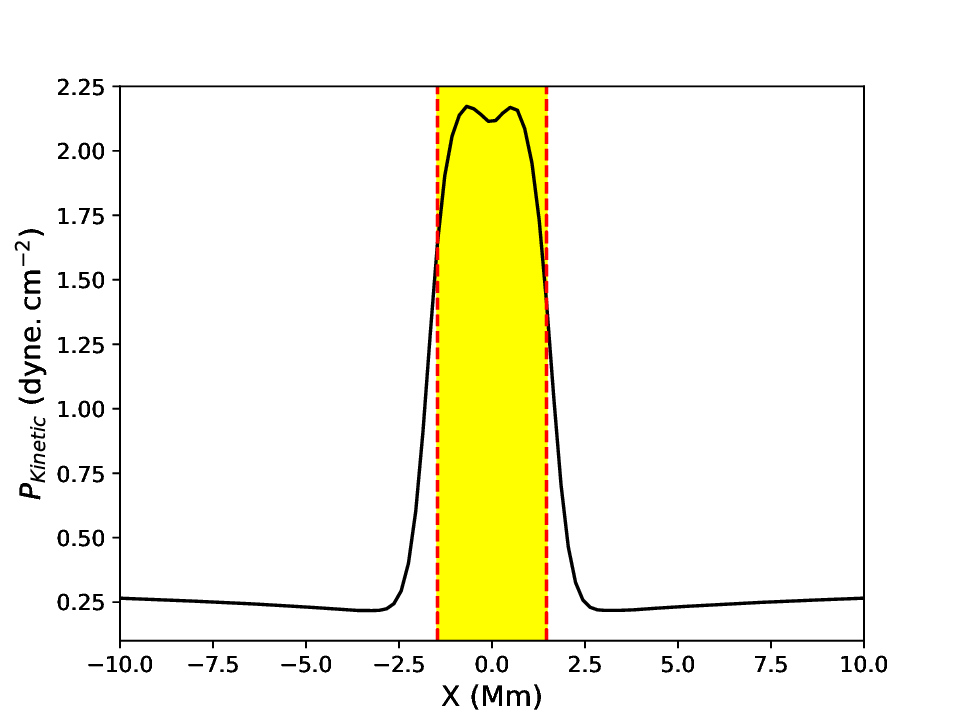}
}
\mbox{
\hspace{1.75 cm}
\includegraphics[height=5 cm, width=7 cm]{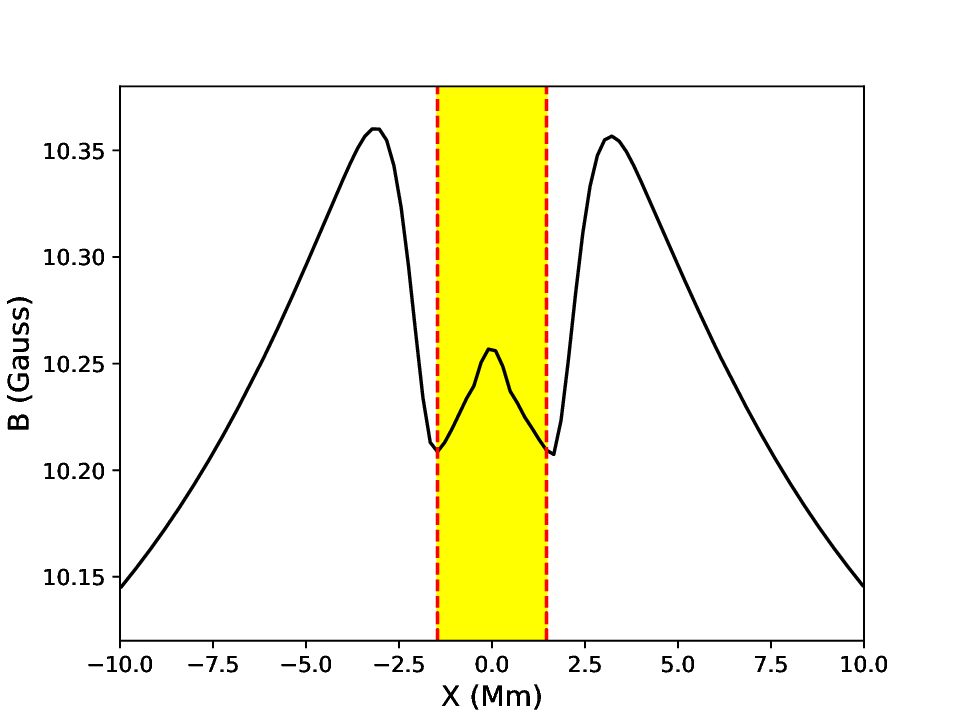}
\includegraphics[height=5 cm, width=7 cm]{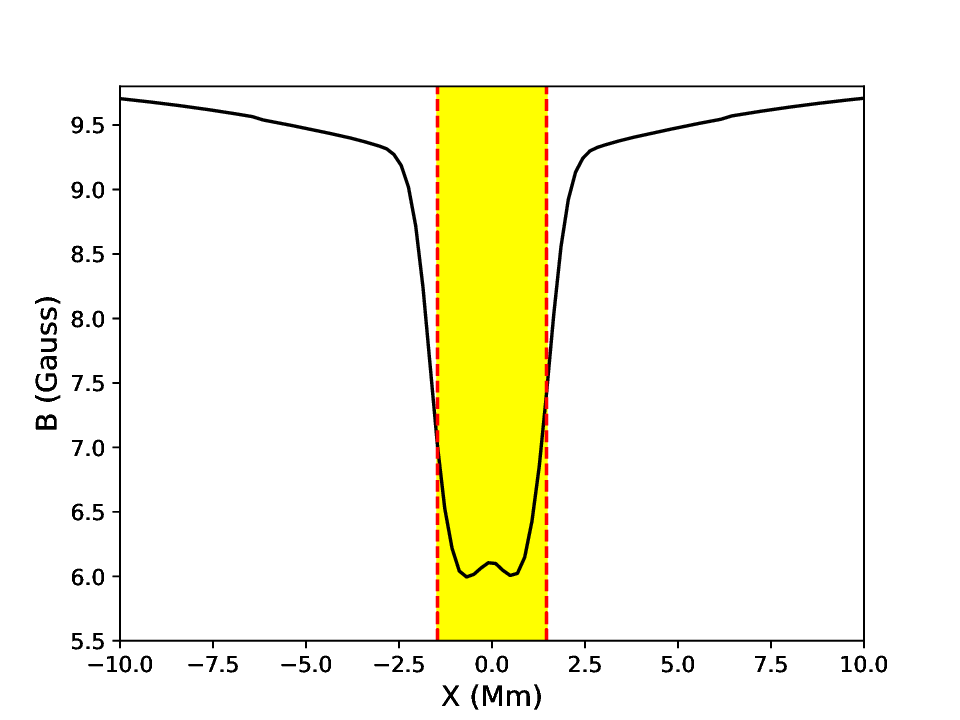}
}

\caption{Profiles of current density, plasma density, plasma pressure and magnetic field  across the CS. Left column: At $y$= 125 Mm at time 384 seconds. Right column:  At $y$= 140 Mm at time 576 seconds just before the start of fragmentation of the CS. All the quantities possess  steep gradients that are the signature of slow-mode MHD shocks. The maximum of the current density gives the approximate location of the shock, which is denoted by red dashed vertical lines in the plots. The yellow shaded region is the outflow region.}
\label{label 4}
\end{figure*}
\begin{figure*}
\mbox{
\includegraphics[height=6cm, width=6 cm]{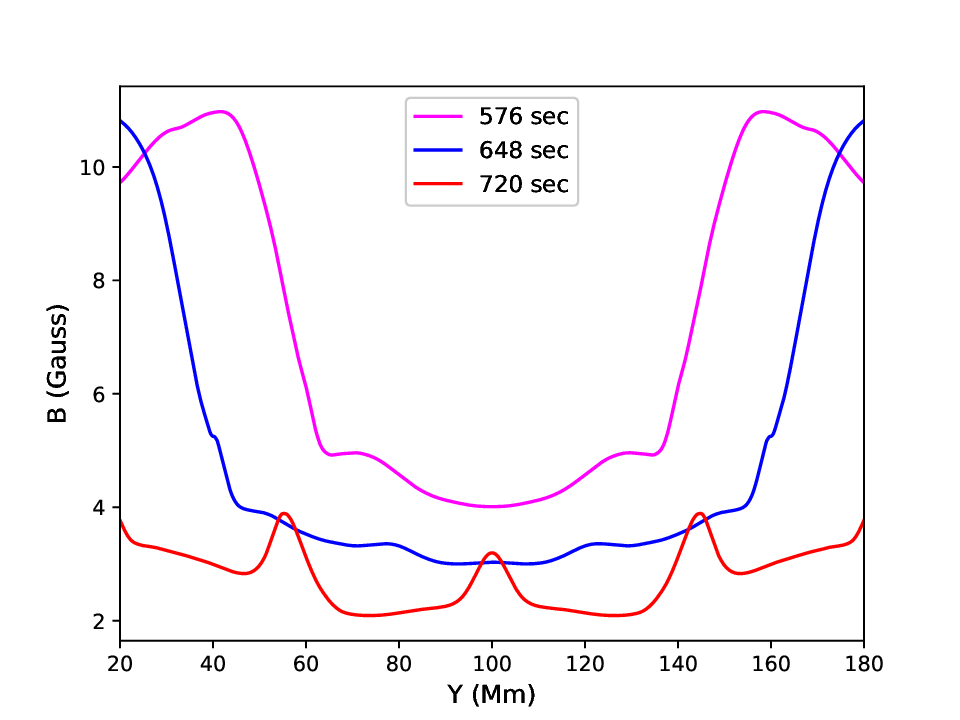}
\includegraphics[height=6cm, width=6 cm]{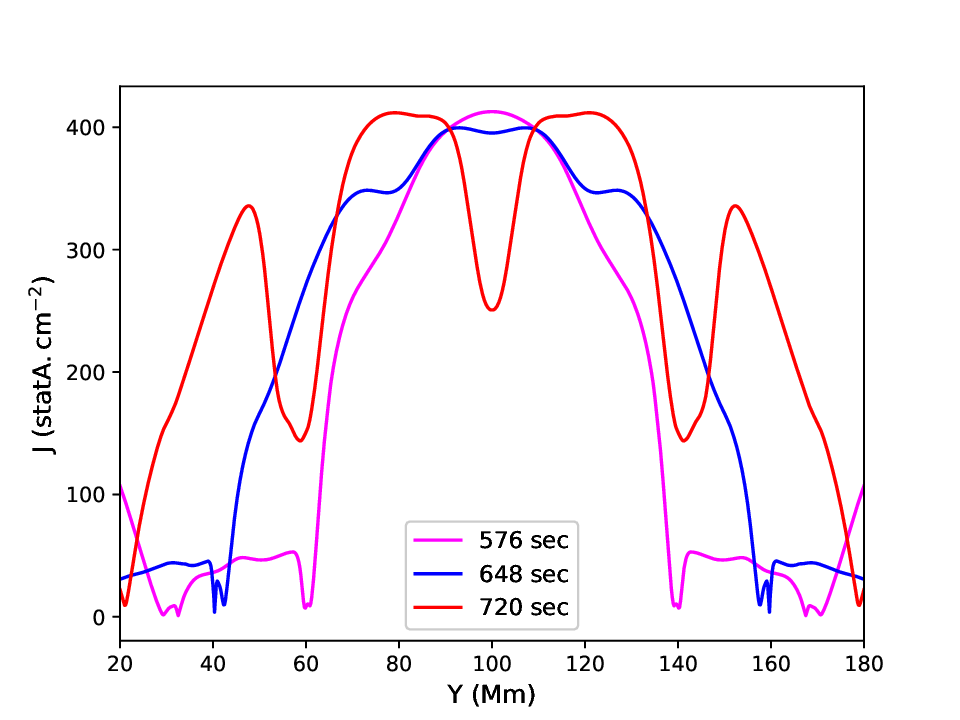}
\includegraphics[height=6cm, width=6 cm]{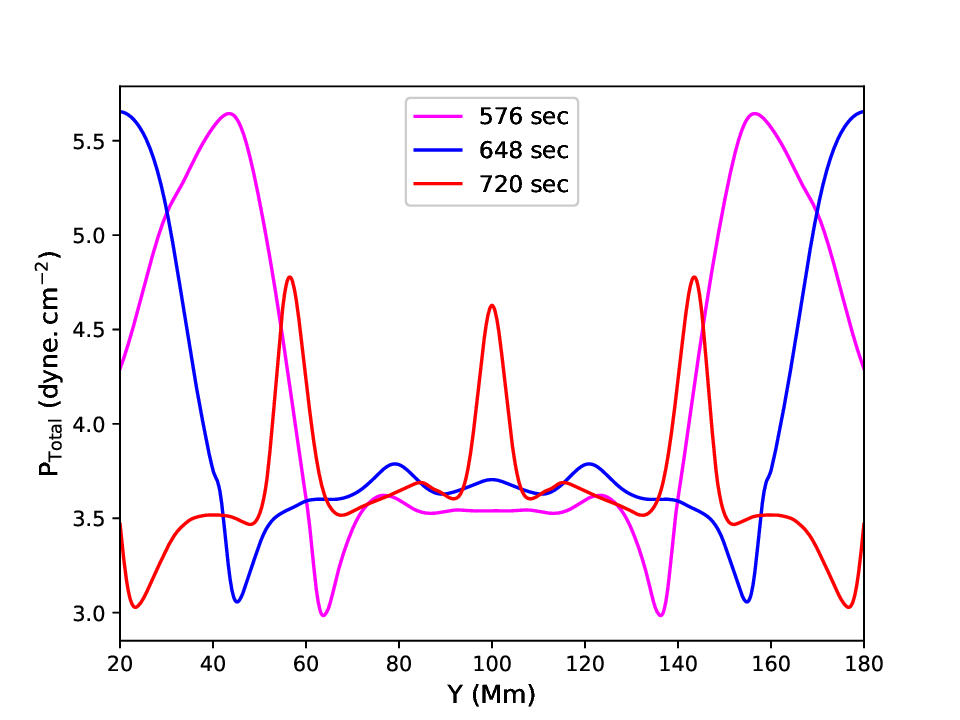}}
\caption{Profiles of various quantities along the CS at three times during its fragmentation. 
Left: Total magnetic field ($B$). As time progresses, the region with low magnetic field \textbf{becomes} stretched along the length of CS while the value of magnetic field decreases. But the flattened profile of $B$  also becomes fragmented and eventually three peaks of $B$ appear, demonstrating the location of plasmoid formation. 
Middle: Total current density ($J$). Low-$J$ regions are  those where plasmoids or magnetic islands are formed. 
Right: Total pressure (kinetic+magnetic). High-pressure regions are formed during the fragmentation of the CS, and the low-pressure region in the middle of the CS indicates a region of plasmoid formation.}
\label{label 5}
\end{figure*}

\section{Results} \label{sec:3}
We start from a force-free vertically elongated CS in initial equilibrium. We first performed a test run (not shown in the paper) to examine whether the CS undergoes thinning and plasmoid formation in the absence of external perturbations or not. We find that due to the presence of a finite uniform resistivity in our simulation domain, the CS undergoes slow diffusion only. That is, no thinning and plasmoid formation occurs without the external velocity perturbation. We also test whether the chosen value of physical resistivity is higher than the numerical resistivity for the considered resolution. We find that we can safely consider the dynamics to be dependent on physical resistivity. The details of comparison between physical and numerical resistivity are discussed in Appendix A in detail. 

As we perturb the CS via a velocity pulse of 350 km \(\mathrm{s^{-1}}\), it undergoes temporal evolution in the form of thinning followed by fragmentation. This action further gives rise to the formation of plasmoids or magnetic islands. The left panel of Figure \ref{label 2} shows the temporal evolution of the maximum value of the outflow velocity \(v_{y}\) along the CS. It shows that the outflow velocity increases with time in an unsteady manner and finally enters an impulsive phase indicative of the plasmoid instability. The middle panel of Figure \ref{label 2} shows the temporal evolution of \(\eta J_{max}\) which we consider as a proxy of the reconnection rate, as often assumed \citep[e.g.,][] {1997ApJ...474L..61Y,2012PhPl...19i2110B}. It also shows an unsteady behaviour as the impulsive bursty phase is approached. We discuss each of the phases of the dynamics of the CS along with their characteristics as follows.

\subsection{Different Phases of the Dynamics of the Current Sheet}

\subsubsection{Thinning of the Current Sheet} As the velocity pulse crosses the CS, it splits into two counter-propagating wavefronts from the CS towards the lateral boundaries. The reflected one traverses to the left of the CS while the transmitted one traverses to the right. These counter-propagating wavefronts produce a fast-mode expansion which results in a decrease of density in the CS region (as shown in top row of Figure \ref{label 1}). Similarly, an inward total pressure gradient is established towards the CS from its immediate surroundings (as shown in the middle row of Figure \ref{label 1}). These inward gradients of density and total pressure lead to an inflow towards the CS from the surrounding region (as shown in the bottom panel of Figure \ref{label 1}) resulting in gradual thinning of the CS with time. These entire dynamics are even better seen in the animations accompanying Figure \ref{label 1}. We now examine the detailed characteristics of this thinning phase. 

[a] To quantify the change in CS configuration during this unsteady phase, we take CS width, CS length and ratio of length to width as the important factors. To estimate the CS width and length, we take a horizontal slit across the CS at $y =$ 100~Mm and a vertical slit along the CS at $x = 0$ Mm, respectively. We discover that the current density distributions across as well as along the CS have maxima at $x = 0$ Mm and $y = 100$ Mm, respectively, and we fit them with Gaussian functions given by
\begin{equation}
   G(s,\sigma) = C~\exp\left(\frac{-(s-\bar{S})^{2}}{2\sigma^{2}}\right).
\end{equation}
The obtained value of \(\sigma\) is then used to calculate the full width half maximum (FWHM) of the distribution as
\begin{equation}
   \mathrm{FWHM = 2~\sqrt{2 \sigma^{2}~ln2}}.
\end{equation}
These FWHMs are taken as estimates of CS width and CS length. We find that, during this unsteady phase of reconnection, the CS undergoes a gradual thinning dynamically as shown in the temporal profile of CS width shown as a red curve in the right panel of Figure \ref{label 2}. The rate of thinning increases with time, resulting in an unsteady character of the thinning process. The right panel of Figure \ref{label 2} also shows the temporal evolution of the CS length (blue curve). It is evident that the measured CS length is not increasing significantly at the beginning of the thinning phase. But during the later phase, it starts to increase rapidly. Nevertheless the most important estimate for characterizing the time-dependent behaviour of a dynamic CS is its aspect ratio. As shown in the right panel of Figure \ref{label 2}, the aspect ratio of the CS increases following a \(t^{2.97}\) profile.
\newline

[b] Now it is interesting to determine whether the reconnection during this thinning phase has Sweet-Parker or Petschek characteristics. Since the CS is dynamic, we check the character of the diffusion region to determine whether it is of Petschek type or Sweet-Parker type at the beginning of the bi-directional outflows and at the beginning of the fragmentation of the CS. \citet{2012PhPl...19i2110B} suggested that for the classical Petschek nature of reconnection to exist, there must be gradual increase of inflow velocity \(v_{x}\) as we move towards the CS from the 
inflow boundaries. Also, the inflow magnetic field \(b_{y}\) and the current density ($J$) should undergo a gradual decrease and increase respectively as we move towards the CS from the inflow domain boundaries for the reconnection to be Petscheck type. Now, in Figure \ref{label 3}, we find that all these three quantities possess the required signature of Petschek-like reconnection (red curves) at the time 384 seconds at the beginning of the bi-directional outflows. But just before the start of CS fragmentation, the profiles of these quantities are more likely to be of Sweet-Parker nature as they show steep gradients across the CS. Therefore, we infer that during the unsteady transition phase before the fragmentation of the CS, the reconnection starts with a Petschek-like nature in a small diffusion region. As time progresses, the length of the diffusion region increases and finally becomes sufficiently elongated (with aspect ratio attaining a value of 48) that it reaches a Sweet-Parker regime.

[c] To support the Petschek-type reconnection scenario, we also look for signatures of the slow mode shocks found in that model. Following \citet{1997ApJ...474L..61Y}, we extract the profiles of plasma density, plasma pressure, magnetic field and current density across the CS just outside the diffusion layer. We take horizontal cuts at y = 125 Mm at time 384 seconds at the beginning of the bi-directional outflows. As shown in the left column of Figure \ref{label 4}, there are steep gradients in all the aforementioned physical parameters. The increase in density and kinetic pressure and decrease in magnetic field strength in the vicinity of the layer denoted by red dashed vertical lines are all evidence of a slow mode MHD shock at that height. Our estimated profiles are similar to those in \citet{1997ApJ...474L..61Y}. So the presence of shocks further confirms that the reconnection at the beginning of {the} bi-directional outflows is of Petschek type. As the CS evolves with time, however, the length of the diffusion layer increases and finally attains that of a Sweet-Parker CS. As previous literature \citep[e.g.,][]{1986JGR....91.5579P,2012PhPl...19i2110B} suggested, even though the diffusion layer becomes Sweet-Parker type in the middle portion of the entire CS,  there are signatures of comparatively shorter slow-mode shocks at the ends of the extended diffusion layers. So we repeat our estimate of the transverse profiles of the physical parameters at a height of $y$ = 140 Mm at time 576 seconds during the beginning of the fragmentation of the diffusion layer at the middle of the CS. We find steep gradients and, more specifically, an increase in density and kinetic pressure and decrease in magnetic field across the layer going into the outflow region, as before (shown in right column of Figure \ref{label 4}). In addition, we notice that the sub-structuring of the plasma density, pressure and magnetic field in the shocked region become less prominent at later times, which may signify a relative weakening of the slow-mode MHD shocks. However, the current density shows the opposite behaviour, i.e., it becomes more sub-structured at later times.

\subsubsection{Fragmentation of the Current Sheet} Once the CS is long enough that the ratio of length to width of the CS reaches around 48 at time 576 seconds -- compared to a value of 5 at the beginning of the thinning of the CS -- 
it starts to fragment into small portions due  to the onset of the tearing mode instability \citep{1963PhFl....6...48F}. This fragmentation is evident from small-scale substructuring of the profile of the total magnetic field magnitude along the CS, as shown in the left panel of Figure \ref{label 5}. Similarly, as shown in the middle panel of Figure \ref{label 5}, the profile of total current density ($J$) starts to form a dip at a height of 100 Mm at time 576 seconds. As time progresses, the profile gets more flattened and the dips in $J$ become even more prominent than earlier. Finally at 720 seconds, just before the visibility of the plasmoids in the mosaics in Figure \ref{label 6}, the dips are clearly showing fragmentation of  current density along the CS (as shown in the middle panel of Figure \ref{label 5}). Likewise, in the right panel of Figure \ref{label 5}, it is evident that the sum of kinetic and magnetic pressures has peaks corresponding to dips in the current density profile. So the profile of total pressure is also showing the signature of fragmentation of the elongated CS into smaller length segments. This fragmentation has been further followed by formation of plasmoids at the dips of the current density or peaks of the total pressure and total magnetic field. The dips of current density at about 60 and 140 Mm are more than that at 100 Mm (as shown in the middle panel of Figure \ref{label 5}) during the onset of the primary stage of plasmoid formation. 

Now let us determine whether the dips or peaks in Figure \ref{label 5} are consistent with the allowable range of wavelength for the tearing mode instability. We note that by this time the velocity disturbance has already passed out of the simulation zone, since there is no prominent reflection of the primary pulse from the boundary. The allowable range of wavelength for the tearing mode instability \citep{1963PhFl....6...48F,2014masu.book.....P} is
\begin{equation}
2\pi l < \lambda < 2\pi l~(R_{m})^{\frac{1}{4}}
\label{equation 14}
\end{equation}
where \(R_{m}\) is the magnetic Reynolds number defined as 
\begin{equation}
R_{m} = \frac{lv_{A}}{\eta}
\end{equation}
with $l$ being the half-width of the CS and \(\eta\) being the magnetic diffusivity. We calculate the average Alfv\'en speed and CS width between $y$ = 20 Mm and $y$ = 180 Mm at a time 720 seconds just before the plasmoids become visible. The average $l$ and  \(v_{A}\) are equal to 0.84 Mm and 528 km \(\mathrm{s^{-1}}\), respectively. This gives us a limit for the allowable  wavelength for the tearing mode as \(\mathrm{5.3~Mm < \lambda < 34.7~Mm}\) from  Equation \ref{equation 14}. Now in our case, the distance between the peaks or dips shown in Alfv\'en \ref{label 5} at 720 seconds is about 41 Mm.  The  slight difference in estimated values may be due to the superposed reconnection outflows, which increase the distance between successive plasmoids. Thus, the distance between  plasmoids is consistent with the theoretical estimate from  the tearing mode instability.

\subsubsection{Multiple stages of plasmoid instability in the Current Sheet} 
\begin{figure*}

\mbox{
\hspace{-1.75 cm}
\includegraphics[height=6.5 cm,trim={1.0cm 0 0 0},clip]{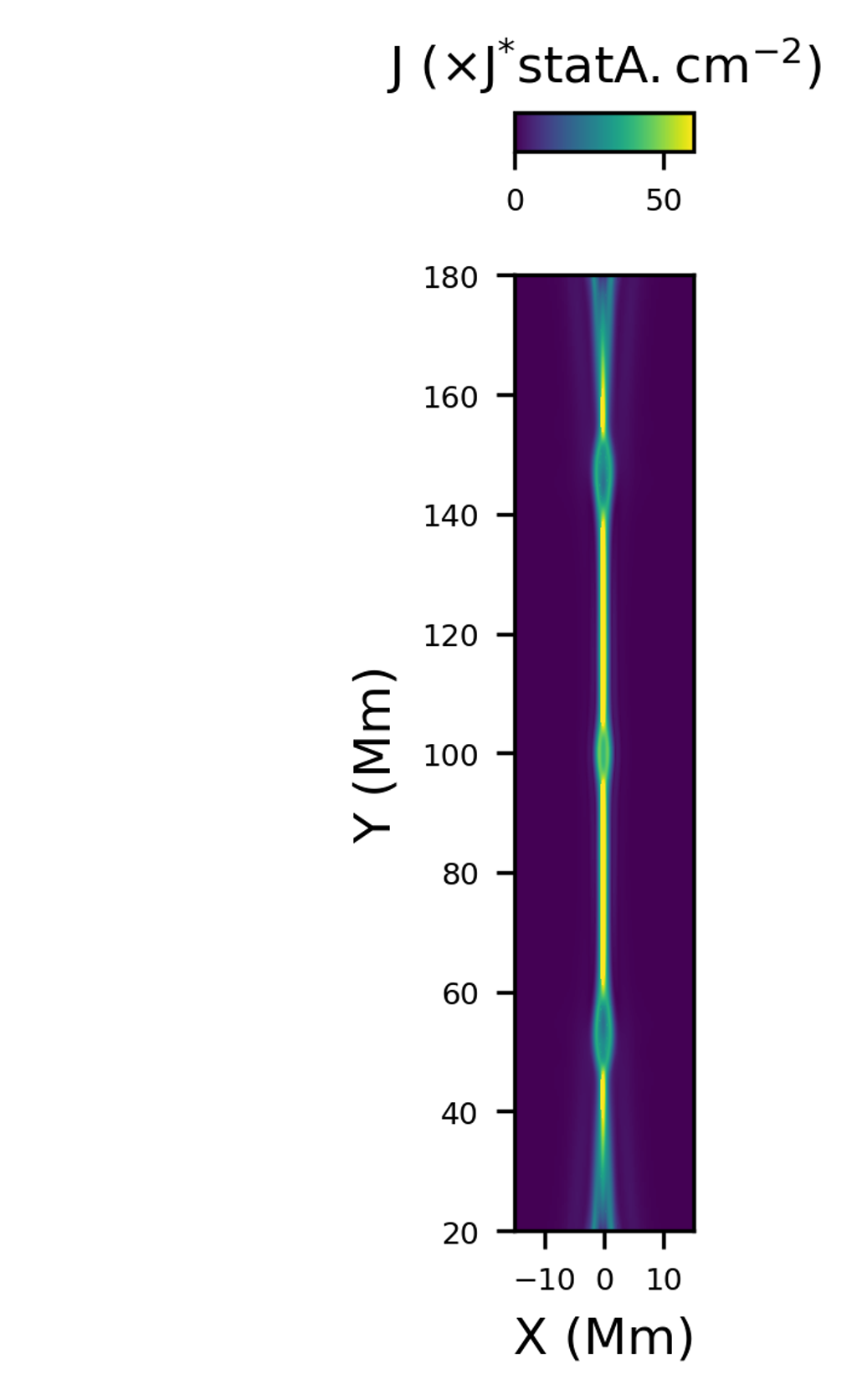}
\includegraphics[height=6.5 cm,trim={6.75cm 0 0 0},clip]{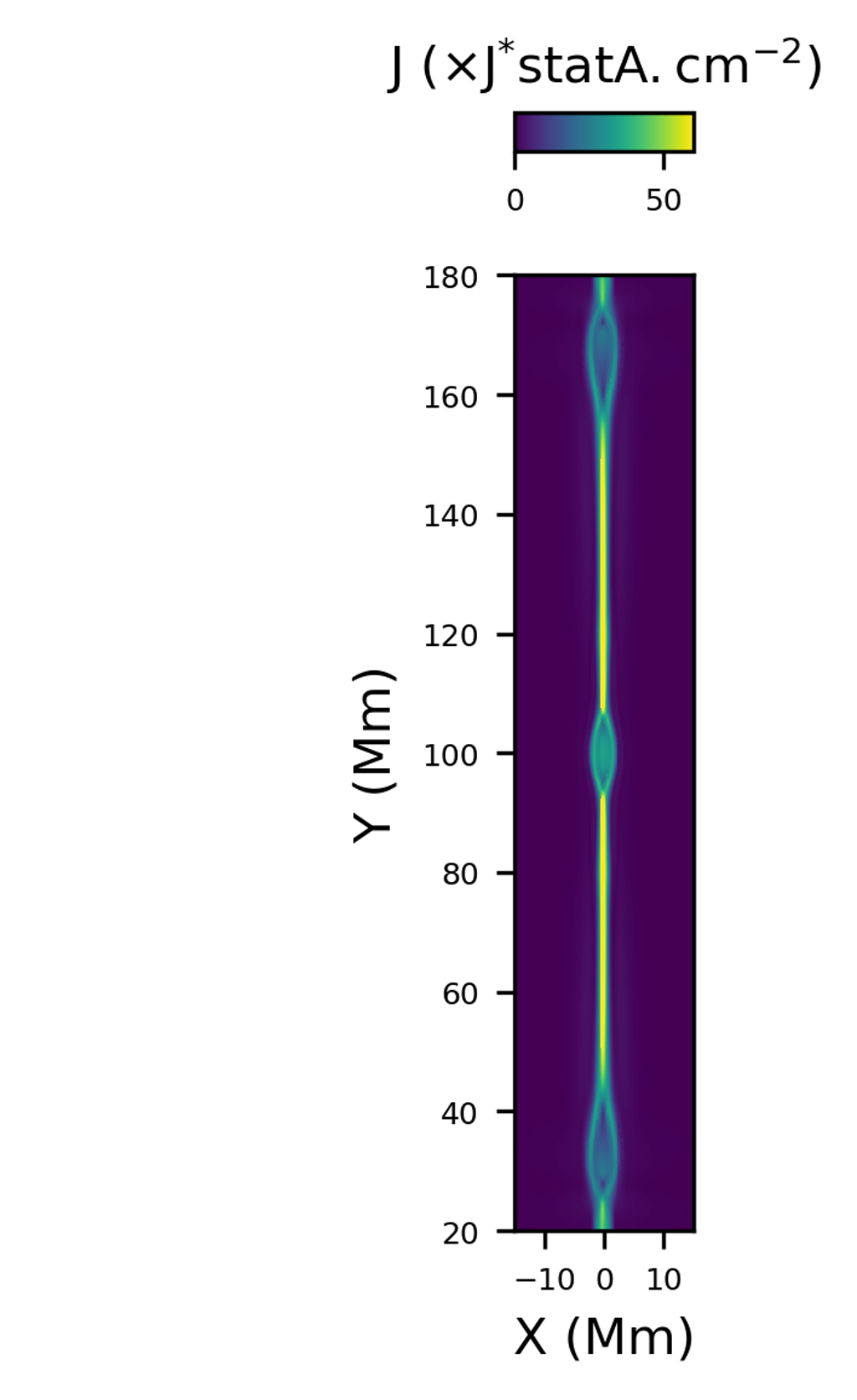}
\includegraphics[height=6.5 cm,trim={6.75cm 0 0 0},clip]{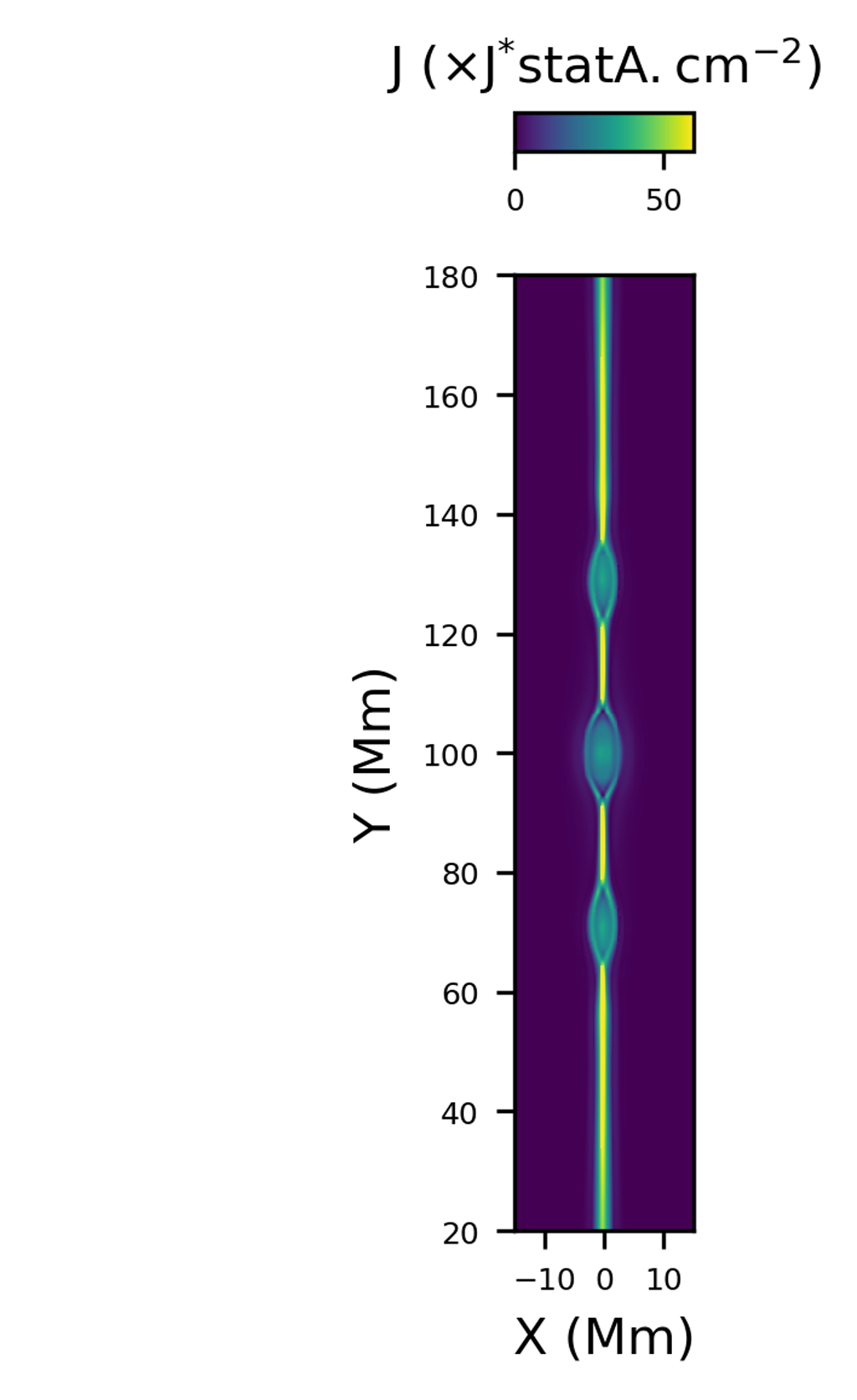}
\includegraphics[height=6.5 cm,trim={6.75cm 0 0 0},clip]{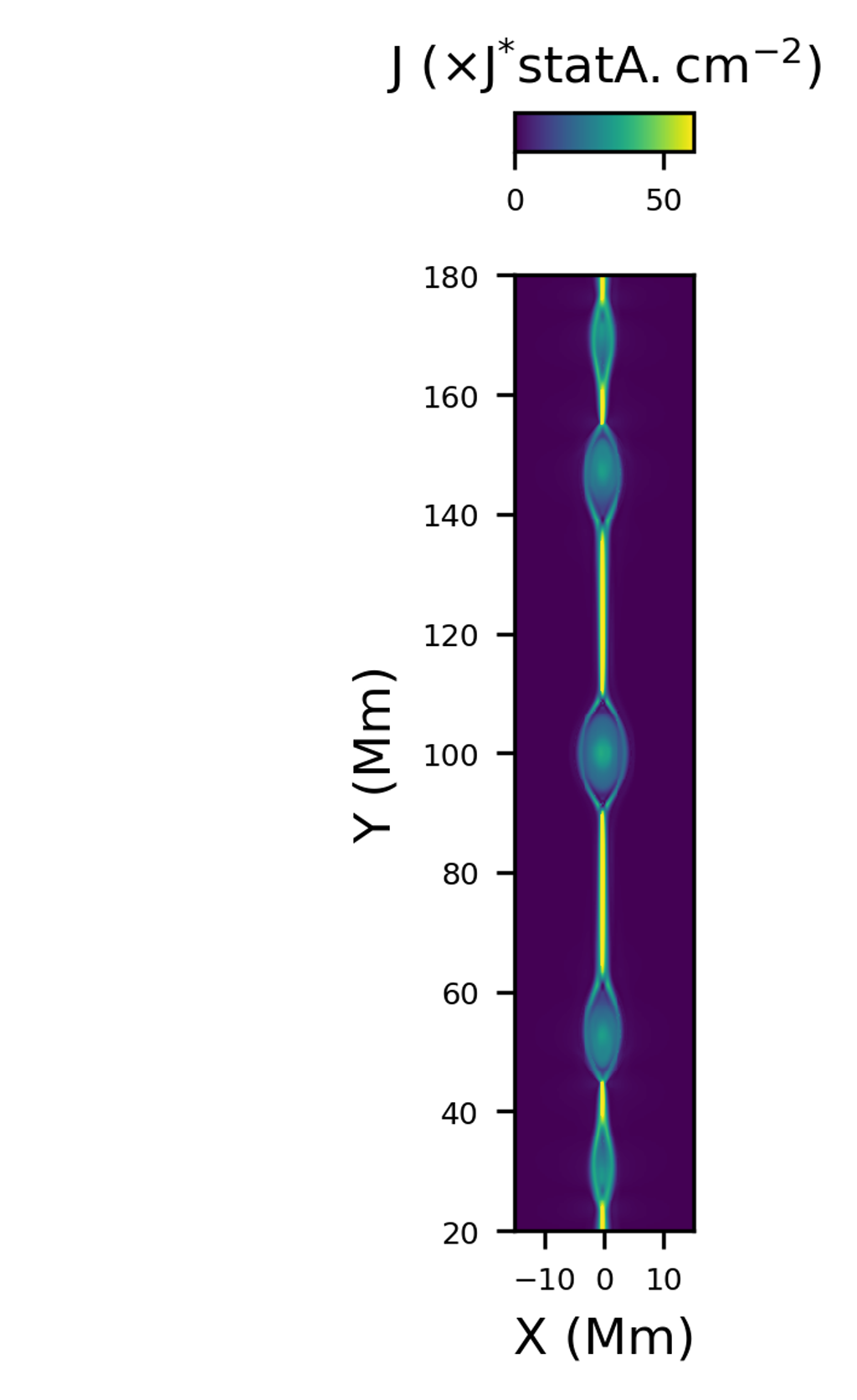}
\includegraphics[height=6.5 cm,trim={6.5cm 0 0 0},clip]{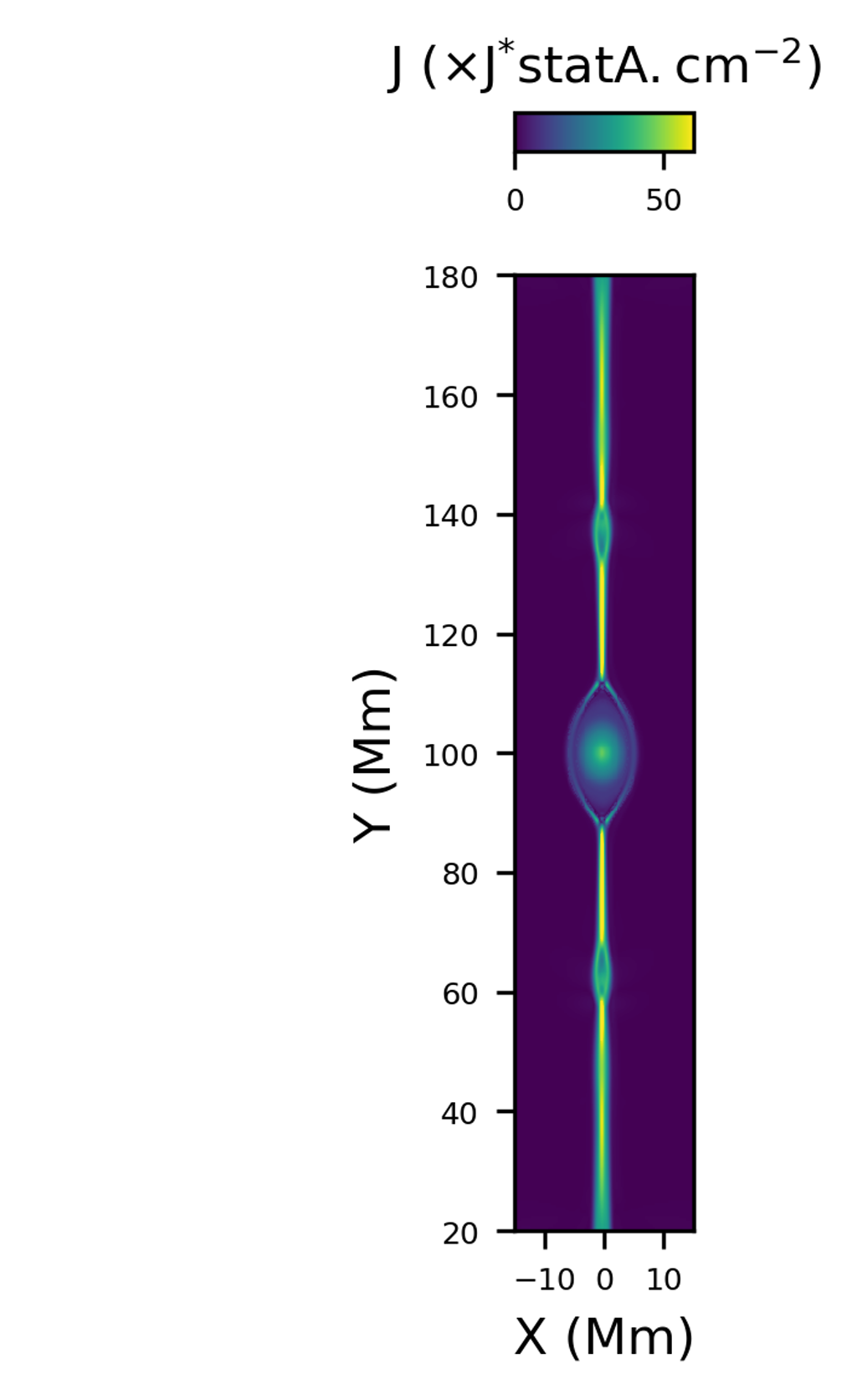}
\includegraphics[height=6.5 cm,trim={6.75cm 0 0 0},clip]{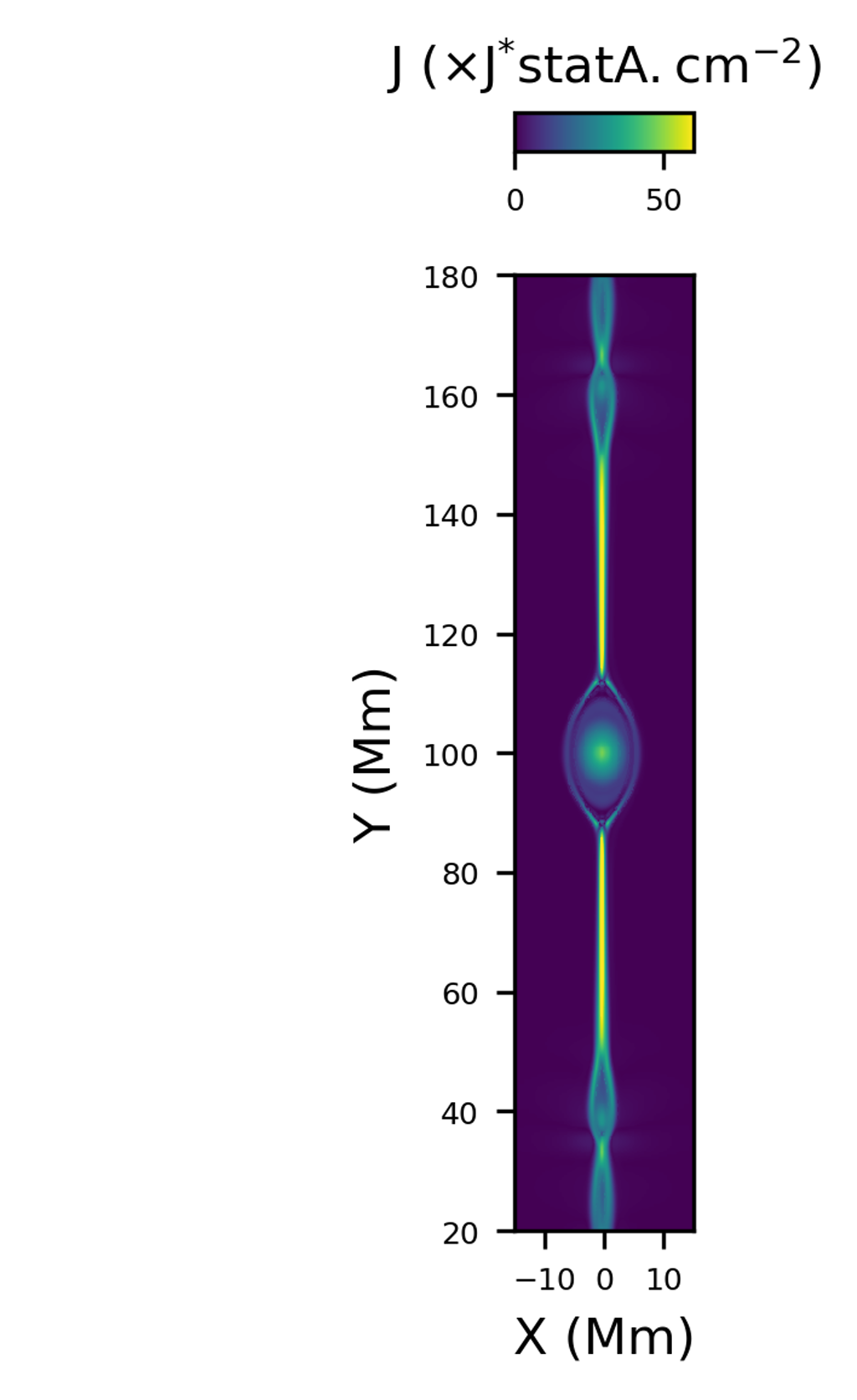}
}

\mbox{
\hspace{-1.75 cm}
\includegraphics[height=6.5 cm,trim={1.0cm 0 0 0},clip]{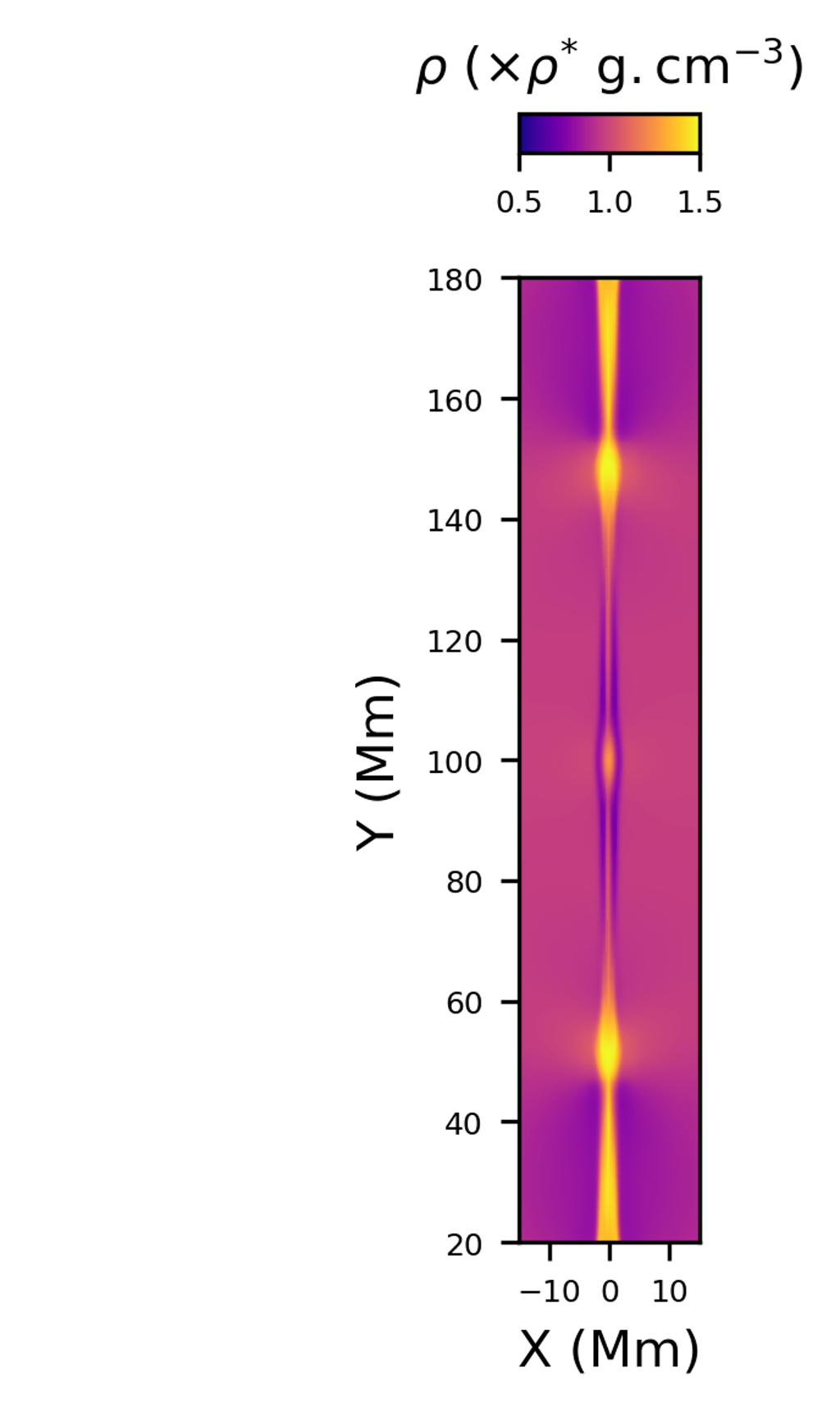}
\includegraphics[height=6.5 cm,trim={6cm 0 0 0},clip]{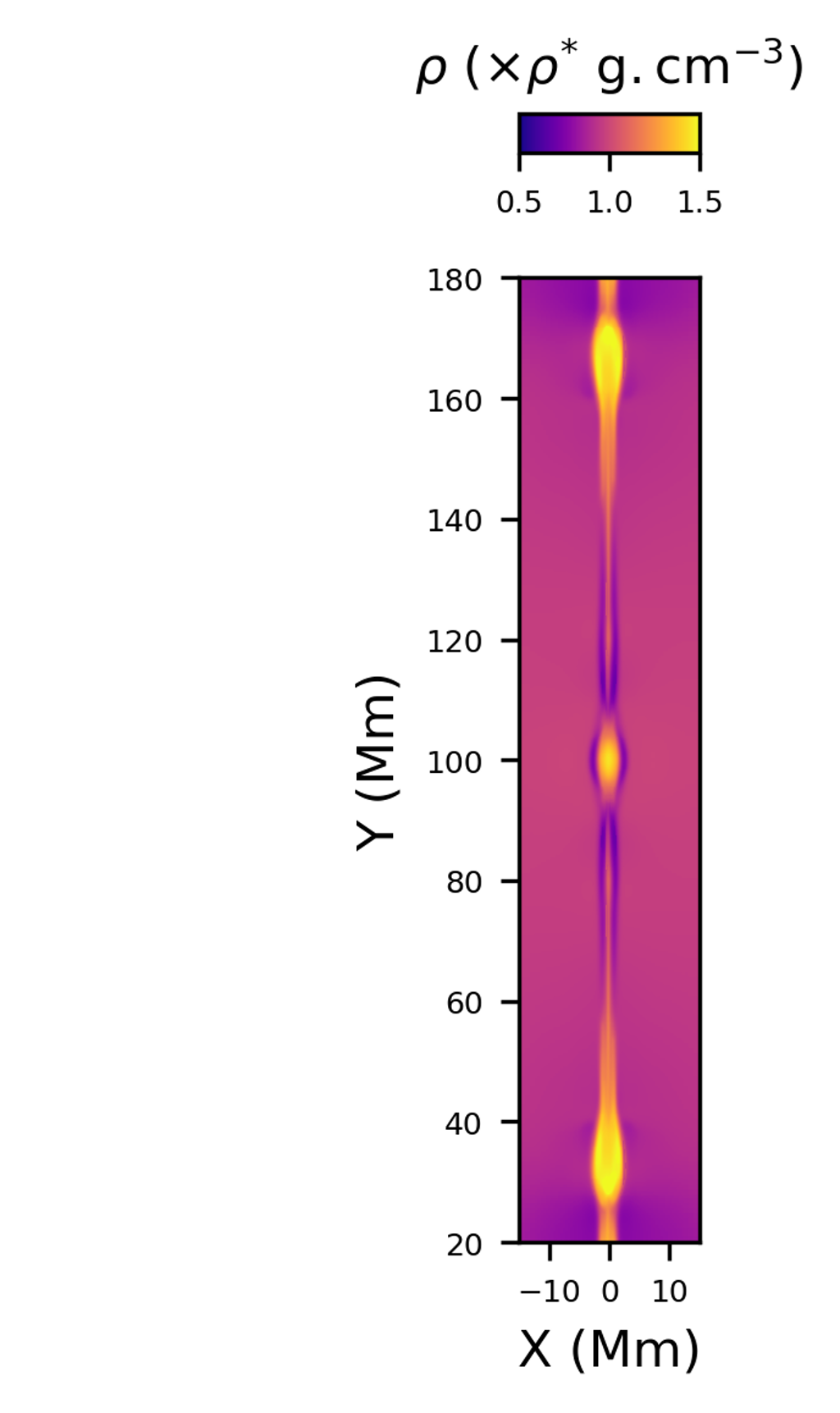}
\includegraphics[height=6.5 cm,trim={6cm 0 0 0},clip]{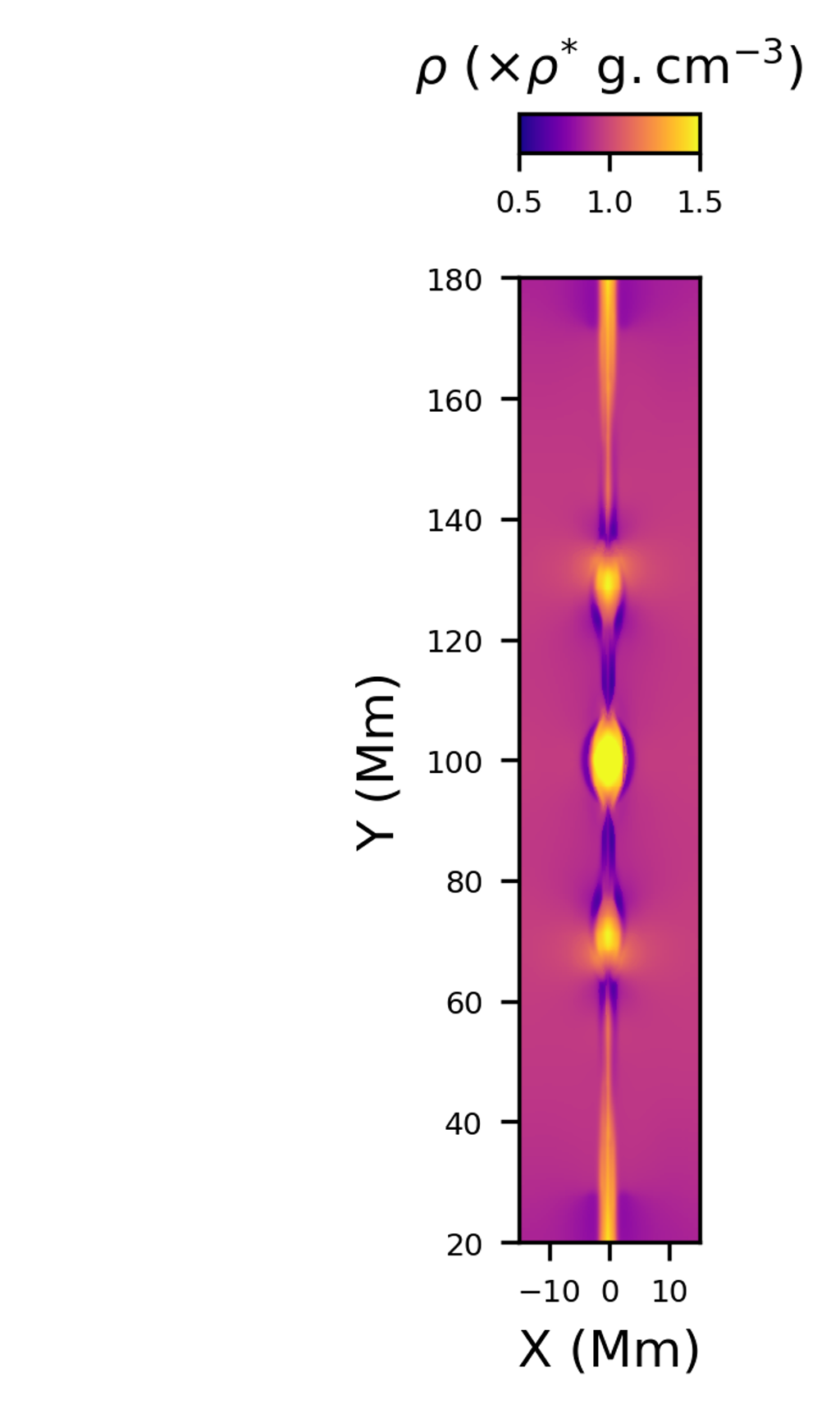}
\includegraphics[height=6.5 cm,trim={6cm 0 0 0},clip]{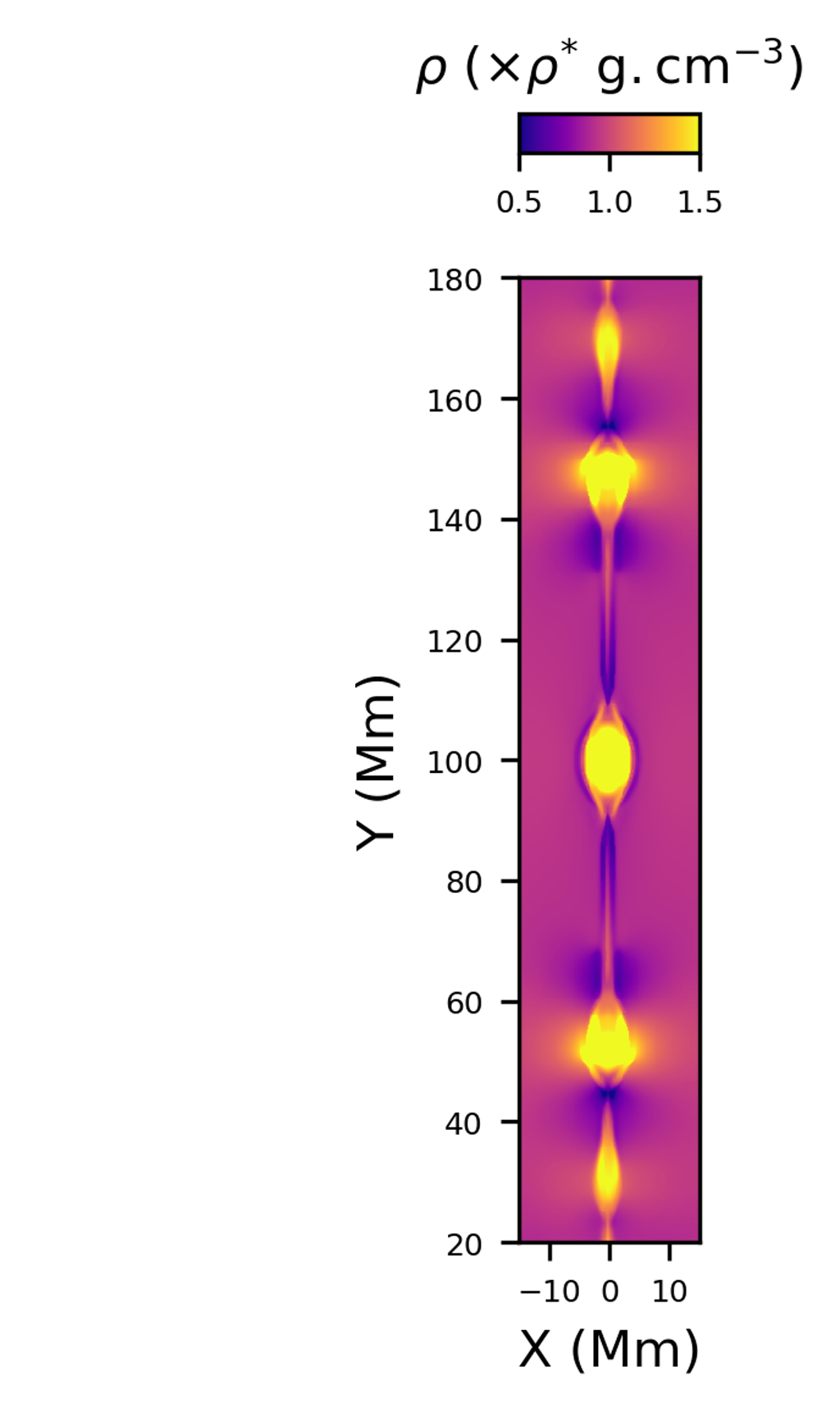}
\includegraphics[height=6.5 cm,trim={5.5cm 0 0 0},clip]{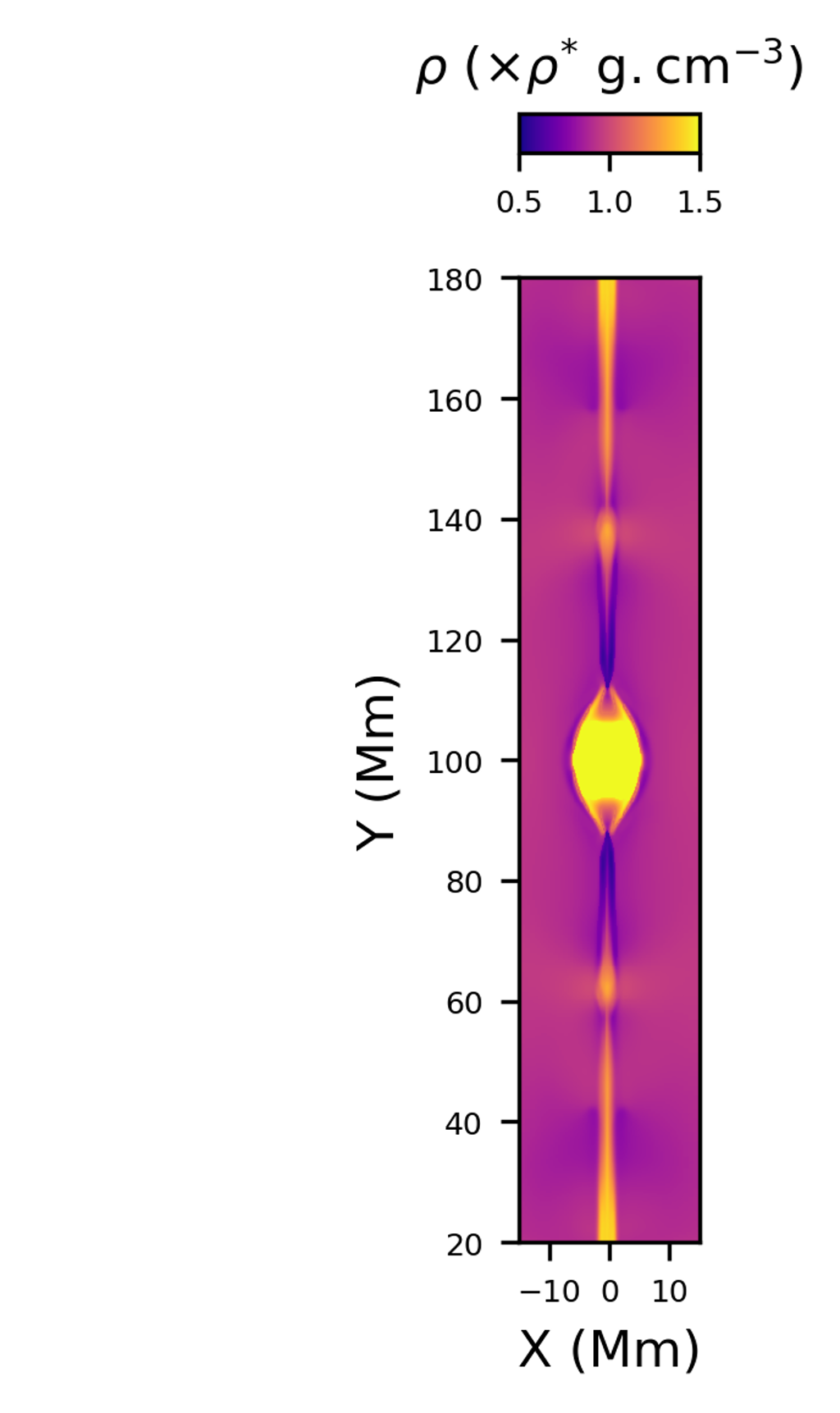}
\includegraphics[height=6.5 cm,trim={5.5cm 0 0 0},clip]{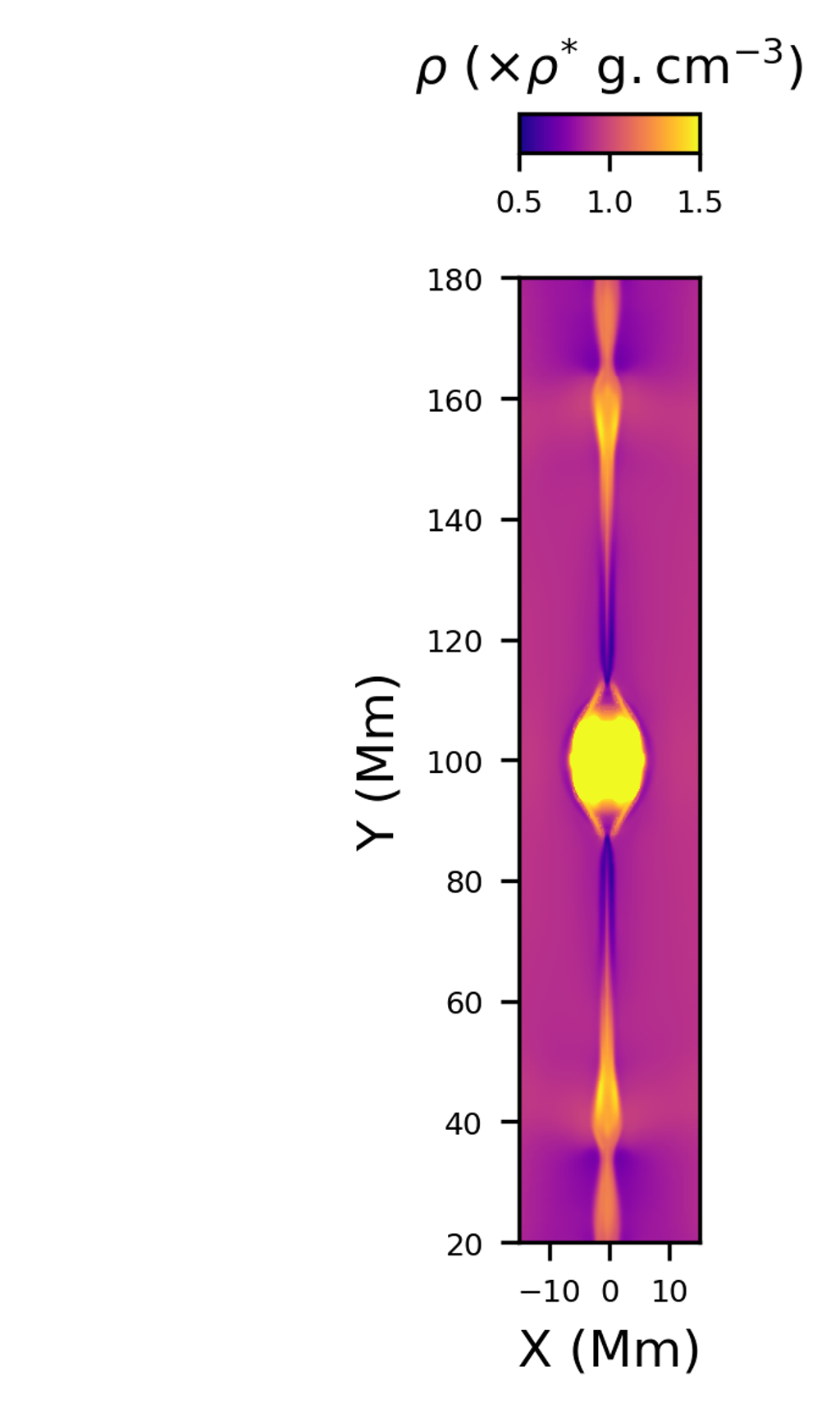}
}

\mbox{
\hspace{-1.75 cm}
\includegraphics[height=6.5 cm,trim={1.0cm 0 0 0},clip]{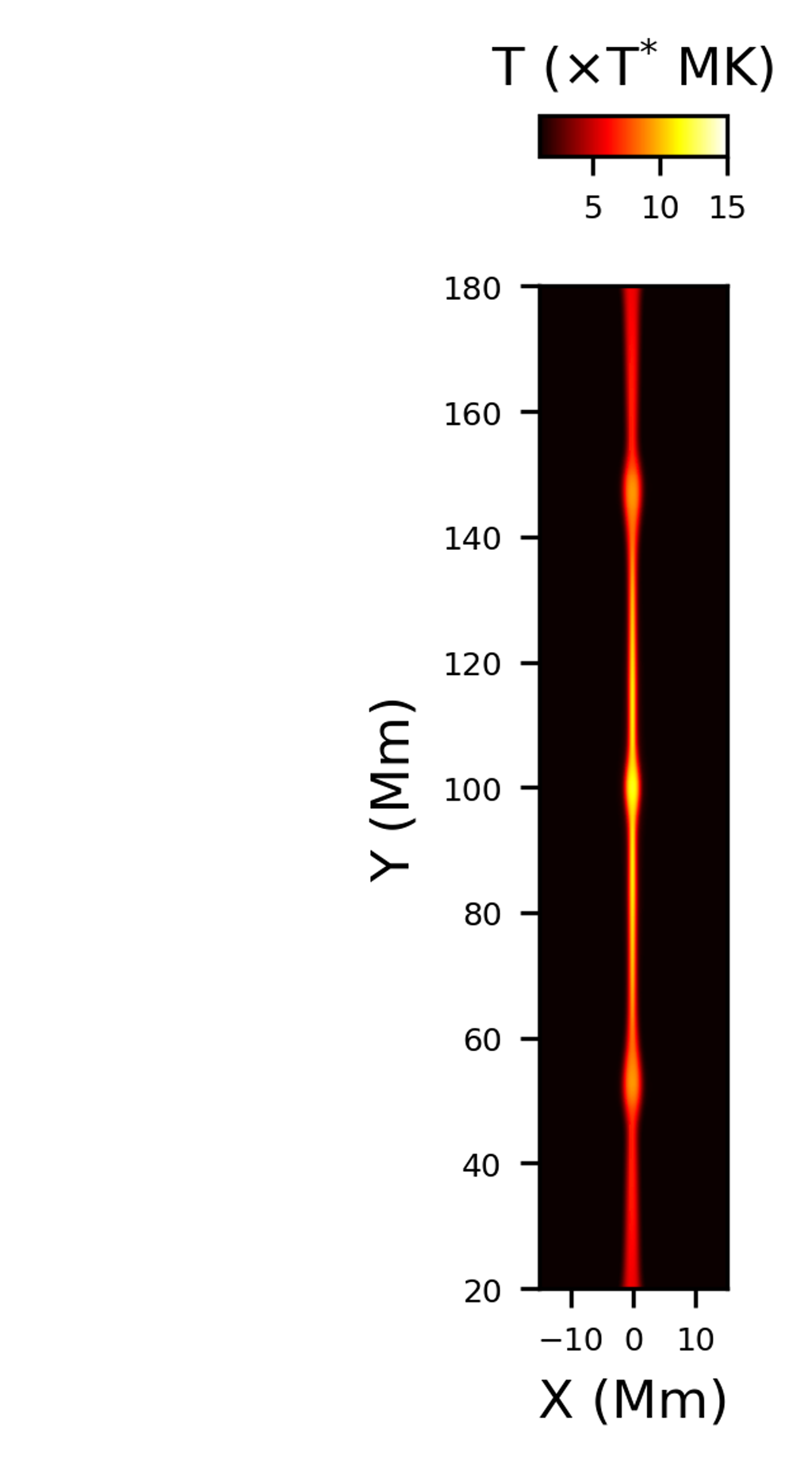}
\includegraphics[height=6.5 cm,trim={4.0cm 0 0 0},clip]{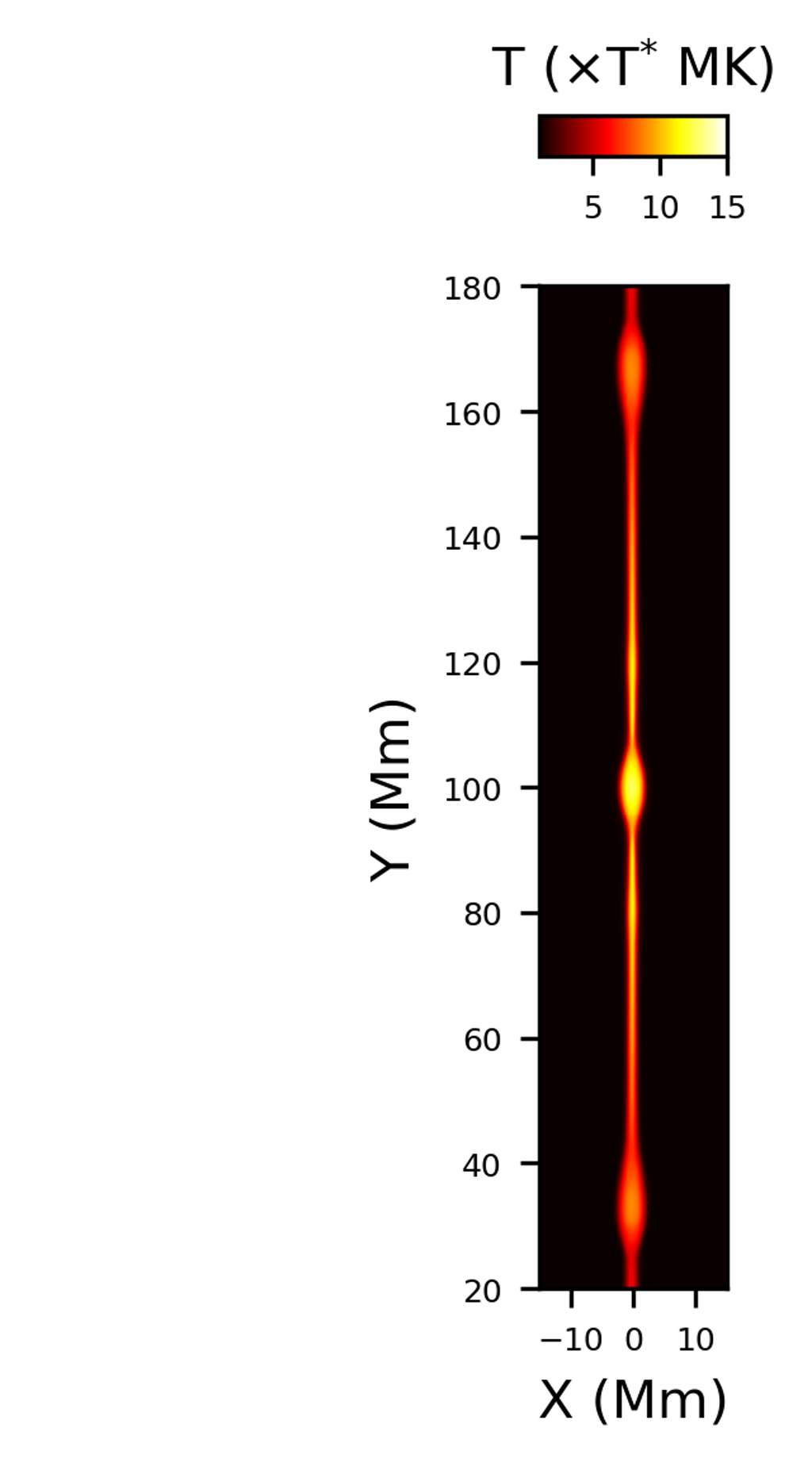}
\includegraphics[height=6.5 cm,trim={4.0cm 0 0 0},clip]{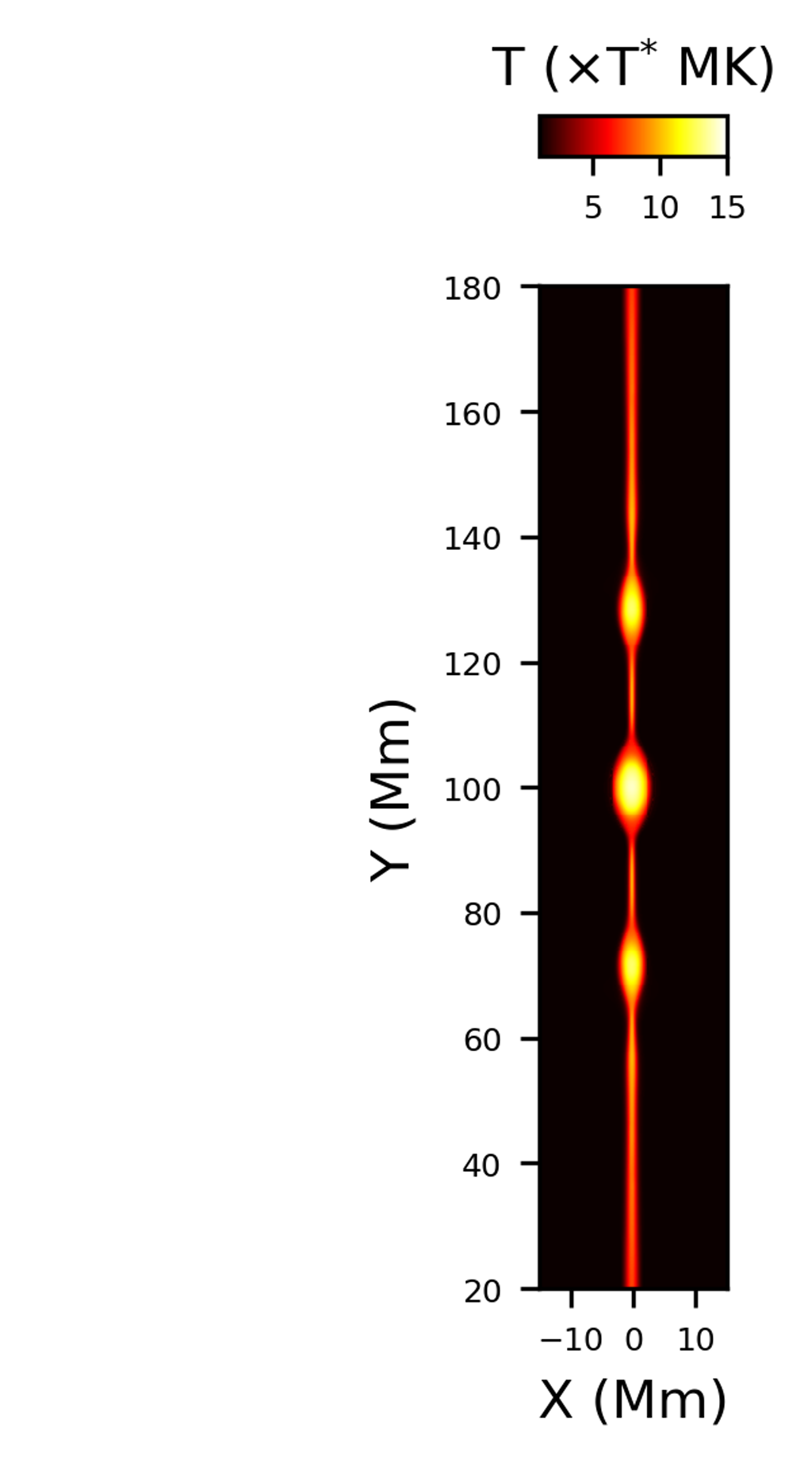}
\includegraphics[height=6.5 cm,trim={3.5cm 0 0 0},clip]{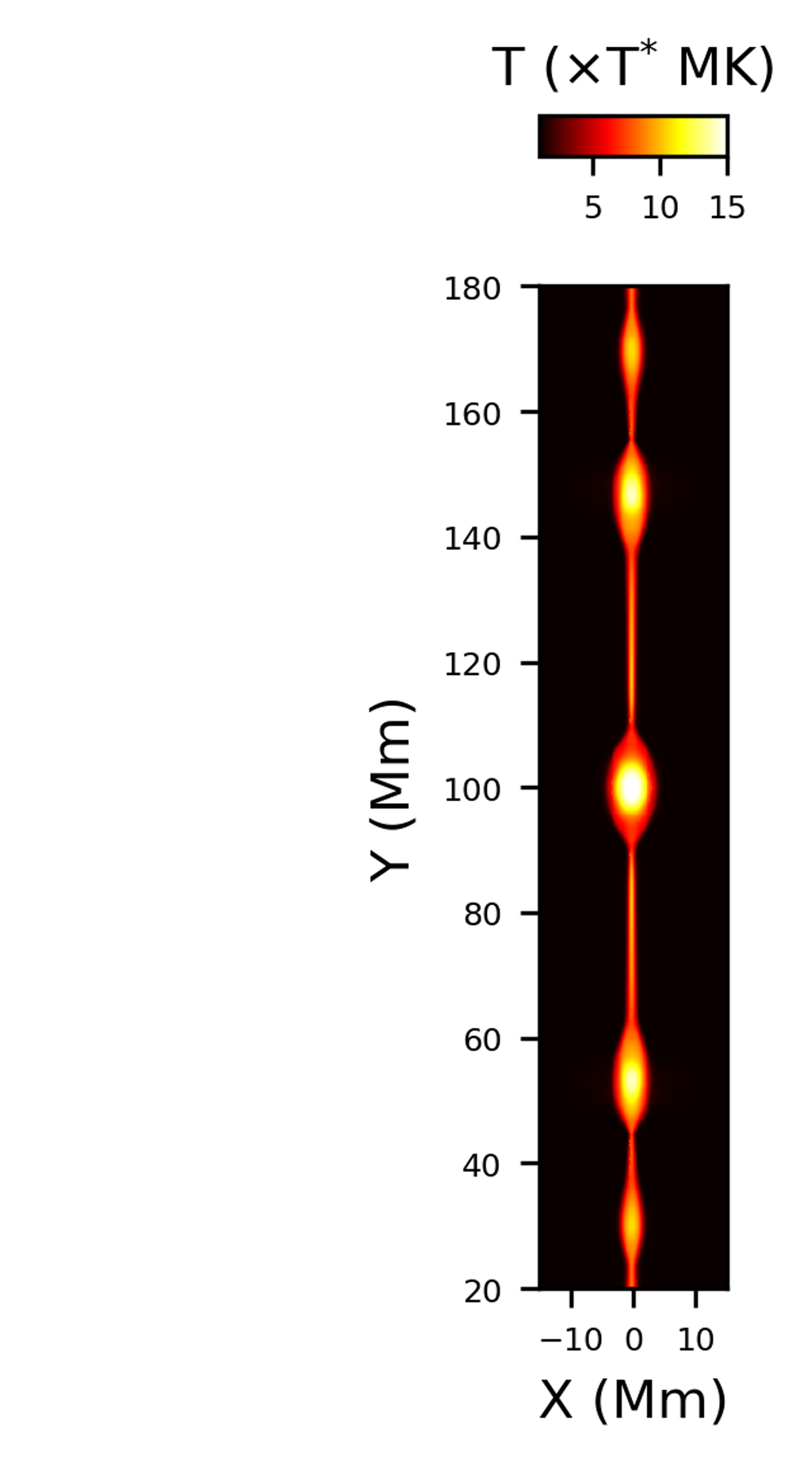}
\includegraphics[height=6.5 cm,trim={3.5cm 0 0 0},clip]{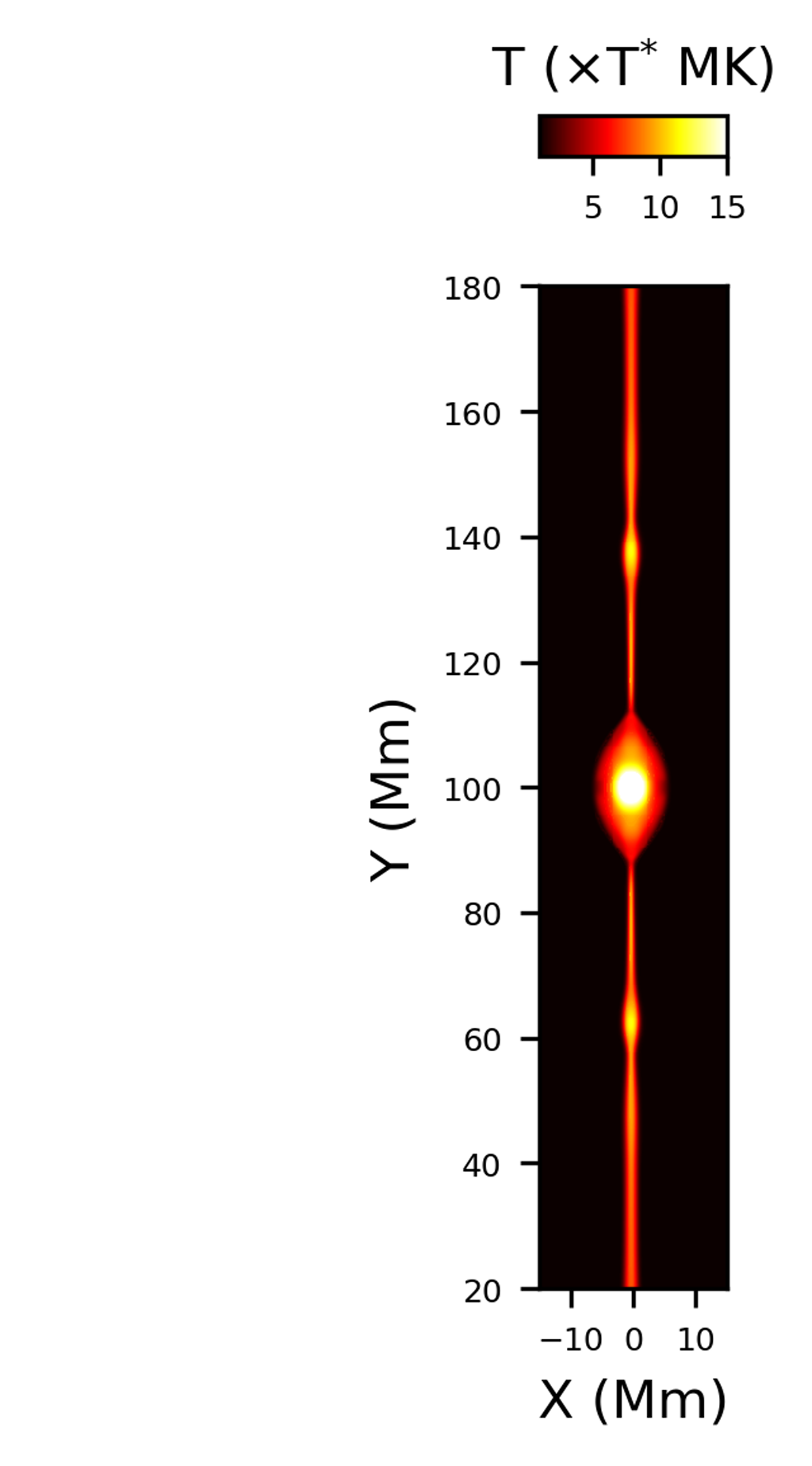}
\includegraphics[height=6.5 cm,trim={3.5cm 0 0 0},clip]{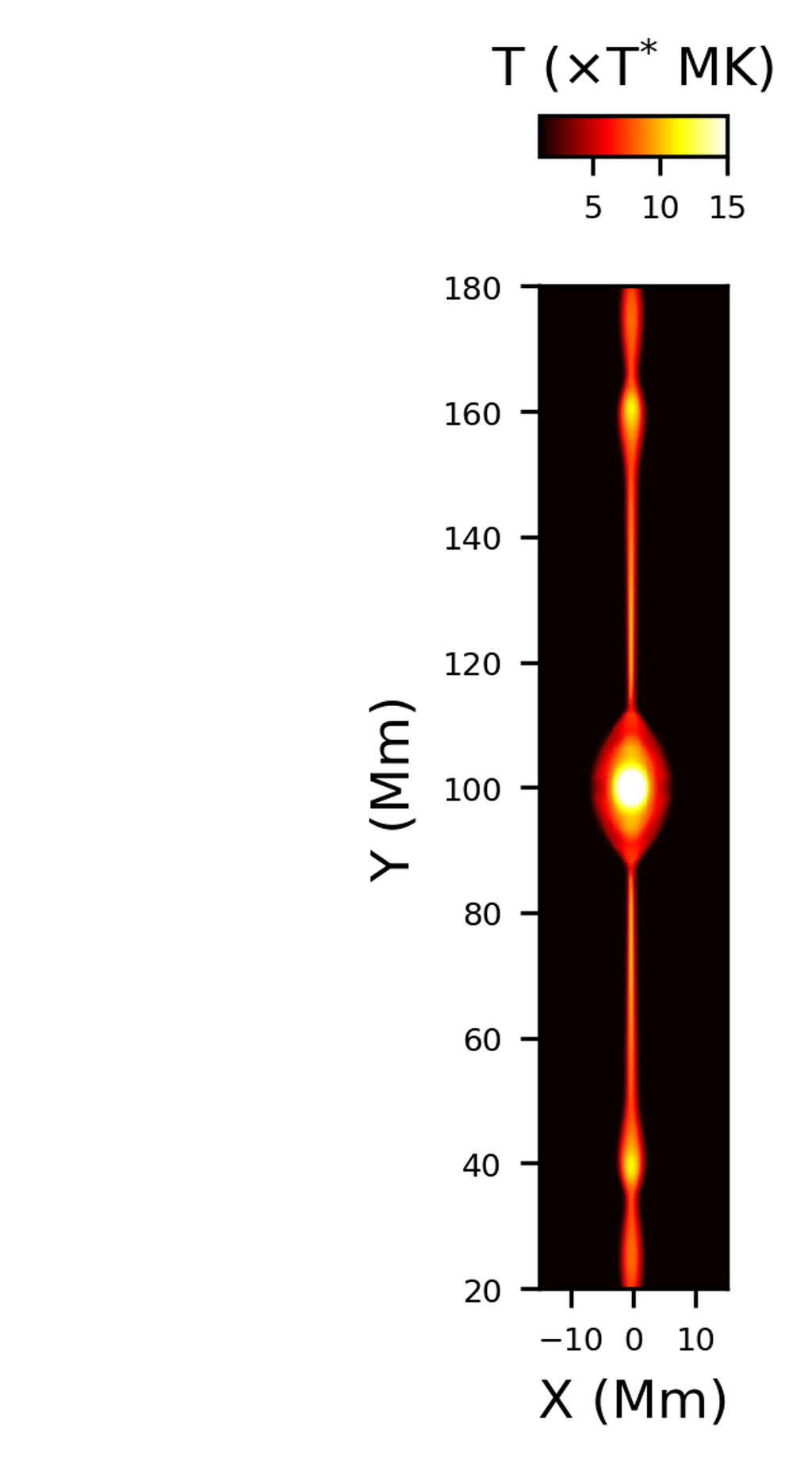}
}

\caption{Top: Snapshots of total current density ($J$) at times 733, 781, 866, 938, 1130 and 1178 seconds, showing three stages of tearing of the CS and plasmoid formation and their outward movements along the CS along with the growth of a steady plasmoid at height of 100 Mm. Middle: Similar snapshots for density at same times. Bottom: Similar snapshots for temperature at the same times. The entire evolution of density and temperature from 613 seconds to 1178 seconds and $J$ from start of the simulation to 1178 seconds are available as animations in the online HTML version. The real-time animation durations for density and temperature are 4 seconds whereas that for $J$ is 10 seconds. The rectangular boxes (cyan for density and temperature; orange for $J$ ) are used in the animated versions to denote the FOV exhibited in this figure. The arrows and annotations in the animation of density maps show the outward movements of the plasmoids formed in multiple stages of tearing.}
\label{label 6}
\end{figure*}

At the primary stage of tearing, three plasmoids become visible around 733 seconds. The plasmoids formed at 60 and 140 Mm are seen to move outward along the CS (as shown in first two columns of  each row of Figure \ref{label 6}). On the other hand, the plasmoid formed at 100 Mm (middle of the CS) remains fixed due to the assumed symmetry of the magnetic  configuration as well as  the external velocity perturbation. This is discussed in detail in Appendix C. Nevertheless due to the outward movement of the plasmoids in the first stage, there is a loss of mass from the CS, and so a tendency for further thinning of the CS, which results in secondary tearing around 866 seconds. This tearing evetually gives rise to the formation of new plasmoids in those regions as shown in the third and fourth columns  of Figure \ref{label 6}. During this stage, two more off-centred plasmoids are formed at larger distances from the center of the CS. Eventually all those off-centred plasmoids move outward with time. Therefore, a third stage of plasmoid formation takes place around 1130 seconds. These plasmoids also eventually move out along the CS (as shown in the last two columns of Figure \ref{label 6}). The temporal evolution of the CS from the initiation of fragmentation followed by multiple stages of tearing and plasmoid formation is clearly exhibited in the animation accompanying Figure \ref{label 6}. The different stages of plasmoid formation and their outward movements are evident also in the time-distance map of Figure \ref{label 7}. The entire temporal regime for the multiple stages of tearing is shown in the animation of density, $J$ and temperature. The plasmoids grow during their outward motion. The central plasmoid also grows in time to a larger size than the other plasmoids. It can absorb plasma from up to the midpoint of the CS layer connecting the other plasmoids and resulting in the rapid growth of its size. Therefore the plasmoid formed at the centre of the CS evolves to become a monster plasmoid with a higher growth rate than the off-centred plasmoids \citep[e.g.,][]{2010PhRvL.105w5002U,2012PhPl...19d2303L,2017ApJ...835..191N}. \citet{2009PhPl...16k2102B} studied tearing without symmetry and found all the plasmoids were swept along the sheet. It will be interesting in future to compare with their results when a lack of symmetry about the centre of the sheet is allowed in the presence of thermal conduction and radiative cooling.

We have so far discussed how the CS undergoes several stages of evolution and eventually reaches an impulsive bursty stage of dynamics in the form of multiple stages of tearing followed by plasmoid formation. Now let us discuss the magnetic, dynamical and thermal characteristics of the formed plasmoids in the next section. Since the off-centred plasmoids are moving, it is difficult to extract their detailed properties. But we have a nearly steady plasmoid formed at a height of 100 Mm as a result of the symmetry of our simulation setup. Therefore we find the detailed thermal characteristics of this central plasmoid. Moreover, the moving plasmoids and their kinematical properties are tracked in the distance-time maps shown in Figure \ref{label 7}.

\subsection{Characteristics of the Plasmoids}

(i) From the top row of the Figure \ref{label 6}, the appearance of plasmoids or magnetic islands with O-points can be seen in the $xy$-plane. In the case of three dimensions, the addition of the $z$-component of the magnetic field means that they will be magnetic flux ropes in three dimensions. The off-centred moving plasmoids are growing as they are moving outward, and the steady central plasmoid also grows with time. Initially, the internal structure of the plasmoid is difficult to quantify due to its small size. But as time progresses, the plasmoid shows a clear internal structuring of $J$ with a magnetic O-point at the centre of the plasmoid having a higher value of $J$ than its surroundings. 

(ii) From the middle row of Figure \ref{label 6}, it can be seen that the plasmoids have higher densities than their surroundings, which results in their presence as bright blobs. Also, a shock moving with the plasmoids is indicated by drastic changes in density across the plasmoids along the CS. The interaction of these shocks with the plasmoids might accelerate them to higher speeds during their outward movement.

(iii) The maximum density at the core of the central plasmoid is 2.8 times the background coronal density during its growth. The average density of this plasmoid is estimated to increase up to twice the background coronal density.

(iv) The temperature within the plasmoids is  higher than their surroundings. As the plasmoids grow bigger in size, the temperature within them also shows an internal structuring. The magnetic O-point is hotter than the ambient neighbourhood within the plasmoid, possibly due to the enhanced Ohmic heating there (as shown in the bottom row of Figure \ref{label 6}).

(v) The maximum temperature at the core of the centred plasmoid is 20 MK during its growth, whereas its average temperature is 8 MK. 

\begin{figure*}[t]
\mbox{
\hspace{-1.0 cm}
\includegraphics[height=6cm, width=10 cm]{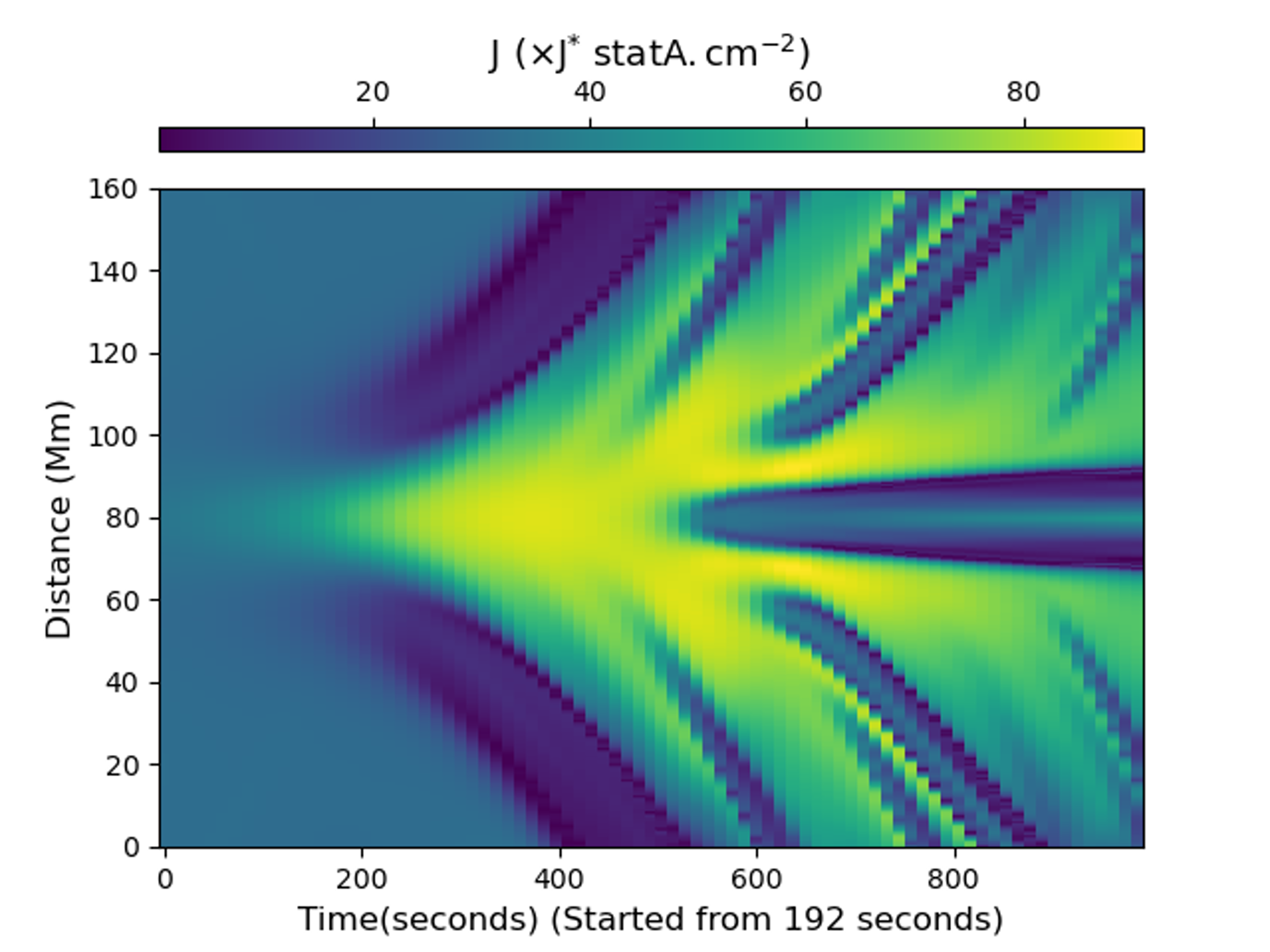}
\includegraphics[height=6cm, width=10 cm]{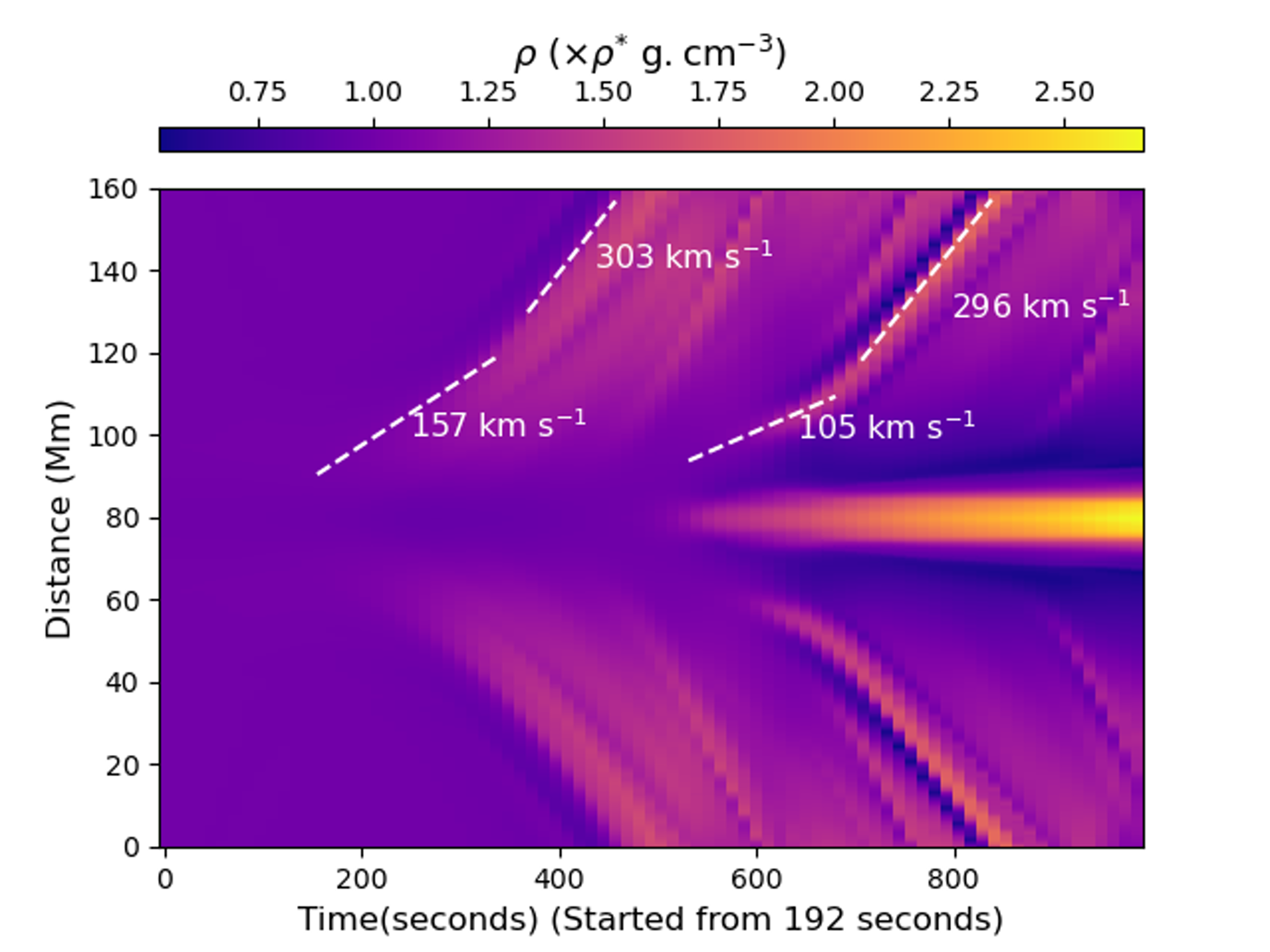}
}
\mbox{
\hspace{-1.0 cm}
\includegraphics[height=6cm, width=10 cm]{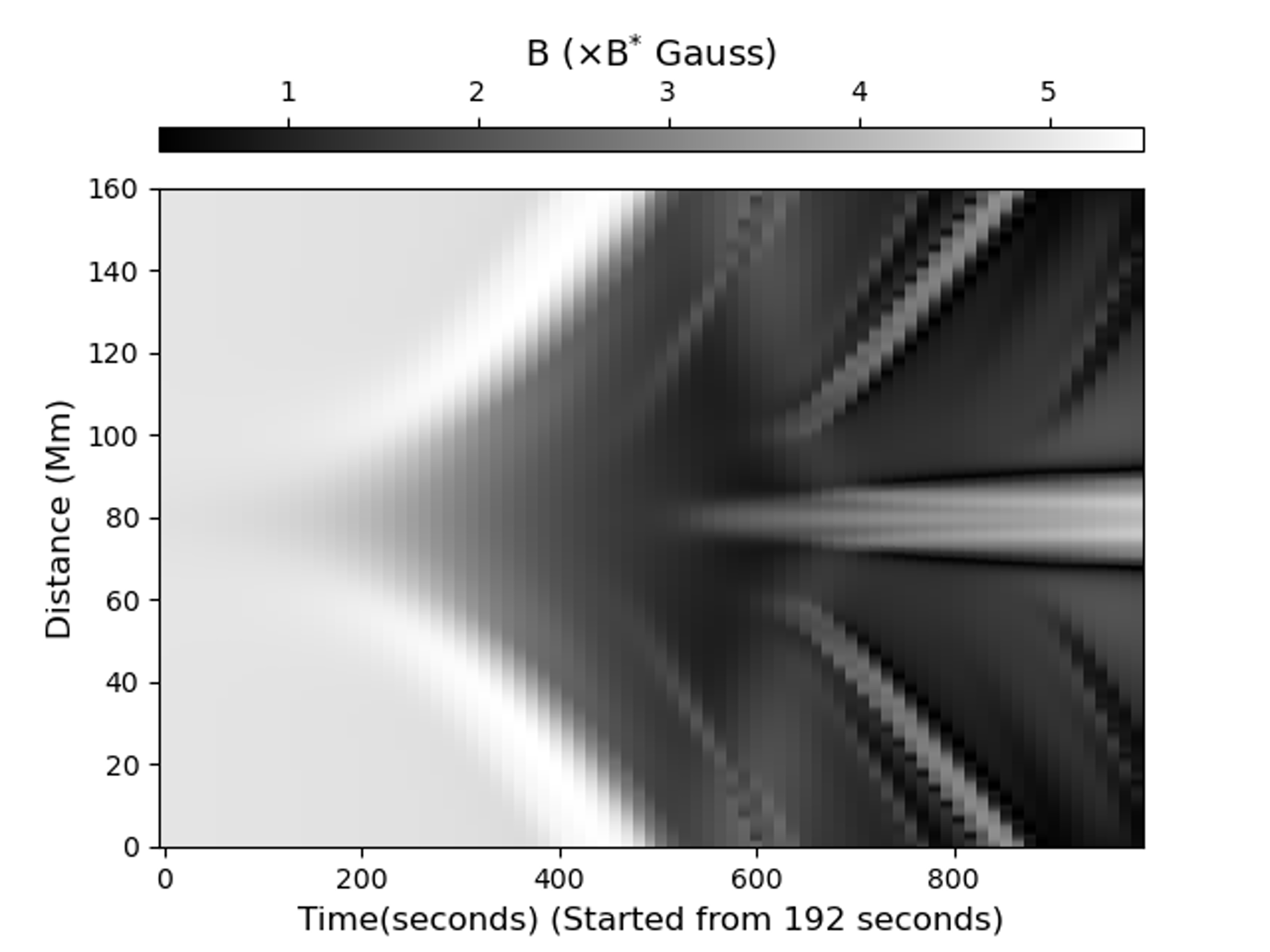}
\includegraphics[height=6cm, width=10 cm]{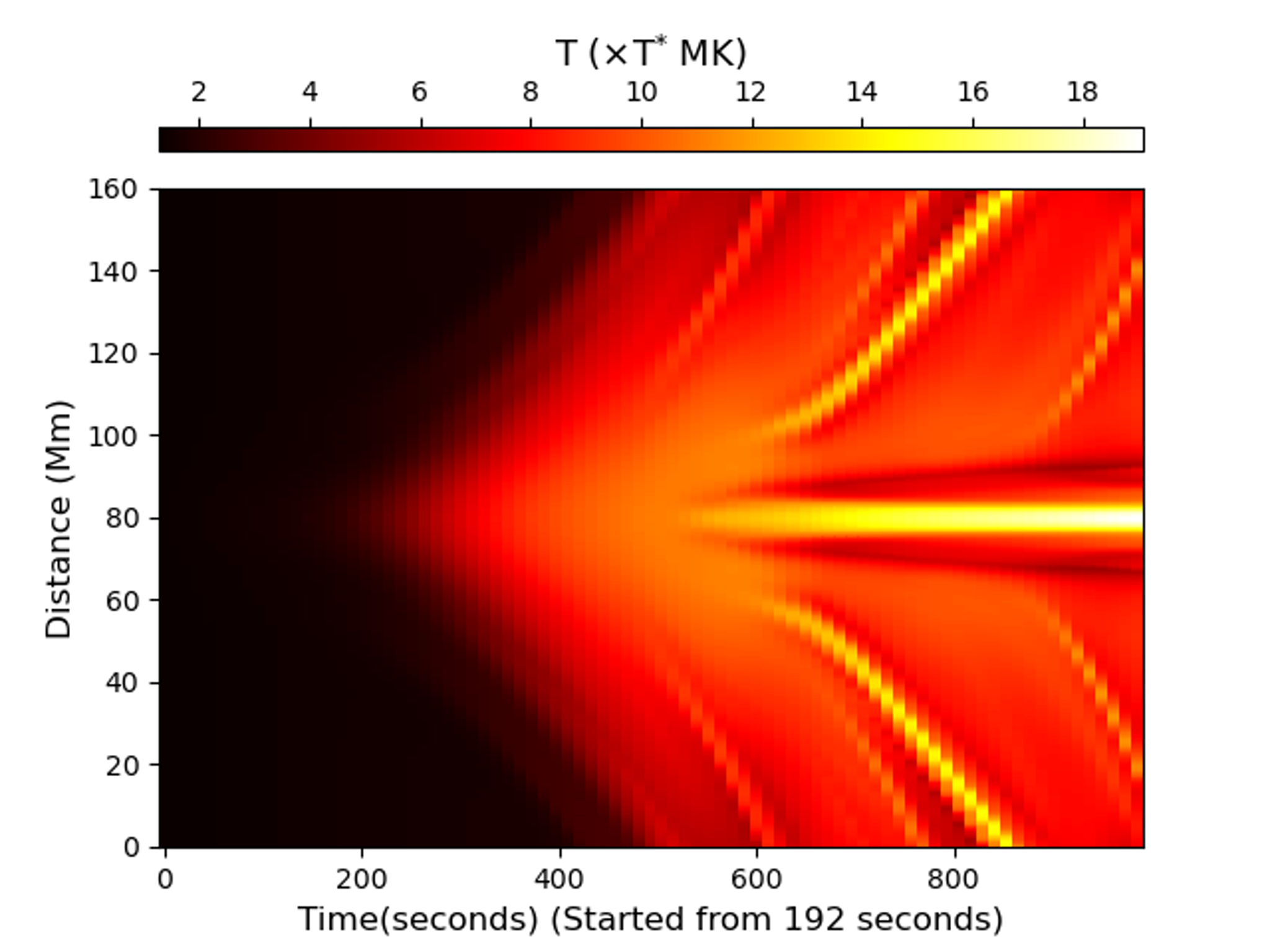}
}
\caption{Time-distance maps of $J$, density, $B$ and temperature extracted from a slit extended from 20 Mm to 180 Mm along the CS at $x = 0$ Mm. These maps show the temporal evolution of the variables along the CS. They clearly show the bi-directional outflows, three stages of bi-directional movement of plasmoids and the growth of the steady central plasmoid.}
\label{label 7}
\end{figure*}

(vi) We estimate the speed of the outflow as well as the outward moving plasmoids from the time-distance map of density. It can be seen that the outflow speed increases from 157 km \(\mathrm{s^{-1}}\) to 303 km \(\mathrm{s^{-1}}\) (\(0.27\mathrm{v_{A}}\) to \(0.52\mathrm{v_{A}}\)) during their outward passage (as shown in the density time-distance map in Figure \ref{label 7}). This acceleration of the reconnection outflow might be due to the magnetic tension force.

Along with this continuous outflowing plasma, there exist distinct chunks of plasma in the form of plasmoids which are moving out along the CS. We find that the speed of the plasmoids is also increasing during their passage from 105 km \(\mathrm{s^{-1}}\) to 296 km \(\mathrm{s^{-1}}\) (0.18 $v_{A}$ to 0.51 $v_{A}$), since they are being accelerated by the magnetic tension force.

\section{Discussions}\label{sec:4}
We impose a Gaussian velocity pulse, as a model of fast magnetoacoustic waves, to disturb an elongated CS under solar coronal conditions. The interaction of the pulse with the CS eventually results in a inward pressure as well as a density gradient towards the CS in the $x$-direction. This gradient results in a bi-directional inflow, which eventually drives  magnetic reconnection in a small diffusion layer. This magnetic reconnection initially possesses a Petschek-type character with the presence of slow-mode MHD shocks at the ends of the diffusion layer, as described in section 3.1.1 (point (b) and (c)) and shown in Figure \ref{label 3} and in left column of Figure \ref{label 4}. Due to an unbalanced magnetic tension force, bi-directional outflow sets in to stretch the effective diffusion layer. Eventually, the CS becomes of Sweet-Parker type for a short period. Gradually, tearing or plasmoid instability sets in via the fragmentation of the physical quantities such as current density, magnetic field and total pressure, as described in section 3.1.2 and depicted in Figure \ref{label 5}. The plasmoids become properly visible around 733 seconds of the simulation. Initially, three plasmoids are formed within the CS. Two off-centred plasmoids move outwards, while the central one remains stationary due to  the symmetry of the imposed velocity disturbance and  the initial state. Actually, the $y$-component of the Lorentz force shows that there are equal and opposite forces acting on the central one, which makes it stationary. 

Nevertheless, the outward movement of the other two plasmoids results in further thinning and secondary tearing and plasmoid formation within the CS. This secondary stage of tearing is followed by two similar  stages of plasmoid formation up to the end of the simulation, i.e., 1178 seconds. Even though the reconnection process finally operates in an impulsive and bursty manner,  we note that the proxy of the reconnection rate, i.e., \(\eta J_{max}\) (middle panel of Figure \ref{label 2}) undergoes a prominent rapid fall  later after the primary onset of tearing. Actually, if a plasmoid grows to large size but is restricted to remain at the center of the computational domain as in our case due to the imposed symmetries, the plasmoid instability hardly reaches a highly nonlinear phase. Even if it attains that phase it will be short-lived and will eventually result in a rapid fall of the reconnection rate \citep{2009PhPl...16k2102B}. Next, let us make a few remarks about the initial simulation setup and its final outcome in the context of magnetic reconnection in the solar corona.

\subsection{Physical Feasibility of the Dynamics}
As noted above, we have performed a numerical simulation of magnetic reconnection under solar coronal conditions, a system with high Lundquist number subjected to external velocity perturbations. Now, when simulating magnetic reconnection in a high-Lundquist number system, if the resolution is not very high,  one needs to assess whether the dynamics is physically viable. Now for Lundquist number \(2.51 \times 10^{5}\) and \(6.28 \times 10^{5}\), \citet{2009PhPl...16k2102B} used [2001, 1001] and [3001, 1001] grid points, respectively. Also, \citet{2022A&A...666A..28S}  have studied plasmoid-related physics with [2048, 2048] grid points with the smallest grid size being 125 km using MPI-AMRVAC and the highest value of their Lundquist number is \(2.34 \times 10^{5}\). Thus, our value of Lundquist number, i.e., \(4.8 \times 10^{5}\) is about twice their value. So, to find out whether the highest resolution with [2048, 2048] grid points with smallest grid size being 97.5 km in our work is sufficient or not, we need to examine whether our physical choice of resistivity is higher than numerical resistivity. Simultaneously, we need to check whether we have more than one grid point within the inner layer of the tearing mode. We find that our model does satisfy these two important criteria, and so it is likely to generate physically acceptable results. These estimates are discussed in detail in Appendix A. In addition, a strong guide field is present in the solar corona and can have significant impact on plasmoid dynamics. Hence we examined whether our choice of magnetic setup provides non-zero guide field in the  vicinity of the CS or not (see the details in Appendix B). It is found that an appropriate guide field is present in the  highly localized coronal current sheets. 

\subsection{Comparison with  Observations}
Here, we compare the dynamics and reconnection in the modelled solar coronal current sheet with features of  current sheets and plasmoid dynamics  seen in various observations. However, an accurate estimate of plasma beta, guide field and Alfv\'en speed is not possible. But we can compare the properties of the plasmoid dynamics qualitatively with those reported in the various solar coronal observations. Basically we intend to compare the morphological  plasma dynamics in observations with our current sheet dynamics. This enables us to infer some physical information regarding the observed current sheet dynamics. \citet{2016SoPh..291..859Z} estimated the number density of observed plasma blobs in EUV jets to be 1.7 to 2.8 times the number density of the background solar corona. So our estimate of the average density of plasmoids to be twice the background density of the solar corona as well as the maximum density at the core of plasmoid being 2.8 times the solar coronal plasma density compare well with the observational estimates. The average temperature of the plasmoid is estimated to be up to 8 MK. \citet{2022ApJ...924L...7L} reported an observation of tearing instability in the CS of a solar flare in which they found the blob temperature to be around 6.4 MK initially which later increased to 7.7 MK. So our estimated average plasmoid temperature is similar to what is measured for plasma blobs in observations. We found the speeds of reconnection outflows to be 157-303 km \(\mathrm{s^{-1}}\) and of outward moving plasmoids to be 105-296 km \(\mathrm{s^{-1}}\). These estimates are very similar to those reported in various coronal observations \citep{2012ApJ...745L...6T,2016SoPh..291..859Z}. We also noticed that both the reconnection outflows and plasmoids accelerate in time. \citet{2008A&A...477..649B} reported similar acceleration characteristics of reconnection outflows and plasmoids. So we see that the dynamical properties of plasmoids as well as their temperature and density are very similar to those from coronal observations. We find that the maximum temperature at the core of the central plasmoid reaches up to 20 MK. Also there exists an internal structuring of density and temperature within all the plasmoids. \citet{1997ApJ...474L..61Y} reported that the presence of thermal conduction does not have a significant effect on the reconnection rate and energy release rate. Nevertheless, this high value of temperature at the core of the plasmoids (due to enhanced Ohmic heating there) as well as the internal structuring of density and temperature are likely to be affected when we include thermal conduction and radiative cooling effects in future. In the next sub-section, we discuss the inference of the velocity amplitude of the disturbances on CS dynamics.

\subsection{Dependence on the Amplitude of the Velocity Pulse}
Fast MHD waves or EUV waves can have different strengths in the solar corona. So we have tried  different amplitudes for the velocity pulse to find out their effect on the reconnection. Although the physics of the different processes are similar, there are some variations in dynamical behaviour as follows--
\newline
[I] We find that the intermediate time between initiation and end of the CS thinning phase depends on the amplitude of the velocity pulse. For higher amplitudes, this CS thinning phase occurs more quickly.
\newline
[II] The transformation from Petschek to Sweet-Parker reconnection also happens earlier in time for higher amplitudes of the velocity pulse.
\newline
[III] The temporal length of Sweet Parker regime of reconnection is similar in all the cases and so independent of the amplitude of velocity pulse.
\newline
[IV] Likewise there is little dependence of the tearing stages on the amplitude of the pulse. Multi-stage formation and evolution of plasmoids takes place in a self-consistent manner independent of the amplitude of the pulse.

Besides, a change in the position of the pulse while preserving symmetry is similar to a change in amplitude. Now if the pulse is positioned asymmetrically about the centre of the simulation box, it has different effects since the disturbance created in the current sheet will no longer be symmetric about its length. It will be interesting to examine this asymmetric interaction of the velocity disturbance with the CS independently in a future study. Indeed, a more extensive parametric study with different CS widths in presence of asymmetric pulses can be carried in future.

\section{Conclusions}\label{sec:5}
There are many studies of fast reconnection and plasmoid formation in the literature in which they initiate the magnetic reconnection by a localized enhancement of resistivity \citep[e.g.,][]{1997ApJ...474L..61Y,2009PhPl...16f0701B,2012PhPl...19i2110B}. On the other hand, \citet{1997A&A...326.1252O} simulated the interaction of a shock wave with a CS resulting in reconnection via the onset of tearing mode instability. \citet{2019A&A...623A..15P} also examined the effect of different forms of periodic external velocity perturbations and the initial configuration of a CS on the dynamics of reconnection  and the energy release process. Usually, they used an anomalous resistivity above a current density threshold to achieve fast reconnection and plasmoid formation. \citet{1997A&A...326.1252O} used pressure perturbations to mimic shock waves. \citet{2019A&A...623A..15P} used periodic perturbations which are present for the entire time of the simulation to drive reconnection. 

Both MHD waves and current sheets are likely to be ubiquitous in the solar corona. Therefore, EUV waves or fast MHD waves may sometimes in turn trigger reconnection in other regions, especially where prominences are located \citep{2019ApJ...887..137S}. Now in the solar corona, eruptions are accompanied by the generation and propagation of EUV waves away from the eruption in a transient manner rather than in a periodic  or continuous process \citep{2015LRSP...12....3W,2017SCPMA..60b9631C,2018ApJ...864L..24L,2018A&A...612A.100C}. In this present paper we mimic such a transient process by a Gaussian velocity pulse at the beginning of the simulation. We consider uniform resistivity instead of a localized enhancement of resistivity or an anomalous resistivity to initiate reconnection. This consideration is of interest in the context of externally driven or forced reconnection in the non-ideally conductive solar corona. Therefore we have provided a detailed scenario for the physical processes such as thinning, fragmentation of the CS and magnetic reconnection  when a localized CS is subjected to the external fast magnetoacoustic perturbations. In addition, the evolution of various dynamical plasma processes such as flows, plasmoids, shocks etc in various phases of the reconnection are also discussed.

This study augments the existing literature on theoretical base-line of impulsive bursty reconnection through the plasmoid instability \citep[e.g.,][]{2010PhPl...17f2104H,2012PhRvL.109z5002H,2013PhPl...20e5702H,2015shin.confE..26H,2016ApJ...818...20H} by (i) exploring the triggering of CS instability by an external driver (specifically, a velocity pulse) in the presence of a uniform resistivity, and (ii) employing plasma parameters relevant to the Sun's corona. Simultaneously, our results are consistent with the latest observations showing onset of reconnection and associated plasma dynamics by external perturbations in the corona \citep[e.g.,][]{2018ApJ...868L..33L,2019ApJ...887..137S,2021ApJ...920...18S}. Therefore this study of the different stages of reconnection which eventually triggers plasmoid formations can be a ready reference to explain observations of reconnection following an external forcing via propagation of transient EUV or fast MHD waves triggered during solar eruptions. Even though this present paper focuses on understanding the dynamics of current sheets formed specifically in the solar corona, the present model may be suitably adopted in future to study the details of CS dynamics and magnetic reconnection in other systems.


\begin{acknowledgments}
We acknowledge the scientific comments of the reviewer that improved our manuscript considerably. S.M. would like to acknowledge the financial support provided by the Prime Minister’s Research Fellowship of India. AKS acknowledges the ISRO grant  (DS 2B-13012(2)/26/2022-Sec.2) for the support of his scientific research. D.I.P. gratefully acknowledges support through an Australian Research Council Discovery Project (DP210100709). D.Y. is supported by the National Natural Science Foundation of China (NSFC; grant numbers 12173012, 12111530078 and 11803005), the Guangdong Natural Science Funds for Distinguished Young Scholar (grant number 2023B1515020049), the Shenzhen Technology Project (grant number GXWD20201230155427003-20200804151658001) and the Shenzhen Key Laboratory Launching Project (grant number ZDSYS20210702140800001). 
\end{acknowledgments}

%

\vspace{5mm}


\software{MPI-AMRVAC, Paraview}



\appendix

\section{Physical vs Numerical Resistivity}

\begin{figure*}
\mbox{
\includegraphics[height=6 cm, width=6 cm]{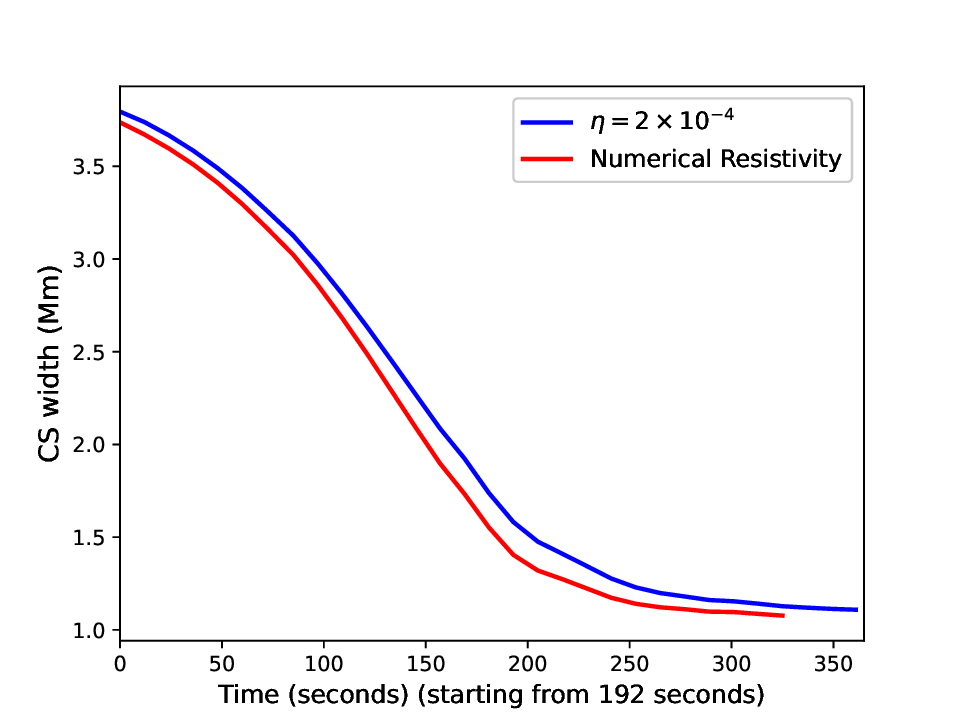}
\includegraphics[height=6 cm, width=6 cm]{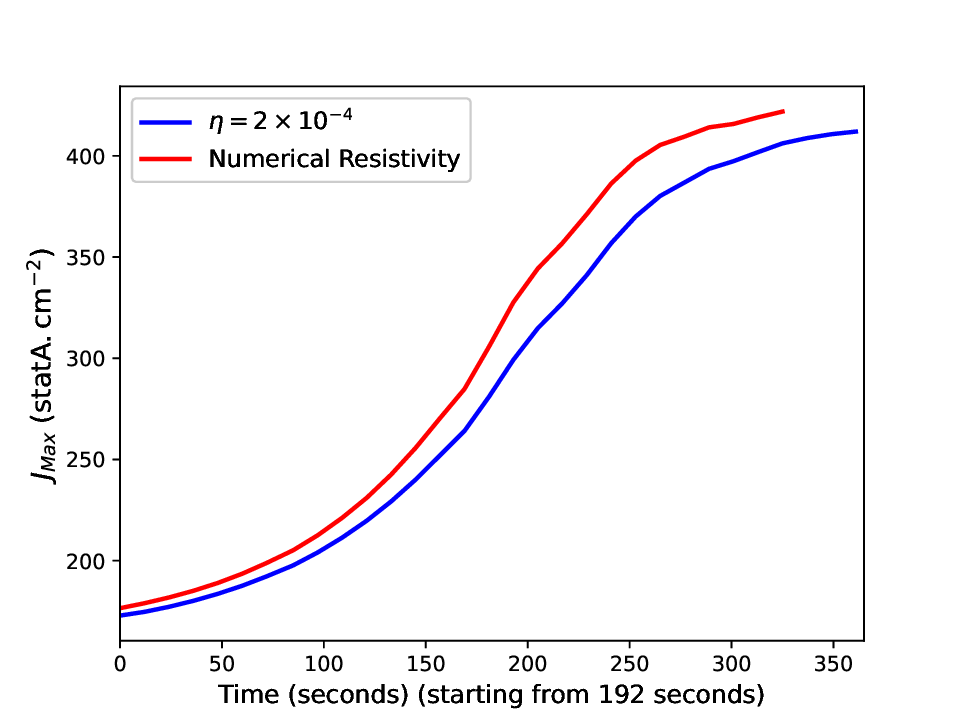}
\includegraphics[height=6 cm, width=6 cm]{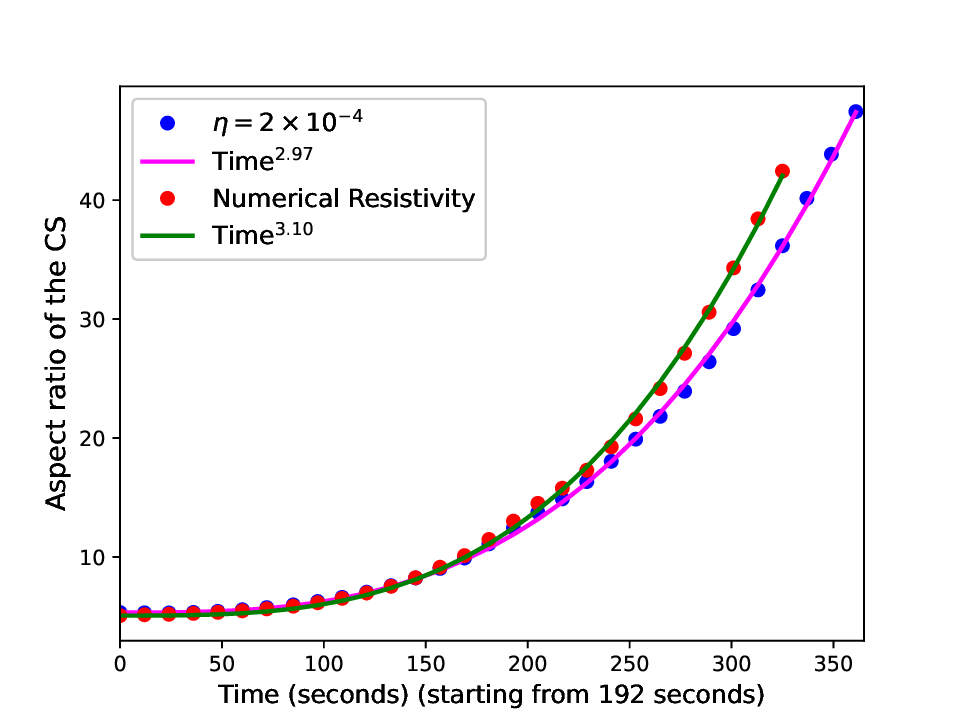}
}
\caption{Left: Temporal evolution of the CS width at a height of 100 Mm for imposed physical resistivity (blue curve) and numerical resistivity (red curve). Middle: Temporal evolution of the maximum $J$ for imposed physical resistivity (blue curve) and numerical resistivity (red curve). Right: Temporal evolution of the aspect ratio of the CS for imposed physical resistivity (blue curve) and numerical resistivity (red curve). It is evident that the exponent of power law fit to the curves of aspect ratios changes from 2.97 to 3.10 when we omit the physical resistivity.}
\label{label 8}
\end{figure*}

\citet{2019MNRAS.485..299R} reported that their simulation of plasmoid formation within the CS at low resolution is predominantly driven by numerical resistivity rather than imposed physical resistivity. They varied the base resolution and maximum level of refinement for various values of physical resistivity to find out the effective resolution beyond which the thickness of the CS no longer depend on the resolution, i.e., numerical resistivity does not play any role in the dynamics. Since in our case the maximum resolution is 97.5 km, we make an attempt to find out whether the dynamics are really happening due to physical resistivity or whether they are influenced by numerical resistivity. So, we carry out a comparative study of the thinning phase of the CS for zero physical resistivity. It is expected that if resistivity is smaller, the diffusion time scale of the CS will be larger. Smaller diffusion results in faster thinning of the CS subjected to the external perturbation. The thinning of the CS is found to be faster in the case of numerical resistivity only, i.e., zero physical resistivity,  compared to that for finite physical resistivity of \(2 \times 10^{-4}\) (shown in left panel of Figure \ref{label 8}). The increase in the maximum of $J$ also follows the same trend (shown in the middle panel of Figure \ref{label 8}). Similarly, the aspect ratio of the CS also increases faster in the case of numerical resistivity,  having a power law exponent in time of 3.10 in comparison to 2.97 in case of physical resistivity of \(2 \times 10^{-4}\) (shown in right panel of Figure \ref{label 8}). So the thinning process of the CS is slower in case of finite physical resistivity. Hence we can infer that the imposed physical resistivity is higher than the numerical resistivity for the resolution used in this simulation. So the dynamics reported in this paper are clearly driven by physical resistivity of \(2 \times 10^{-4}\) as opposed to numerical resistivity. Even though the internal substructuring within the plasmoids is resolved, the resolution and physical resistivity employed in this work preclude the appearance of some smaller-scale dynamics. This is expected to include, for example, the formation of small-scale plasmoids and their coalescence with larger plasmoids. 

To further justify our chosen value of imposed resitivity, we calculate the diffusion layer width (i.e., the width in which the diffusion is really important) to determine whether the inner layer of the tearing mode is fully resolved. We use \(\epsilon l\) as the estimate of diffusion layer width with \(\epsilon\) being calculated as \(R_{m}^{-1/4}\) \citep{2014masu.book.....P}. The magnetic Reynolds number (\(R_{m}\)) is measured to be of the order of \(10^{4}\) using the initial CS width, i.e., 3 Mm, and initial \(v_{A}\) and the imposed uniform physical resistivity. So we find that diffusion layer width is around 325 km. Since the maximum resolution in this present work is 97.5 km, we infer that there is are at least three grid points within the inner layer of tearing. Therefore the resolution of the simulation is capable to resolve the diffusion layer connected to the tearing mode. 
\section{Relevance of Initial Magnetic Field Configurations}

\begin{figure*}
\mbox{
\hspace{1.0 cm}
\includegraphics[height=6.5 cm, width=8 cm]{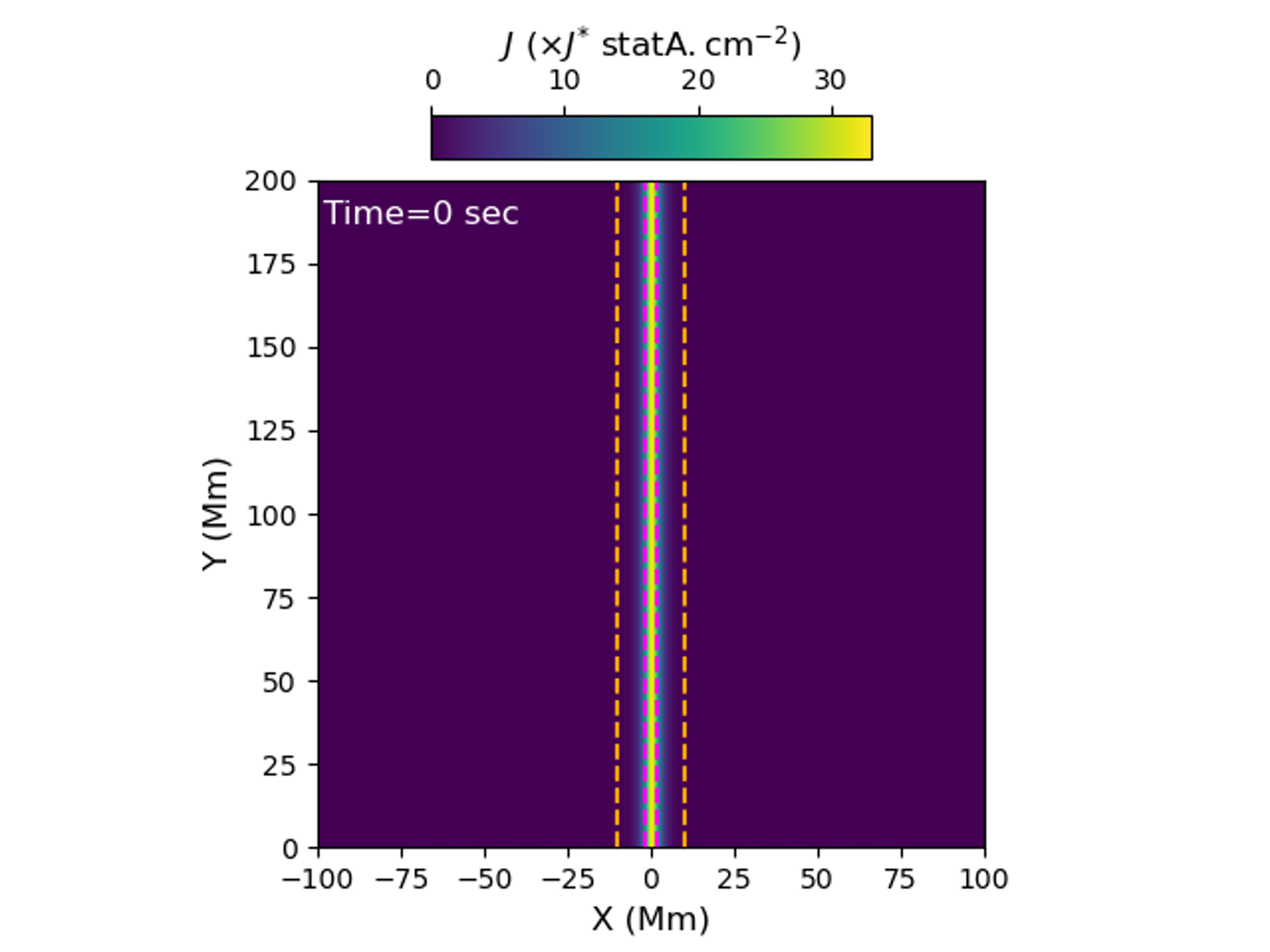}
\includegraphics[height=6.5 cm, width=6.5 cm]{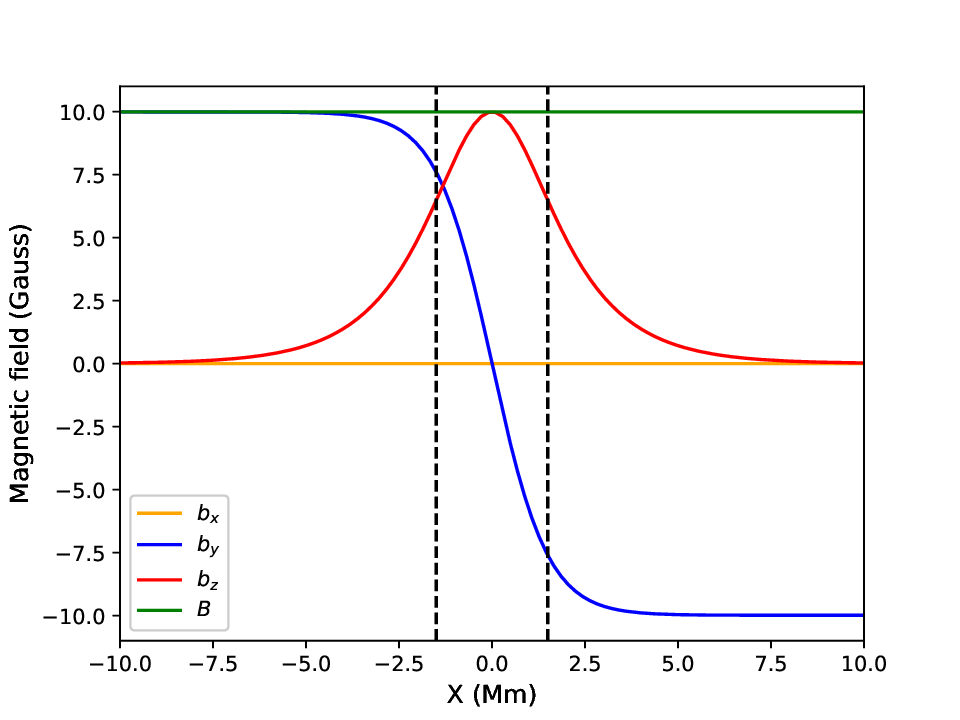}
}
\caption{Left: Map of current density ($J$) showing the CS as a region of higher $J$ whose extents are bounded by magenta dashed lines. The region within the pair of orange dashed lines is taken under consideration to show the profile of magnetic field magnitude and its components. Right: Profiles of magnetic field components. The red curve gives the profile of the guide field which shows that the guide field vanishes at a distance more than 5 Mm from the center of the CS. The pair of black dashed vertical lines denote the extent of the CS having a half-width of 1.5 Mm.} 
\label{label 9}
\end{figure*}

We are simulating externally driven or forced magnetic reconnection and associated dynamics in  solar coronal conditions. So we must ensure that the numerical setup is initially in magnetohydrostatic equilibrium before giving any external perturbations like velocity pulse in our case. We had two choices to establish an equilibrium initial state as follows-- 
\newline
[I] We may adopt the field of a normal simple current sheet with, say, a tanh profile for $b_y(x)$ balanced by a plasma pressure $p(x)$ without or with a uniform guide field $b_z$. 
\newline
[II] We may consider the same profile for $b_y(x)$ balanced by a force-free guide field $b_z(x)$ with pressure, density and magnetic field being uniform. 
\newline
Since in practice there is usually a strong guide field present in the corona, we decided to adopt a force-free current sheet with uniform plasma pressure instead of a magnetostatic current sheet with a strong central plasma pressure. Importantly, this incorporates the effect of a guide field inside the central part of the current sheet and extending quite some distance outside it. It is worth mentioning that the half width of the CS is 1.5 Mm whereas the guide field tends to zero at more than 5 Mm from the center of the CS, as shown in the right panel of Figure \ref{label 9}. Therefore non-zero guide field is present in the nearest vicinity of the CS in which all the dynamics are going on. Hence it ensures the force-freeness of the initial system for homogeneous and isotropic plasma parameters in the solar corona. A stronger guide field than considered here may be more relevant to many solar flares, and its inclusion can be carried out as a separate study in the future.

\section{Reason Behind Steady Behaviour of Centred Plasmoid}

\begin{figure*}

\mbox{
\hspace{-1.5 cm}
\includegraphics[height=6.5 cm,trim={3.0cm 0 0 0},clip]{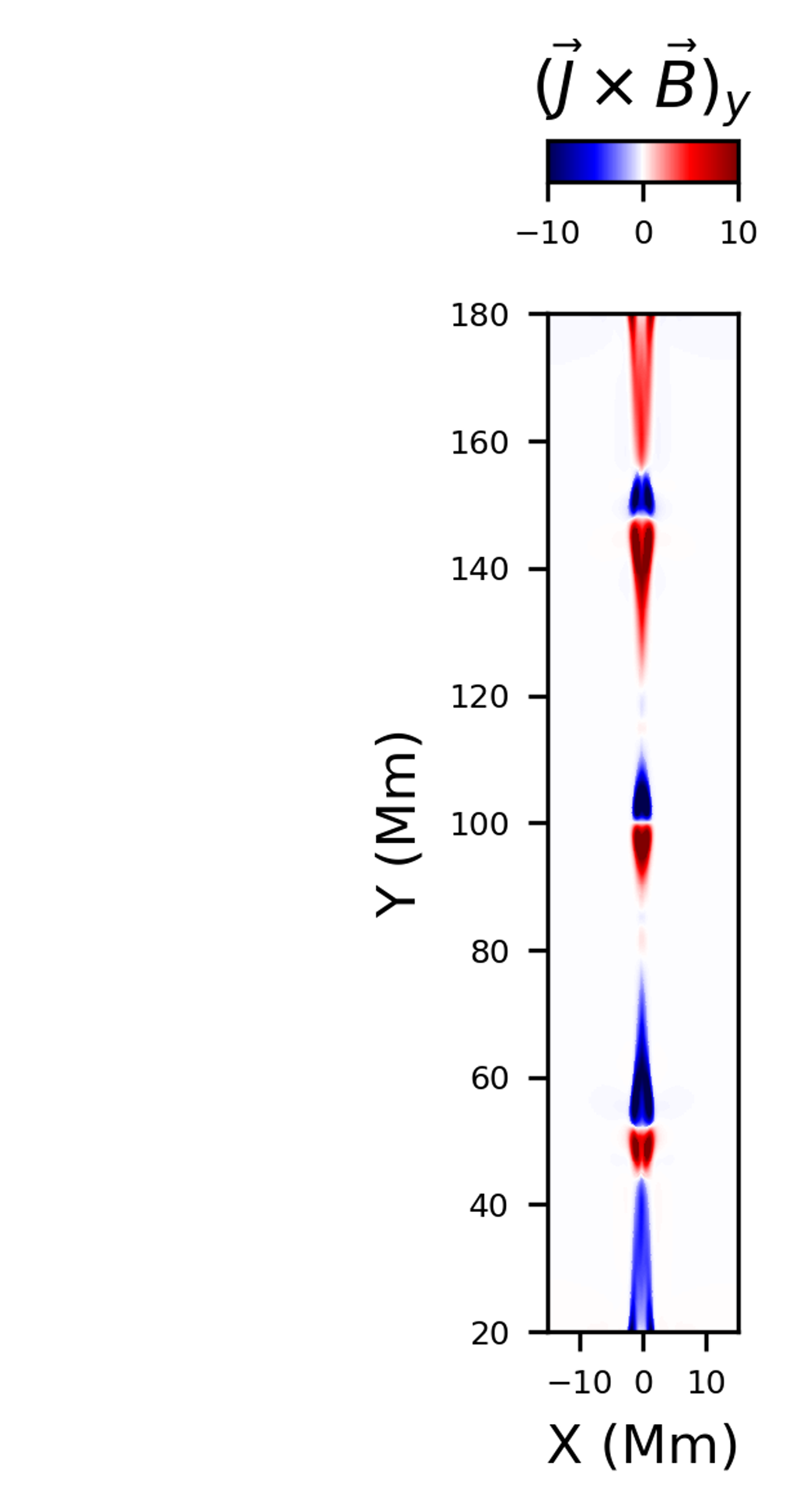}
\includegraphics[height=6.5 cm,trim={3.0cm 0 0 0},clip]{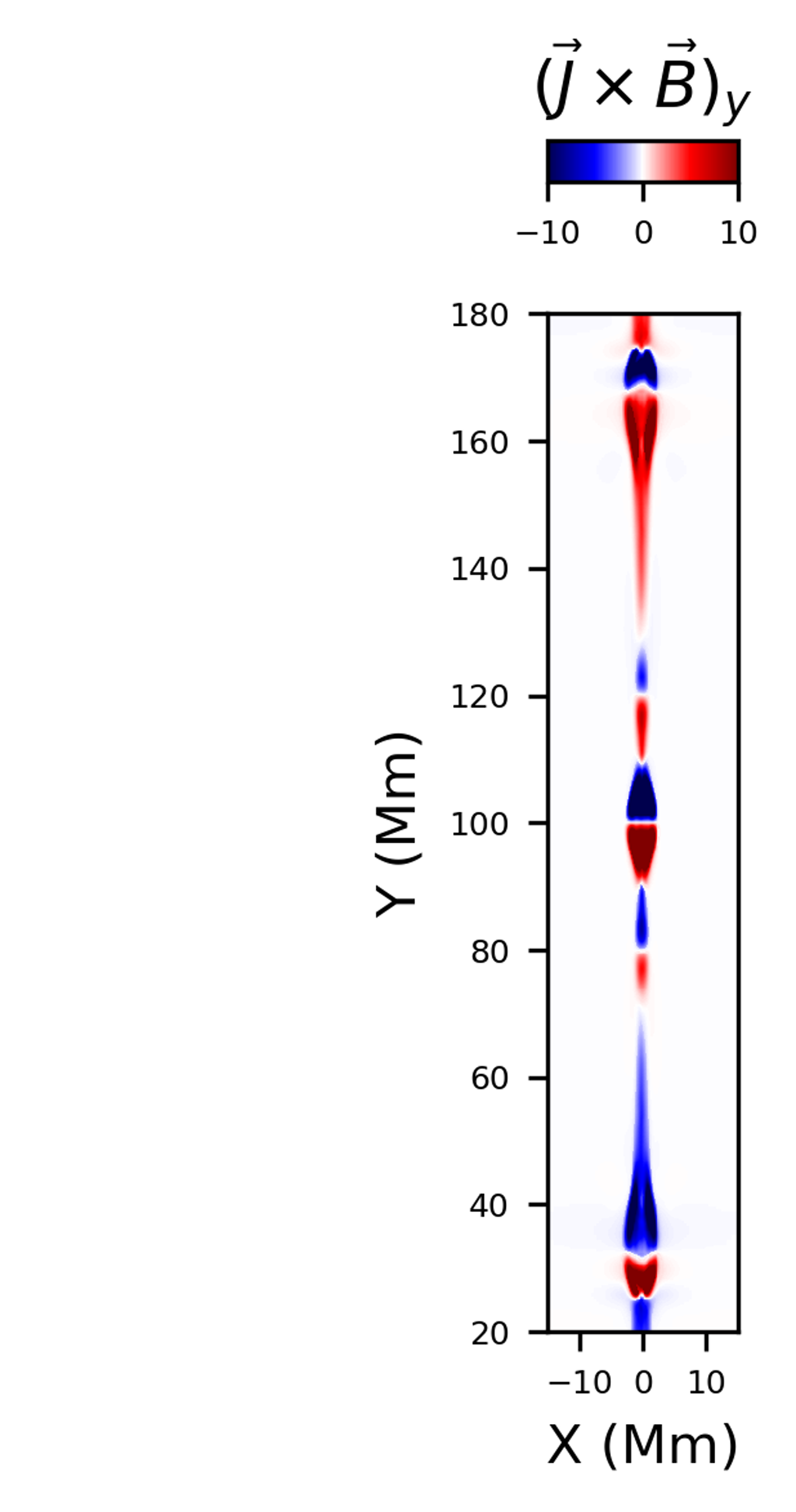}
\includegraphics[height=6.5 cm,trim={3.0cm 0 0 0},clip]{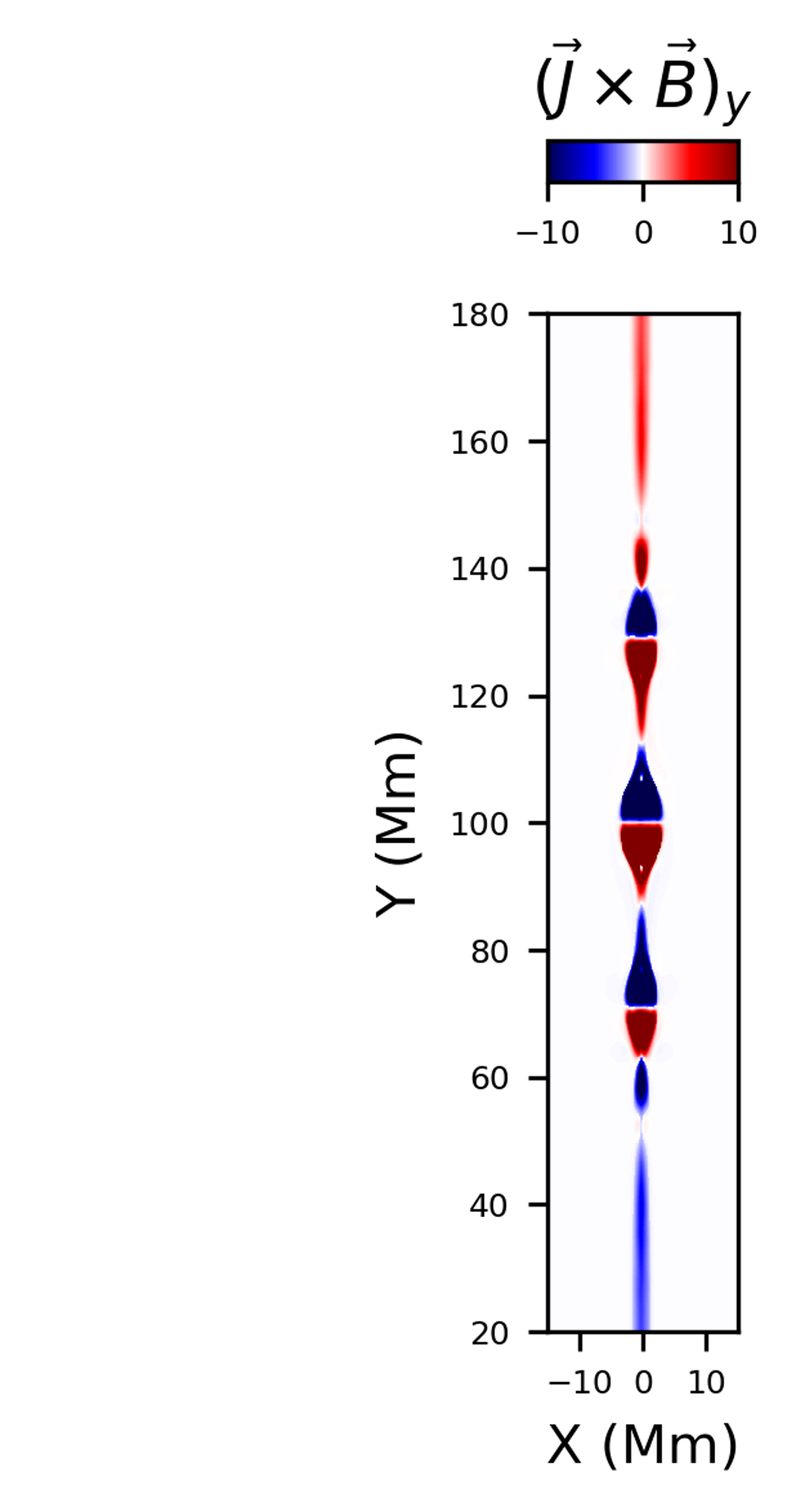}
\includegraphics[height=6.5 cm,trim={3.0cm 0 0 0},clip]{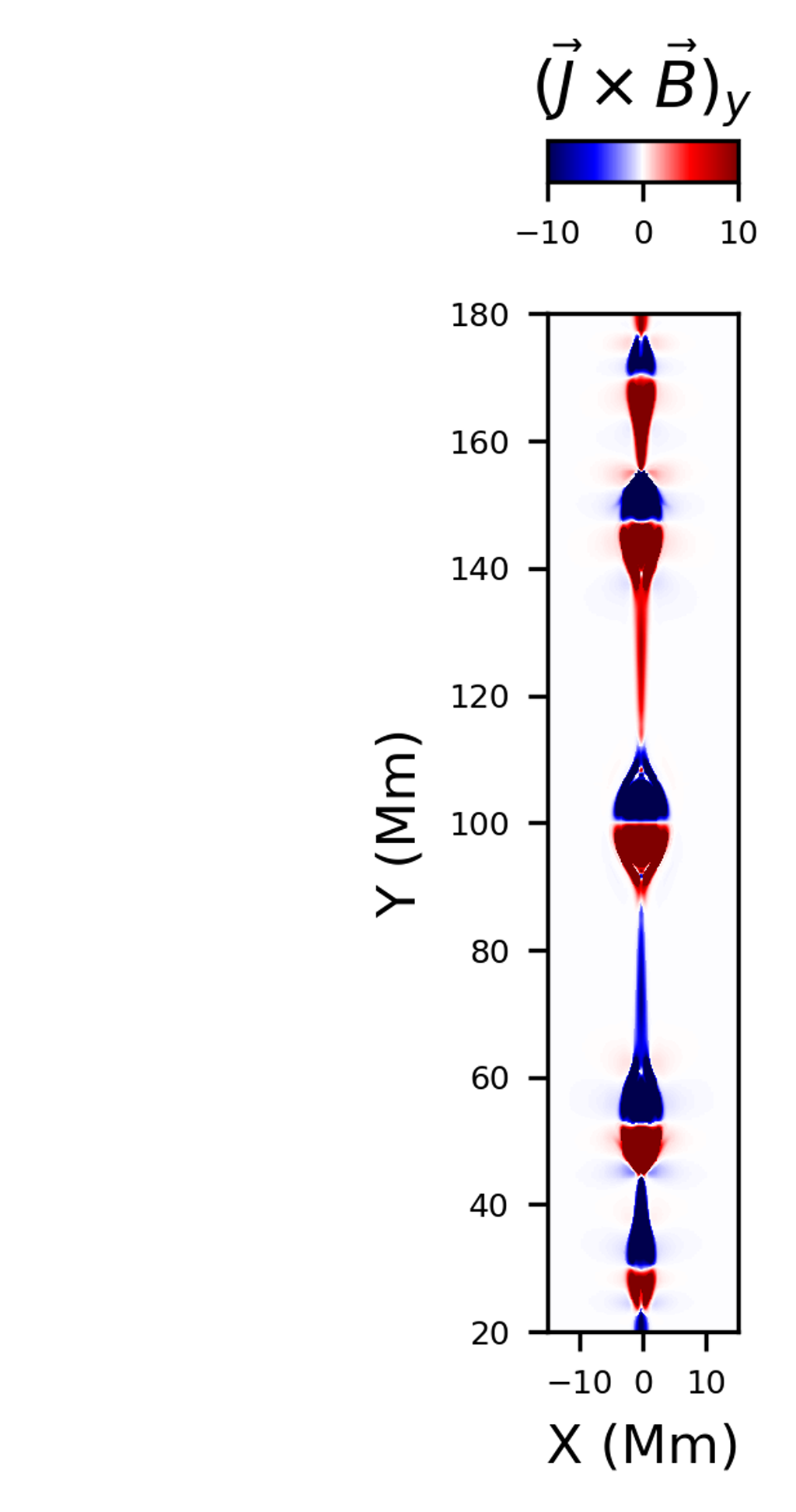}
\includegraphics[height=6.5 cm,trim={3.0cm 0 0 0},clip]{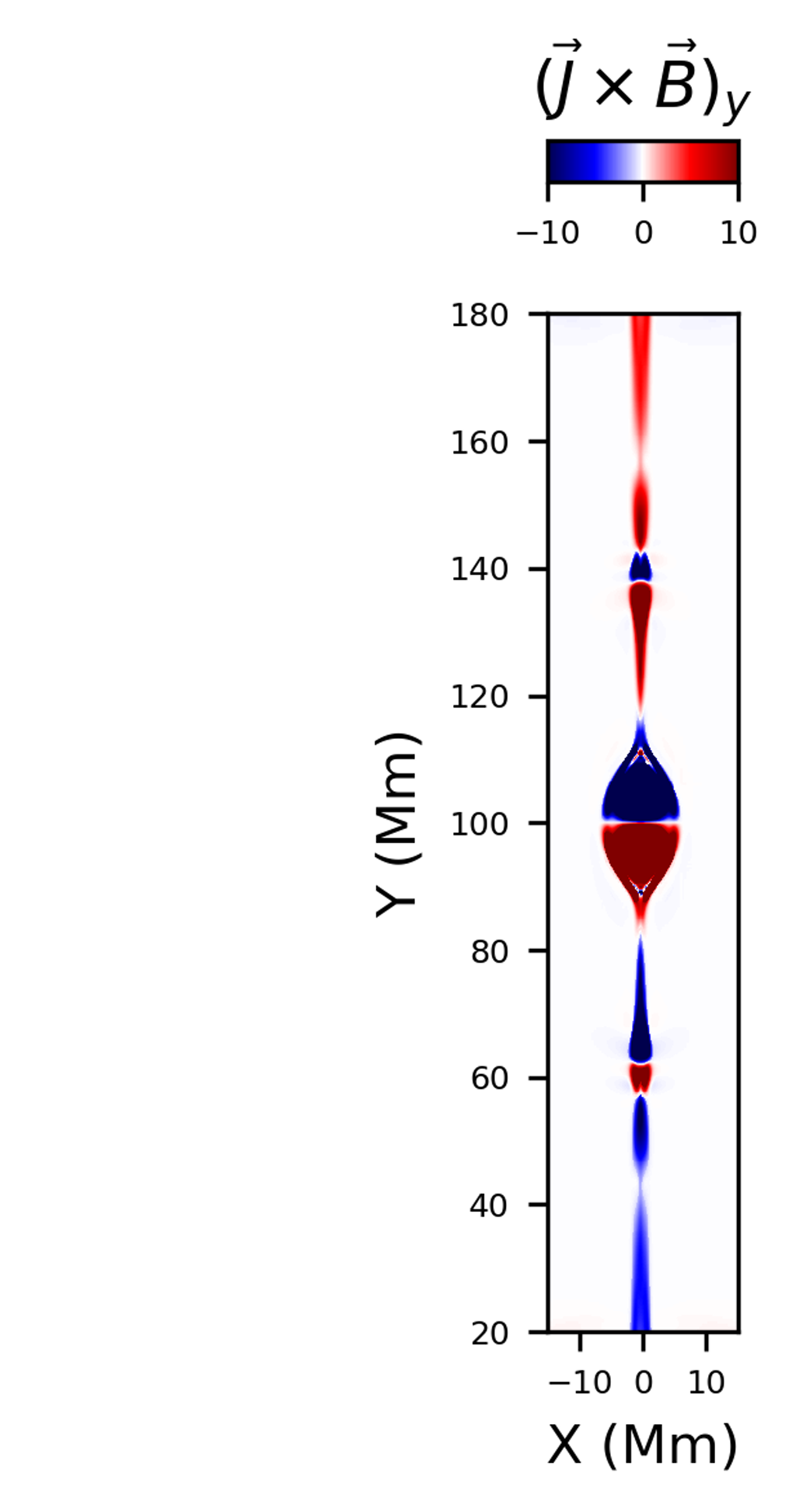}
\includegraphics[height=6.5 cm,trim={3.0cm 0 0 0},clip]{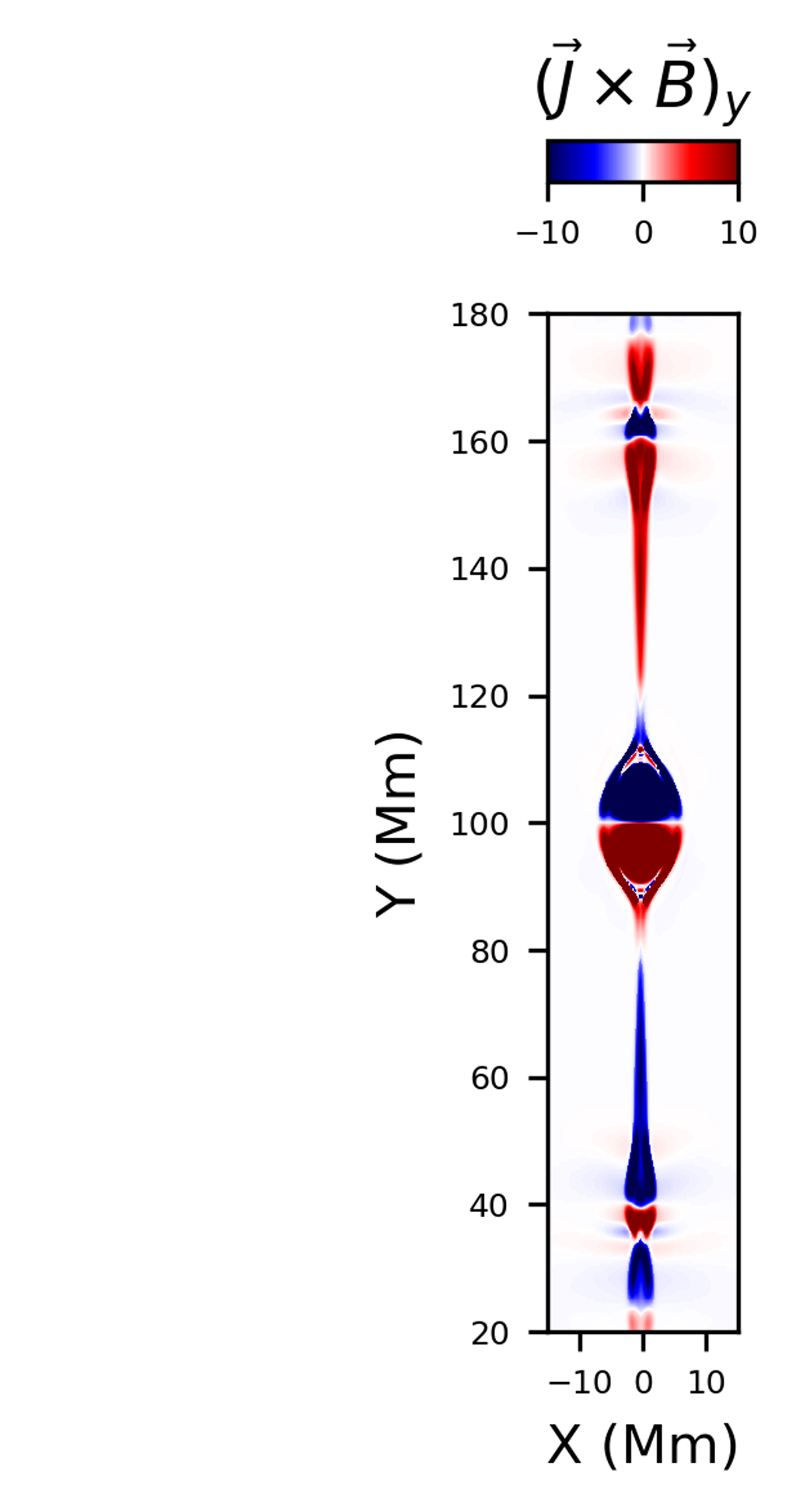}
}

\caption{Snapshots of the $y$-component of the Lorentz force \((\vec{J}\times\vec{B})_{y}\) at times 733, 781, 866, 938, 1130 and 1178 seconds of the resistive MHD simulation. It is evident that \((\vec{J}\times\vec{B})_{y}\) has same direction above and below the off-centred plasmoids which drives the plasmoids out. But the direction of \((\vec{J}\times\vec{B})_{y}\) just above and below the central plasmoid is oppositely directed giving no resultant net force. The entire evolution of this quantity from 601 to 1178 seconds is available as animation in the online HTML version. The real-time animation duration is 5 seconds. Green rectangular box is used in the animation to show the FOV which is exhibited in this figure.}
\label{label 10}
\end{figure*}

For the present case, the initial magnetic field does not possess any complex asymmetrical geometry. In addition, the initial pulse that we imposed is symmetrical in the Y-direction. It originates from $y =100$ Mm with an isotropic standard deviation (SD), i.e., direction independent SD. The magnetic field lines are purely vertically directed without any curvature or discontinuity except at the CS itself. Therefore, the pulse interacts with the CS in a symmetrical manner about the midpoint at y = 100 Mm. Now the dynamics of the plasmoids -- i.e., whether it will move or remain steady -- will depend upon the resultant force acting on it. So we investigate the y-component of Lorentz force during different stages of plasmoid dynamics, as exhibited in Figure \ref{label 10} and its animated version.

From Figure \ref{label 10}, it can be noticed that in the cases of plasmoids moving along the CS, the y-component of Lorentz force just above and below the plasmoids have the same orientation, i.e., in the outward direction. On the contrary, the y-component of Lorentz force has much lower magnitude in the close proximity of the centred plasmoid. Besides those are directed in opposite direction at the opposite extents of the centred plasmoid. Hence the net resultant force on the centred plasmoid is almost zero or relatively negligible keeping it steady at the middle of the CS during its growth with time. So it is the initial symmetric magnetic configuration without any curvature leading to isotropic propagation of initial velocity pulse before its interaction with the CS which results in this steady growth of the centred plasmoid. However, it should be noted that in some previous numerical models, by imposing random perturbations the central monster plasmoid is not formed \citep[e.g.,][]{2010PhPl...17f2104H,2012PhRvL.109z5002H,2013PhPl...20e5702H,2015shin.confE..26H,2016ApJ...818...20H}. It will be a matter of future investigation that how  the plasmoids will be formed under the initial perturbation that is not completely symmetric, and when the non-ideal plasma conditions are switched-on.





\end{document}